\documentclass{cernrep}
\usepackage{latexsym,graphicx,epsfig,psfrag,here}
\usepackage{bbm}
\usepackage{pifont} 
\newcommand{\no}{\nonumber}
\newcommand{\nn}{\nonumber \\}
\newcommand{\cL}{{\cal L}}
\newcommand{\cM}{{\cal M}}
\newcommand{\lra}{\longrightarrow}
\newcommand{\ve}{\varepsilon}

\newcommand{\dg}{\dagger}

\newcommand{\st}{\stackrel}

\newcommand{\IM}{\mbox{\rm Im}}
\newsavebox{\fmbox}
\newenvironment{fmpage}[1]
  {\begin{lrbox}{\fmbox}\begin{minipage}{#1}}
  {\end{minipage}\end{lrbox}\fbox{\usebox{\fmbox}}}
\begin{document}
\begin{flushright}
UWThPh-2006-9
\end{flushright}
\title{QUANTUM CHROMODYNAMICS}
\author{G. Ecker}
\institute{Inst. Theor. Physics, Univ. of Vienna, Austria}
\maketitle

\begin{abstract}
After a brief historical review of the emergence of QCD as the quantum 
field theory of strong interactions, the basic notions of colour and
gauge invariance are introduced leading to the QCD Lagrangian. The
second lecture is devoted to perturbative QCD, from tree-level
processes to higher-order corrections in renormalized perturbation
theory, including jet production in $\displaystyle{e^+ e^-}$ 
annihilation, hadronic
$\tau$ decays and deep inelastic scattering. The final two lectures  
treat various aspects of QCD beyond perturbation theory. The main
theme is effective field theories, from heavy quarks to the light
quark sector where the spontaneously broken chiral symmetry of QCD
plays a crucial role.
\end{abstract}

\vspace*{.4cm} 
\begin{center}
Table of contents
\end{center} 
\contentsline {section}{\numberline {1}\upshape Introduction}{2}
\contentsline {subsection}{\numberline {1.1}Historical background}{2}
\contentsline {subsection}{\numberline {1.2}Colour}{5}
\contentsline {subsection}{\numberline {1.3}Gauge invariance}{6}
\contentsline {subsection}{\numberline {1.4}$SU(3)_c$ and the QCD 
Lagrangian}{7}
\contentsline {section}{\numberline {2}\upshape Perturbative QCD}{10}
\contentsline {subsection}{\numberline {2.1}QCD at tree level}{10}
\contentsline {subsection}{\numberline {2.2}Higher-order corrections 
and renormalization}{13}
\contentsline {subsection}{\numberline {2.3}Measurements of $\alpha
_s$}{16} 
\contentsline {subsection}{\numberline {2.4}Hadronic $\tau $ decays}{18}
\contentsline {subsection}{\numberline {2.5}Deep inelastic scattering}{19}
\contentsline {section}{\numberline {3}\upshape Heavy and light quarks}{25}
\contentsline {subsection}{\numberline {3.1}Effective field theories}{25}
\contentsline {subsection}{\numberline {3.2}Heavy quarks}{26}
\contentsline {subsection}{\numberline {3.3}QCD sum rules}{29}
\contentsline {subsection}{\numberline {3.4}Chiral symmetry}{33}
\contentsline {subsection}{\numberline {3.5}Chiral perturbation theory}{37}
\contentsline {subsection}{\numberline {3.6}Light quark masses}{40}
\contentsline {subsection}{\numberline {3.7}Pion pion scattering}{42}
\contentsline {subsection}{\numberline {3.8}$K_{l3}$ decays and $V_{us}$}{43}
\contentsline {section}{\numberline {4}\upshape Summary and
epilogue}{46}
\contentsline {section}{\upshape References}{46}

\newpage
\section{Introduction}
Why do we still study QCD after more than 30 years?
\begin{itemize} 
\item[$\bullet$] By decision of the Nobel Prize Committee in 2004 
\cite{Nobel04}, QCD is the correct theory of the
strong interactions.
\item[$\bullet$] The parameters of QCD, the coupling strength
$\alpha_s$ and the quark masses, need to be measured as precisely as
possible. 
\item[$\bullet$] Electroweak processes of hadrons necessarily involve
the strong interactions.
\item[$\bullet$] In searches for new physics at present and future
accelerators, the ``QCD background'' must be understood quantitatively.
\item[$\bullet$] Although QCD is under control for high-energy
processes, many open questions remain in the nonperturbative domain
(confinement, chiral symmetry breaking, hadronization, \dots)
\item[$\bullet$] Last but not least, QCD is a fascinating part of
modern physics. The lectures will therefore start with a brief
historical review of the developments in particle physics in the
sixties and early seventies of the last century.
\end{itemize} 

\noindent 
The following lectures were given to an audience of young experimental
particle physicists. Although the lectures emphasize some of the
theoretical aspects of QCD, the mathematical level was kept
reasonably low. The first two lectures cover the basics of QCD,
from the concepts of colour and gauge invariance to some
applications of perturbative QCD. The last two lectures treat
aspects of QCD beyond perturbation theory. The main theme is effective
field theories, from heavy quarks to the light quark sector where the
spontaneously broken chiral symmetry plays a crucial role.

\subsection{Historical background}
\label{sec:hist}
Particle physics in the early sixties of the last century was not in a
very satisfactory state. Only for the electromagnetic interactions of
leptons a full-fledged quantum field theory (QFT) was available. Quantum
electrodynamics (QED) produced increasingly precise predictions that
were confirmed experimentally. Nevertheless, the methodology of
renormalization, an essential aspect of the perturbative treatment of
QED, was not universally accepted. Even among the founding fathers of
QFT, the dissatisfaction with ``sweeping the
infinities under the rug'' was widespread. At the Solvay Conference of
1961, Feynman confessed \cite{Feynman} that he did not ``subscribe to 
the philosophy of renormalization''. 

What is the essence of this controversial procedure of renormalization
that has turned out to be crucial for the shaping of QCD and of
the Standard Model altogether? Specializing to QED for definiteness,
three main steps are important.
\begin{itemize} 
\item Amplitudes $A(p_i;e_0,m_0;\Lambda)$ depend on the momenta $p_i$ 
of the particles involved, on the parameters $e_0,m_0$ of the QED 
Lagrangian and on a cutoff $\Lambda$ that cuts off the high-momentum
modes of the theory. The cutoff is essential because
$A(\displaystyle{p_i};e_0,m_0;\Lambda)$ diverges for $\Lambda \to
\infty$, rendering the result meaningless.
\item With the help of measurable quantities (cross sections, particle
four-momenta) one defines physical parameters $e(\mu),m(\mu)$ that
depend in general on an arbitrary renormalization scale $\mu$. One 
then trades  $e_0,m_0$ for the physical $e(\mu),m(\mu)$ to a given
order in perturbation theory.
\item The limit $\displaystyle\lim_{\Lambda \to \infty} \, 
A(\displaystyle{p_i};e_0(e,m,\Lambda),m_0(e,m,\Lambda);\Lambda) =
\hat{A}(\displaystyle{p_i};e(\mu),m(\mu))$ is now finite and unambiguous for 
the chosen definitions of $e(\mu),m(\mu)$.
\end{itemize} 
Based on this procedure, the agreement between theory and experiment
was steadily improving.

For the weak interactions, the Fermi theory (in the $V-A$ version) was
quite successful for weak decays, but     
\begin{itemize} 
\item[$\bullet$] higher-order corrections were not calculable;
\item[$\bullet$] for scattering processes the theory became
inconsistent for energies $E \gtrsim 300$ GeV (unitarity problem).
\end{itemize}
The rescue came at the end of the sixties in the form of the
electroweak gauge theory of the Standard Model.

Of all the fundamental interactions, the strong interactions were in
the most deplorable state. Although the rapidly increasing number of
hadrons could be classified  successfully by the quark model of
Gell-Mann and Zweig \cite{QM}, the dynamics behind the quark model was
a complete mystery. A perturbative treatment was clearly hopeless and
the conviction gained ground that QFT might not be
adequate for the strong interactions.

This conviction was spelled out explicitly by the proponents of
the bootstrap philosophy (Chew et al.). Under the banner of nuclear
democracy, all hadrons were declared to be equal. Instead of looking
for more fundamental constituents of hadrons, the S-matrix for strong 
processes was investigated directly without invoking any quantum field
theory. Although the expectations were high, nuclear democracy shared
the fate of the student movement of the late sixties: the promises
could not be fulfilled. 

A less radical approach assumed that QFT could still
be useful as a kind of toy model. The main proponent of this approach
was Gell-Mann who suggested to abstract algebraic relations from
a Lagrangian field theory model but then throw away the model (``French
cuisine program'' \cite{cuisine}). The usefulness of this approach had
been demonstrated by Gell-Mann himself: current algebra and the quark
model were impressive examples. Until the early seventies,
Gell-Mann took his program seriously in declaring the quarks to be
purely mathematical entities without any physical reality, a view
shared by many particle physicists of the time.

The decisive clue came from experiment. Started by the MIT-SLAC
collaboration at the end of the sixties, deep inelastic scattering of 
leptons on nucleons and nuclei produced
unexpected results. Whereas at low energies the cross sections were
characterized by baryon resonance production, the behaviour at large
energies and momentum transfer was surprisingly simple: the nucleons
seemed to consist of noninteracting partons (Feynman). Obvious
candidates for the partons were the quarks but this idea led
to a seeming paradox. How could the quarks be quasi-free at high
energies and yet be permanently bound in hadrons, a low-energy
manifestation?

That the strength of an interaction could be energy dependent was not
really new to theorists. In QED, the vacuum acts like a
polarisable medium leading to the phenomenon of charge screening. 
However, contrary to what the deep inelastic experiments seemed
to suggest for the strong interactions, the effective charge in QED 
increases with energy: QED is ultraviolet unstable.

To understand the phenomenon of an energy dependent interaction, we
consider the dimensionless ratio of cross sections
\begin{equation}
R_{e^+e^-} = \displaystyle\frac{\sigma(e^+ + e^- \to {\rm hadrons})}
{\sigma(e^+ + e^- \to \mu^+ + \mu^-)}~.
\end{equation}
Beyond the leading-order value $R_0$ (cf. Sec.~\ref{sec:QCDtree}), one
finds to lowest order in the strong coupling constants $g_s$:
\begin{equation}
R_{e^+e^-} = R_0 \left(1 + \displaystyle\frac{g_s^2}{4\pi^2} \right)~.
\end{equation}
The general form to any order in $g_s$ (neglecting quark masses) is
\begin{equation}
R_{e^+e^-}=R_{e^+e^-}(E,\mu,g_s(\mu))
\end{equation}
where $E$ is the center-of-mass energy and $\mu$ is the
renormalization scale. Since $\displaystyle{R_{e^+e^-}}$ is a
measurable quantity, it must be independent of the  arbitrary scale
$\mu$: 
\begin{eqnarray}
\mu \displaystyle\frac{d}{d\mu}R_{e^+e^-}(E,\mu,g_s(\mu))=0 \quad &\lra& 
\quad \left(\mu
\displaystyle\frac{\partial}{\partial \mu} + \beta(g_s)
\displaystyle\frac{\partial}{\partial g_s} \right) R_{e^+e^-}=0~,
\end{eqnarray}
with the beta function
\begin{equation}
\beta(g_s) = \mu \displaystyle\frac{d g_s(\mu)}{d\mu}~.
\end{equation}
Dimensional analysis tells us that the dimensionless ratio 
$\displaystyle{R_{e^+e^-}}$ must be of the form
\begin{equation}
R_{e^+e^-}(E,\mu,g_s(\mu)) = f(\displaystyle\frac{E}{\mu},g_s(\mu))~.
\end{equation}
The seemingly uninteresting dependence on $\mu$ can therefore be
traded for
the dependence on energy or on the dimensionless ratio $z=E/\mu$\,:
\begin{equation}
\left(- \displaystyle\frac{\partial}{\partial \log z} +
\beta(g_s) \displaystyle\frac{\partial}{\partial g_s} \right) 
f(z,g_s(\mu)) = 0 ~.
\end{equation}
The general solution of this renormalization group equation is
\begin{equation}
f(z,g_s(\mu)) = \hat f(\overline{g_s}(z,g_s))~, 
\end{equation}
i.e., a function of a single variable, the energy dependent (running)
coupling constant $\overline{\displaystyle{g_s}}(z,g_s)$ satisfying
\begin{equation}
\displaystyle\frac{\partial \overline{g_s}}{\partial \log
    z}=\beta(\overline{g_s})
\label{eq:beta_gs}
\end{equation}
with the boundary condition $\overline{g_s}(1,g_s)=g_s$. For any gauge
coupling, the leading one-loop result for the $\beta$ function is
\begin{equation}
\beta(x) = -\displaystyle\frac{\beta_0}{(4\pi)^2} x^3 
\label{eq:b0}
\end{equation}
implying
\begin{equation}
\overline{g_s}^2(E/\mu,g_s(\mu)) =
\displaystyle\frac{g_s^2(\mu)}{1+
\displaystyle\frac{\beta_0}{(4\pi)^2}g_s^2(\mu) \log E^2/\mu^2} ~.
\end{equation}
Expanding the denominator, we observe that the renormalization
group equation has allowed us to sum the leading logs
$\left(g_s^2(\mu) \log E^2 /\mu^2 \right)^n$ of all orders in
perturbation theory. Even more importantly, the energy dependence of
the running coupling constant is determined by the sign of
$\displaystyle{\beta_0}$ in Eq.~(\ref{eq:b0}):
\begin{center} 
\begin{fmpage}{0.7\textwidth}
\begin{center}
\vspace*{.1cm}   
\begin{tabular}{lll}
$\beta_0 < 0$: & \hspace*{1cm} $\displaystyle\lim_{E\to 0} \overline{g}(E) =
  0$ & \hspace*{.5cm} infrared stable (QED) \\[.2cm]
$\beta_0 > 0$: & \hspace*{1cm} $\displaystyle\lim_{E\to \infty}
  \overline{g}(E) = 0$ & \hspace*{.5cm} ultraviolet stable (QCD)
\vspace*{.1cm}   
\end{tabular} 
\end{center} 
\end{fmpage} 
\end{center} 
For the cross section ratio $R_{e^+e^-}$ we get finally
\begin{equation}
R_{e^+e^-} = R_0 \left(1 + \displaystyle\frac{g_s^2(E)}{4\pi^2} + 
O(g_s^4(E)) \right)
\end{equation}
in terms of $g_s^2(E)\equiv \overline{g_s}^2(E/\mu,g_s(\mu))$. 

The crucial question in the early seventies was therefore whether
QFT was compatible with ultraviolet stability
(asymptotic freedom)? The majority view was expressed in a paper by
Zee \cite{Zee:1973gn}: `` \dots we conjecture that there are no
asymptotically free quantum field theories in four dimensions.'' While
Coleman and Gross set out to prove that conjecture their graduate
students Politzer and Wilczek (together with Gross) tried to close a
loophole: the $\beta$ function for nonabelian gauge theories
(Yang-Mills theories) was still unpublished and probably unknown to
everybody except t'Hooft. In the spring of 1973, the Nobel prize
winning work of Politzer and Gross and Wilczek \cite{Gross:1973id}
demonstrated that
Yang-Mills theories are indeed asymptotically free.

The crucial difference between QED and QCD is that photons are
electrically neutral whereas the gluons as carriers of the strong
interactions are coloured. Further physical insight can be obtained by
taking up an analogy with the electrodynamics of continuous media
\cite{Nielsen:1980sx}. Because of Lorentz invariance, the vacuum of a
relativistic QFT is characterized by
\begin{equation}
\ve \mu =1
\label{eq:epsmu}
\end{equation}
for the product of permittivity $\ve$ and permeability $\mu$. In QED
charge screening implies $\ve > 1$ so that the vacuum of QED acts like
a diamagnet ($\mu < 1$). In QCD the colour charge screening of quarks
($\ve > 1$) is overcompensated by gluons (spin 1) acting as permanent 
colour dipoles ($\mu > 1$). Because
\begin{equation}
\beta_0^{\rm QCD} = \displaystyle\frac{1}{3}\left( 11 N_c -  2 N_F
\right)~, 
\end{equation}
the QCD vacuum is a (colour) paramagnet for $N_F < 11\,N_c/2 < 17$
quark flavours (for $N_c=3$). Because of the general relation
(\ref{eq:epsmu}) this can also be interpreted as anti-screening
($\ve < 1$). 

The existence of three colours was already widely accepted at that
time. Gell-Mann and collaborators had been investigating a model of
coloured quarks interacting via a singlet gluon (not asymptotically
free). In a contribution of Fritzsch and Gell-Mann
in the Proceedings of the High Energy Conference in Chicago in 1972
\cite{Fritzsch:1972jv}
one finds the probably first reference to nonabelian gluons: ``Now the
interesting question has been raised lately whether we
should regard the gluons as well as the quarks as being 
non-singlets with respect to colour (J. Wess, private
communication to B. Zumino).'' Although Gell-Mann is generally credited
for the name QCD, the first published occurrence of QCD is much less
known (cf., e.g., Refs.~\cite{Nobel04}). My own investigation of the
early literature has produced a footnote in a paper of Fritzsch,
Gell-Mann and Minkowski  in 1975 \cite{Fritzsch:1975sr} suggesting 
``A good name for this theory is quantum chromodynamics.''

\subsection{Colour}
\label{sec:col}
Already before the arrival of QCD, there were a number of indications
for the colour degree of freedom.
\begin{dinglist}{108} 
\item Triality problem \\[.1cm]
In the original quark model often called the naive quark model, 
the three quarks
$u$, $d$, $s$ give rise to mesonic bound states of the form
$\overline{q}q$. All of the nine expected bound states had already
been observed suggesting an attractive force between all quarks and
antiquarks. The baryons fit nicely into $q q q$ bound states. If the
strong force is purely attractive why do antiquarks not bind to
baryons? The resulting objects of the form $\overline{q} q q q$
would have fractional charge and have never been
observed. Introducing three colours for each quark and antiquark
allows for $9 \times 9=81$
combinations of $\overline{q}q$ only nine of which had been
found. The remaining 72 combinations are not bound states invalidating
the previous argument.
\item Spin-statistics problem \\[.1cm]
Consider the state
\begin{equation} 
|\Delta^{++}(S_z=3/2)\rangle \sim |\,u \uparrow
\,u \uparrow \,u \uparrow\rangle ~.
\label{eq:Delta}
\end{equation}
Since the spin-flavour content is completely symmetric, Fermi
statistics for quarks seems to require an antisymmetric spatial wave
function. On the other hand, for every reasonable potential the ground
state is symmetric in the space variables. Colour solves this
problem because the state (\ref{eq:Delta}) is totally antisymmetric in 
the colour indices respecting the generalized Pauli principle with a
spatially symmetric wave function.
\item Renormalizability of the Standard Model \\[.1cm]
With the usual charges of quarks and leptons, the Standard Model is
a consistent gauge invariant QFT only if there are
three species of quarks in order to cancel the so-called gauge
anomalies.
\item $\pi^0 \to 2\,\gamma$ decay \\[.1cm]
The by far dominant contribution to the decay amplitude is due to the
chiral anomaly (exact for massless quarks). The observed rate can only
be understood if there are again three species of quarks.
\item Momentum balance in deep inelastic scattering \\[.1cm]
The momentum sum rule indicates that only about 50 \% of the nucleon
momentum is carried by valence quarks. The remainder is mainly
carried by gluons.
\item Quark counting \\[.1cm]
After the advent of QCD, many more direct confirmations of the colour
degree of freedom were obtained. One of the first 
confirmations was provided by the total cross section 
$\displaystyle{\sigma(e^+ + e^-} \to {\rm hadrons})$ already 
discussed in the previous subsection.
\item Hadronic $\tau$ decays \\[.1cm]
As we shall discuss in the next lecture, hadronic $\tau$ decays not
only give clear evidence for $\displaystyle{N_c=3}$ but they also 
provide an excellent opportunity for extracting 
$\displaystyle{\alpha_s=g_s^2/4\pi}$. \\
\parbox[c]{0.45\textwidth}{
\begin{figure}[H]
\centering\includegraphics[width=.9\linewidth]{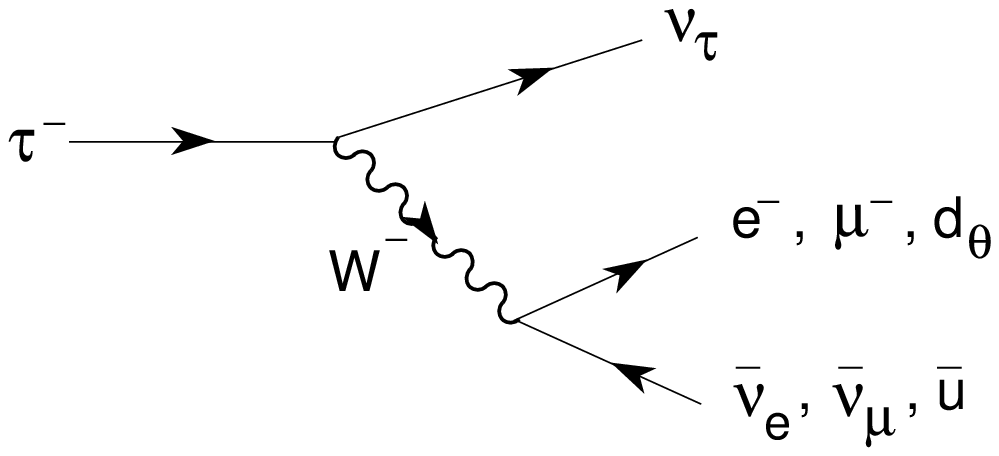}
\caption{Feynman diagram for $\tau^- \to \nu_\tau + X$.}
\label{fig:tau1}
\end{figure}
}

\vspace*{-4cm}
\hspace*{6.5cm} 
\parbox[c]{0.55\textwidth}{
\begin{eqnarray} 
R_\tau &=& \displaystyle\frac{\Gamma(\tau^- \to \nu_\tau + {\rm
    hadrons})}{\Gamma(\tau^- \to \nu_\tau e^- \overline{\nu_e})}\\
&=& N_c \left(|V_{ud}|^2+|V_{us}|^2 \right)\left(1 +
    O(\alpha_s)\right) ~.\no
\end{eqnarray}  
}

\end{dinglist} 
   
\vspace*{.8cm}

\subsection{Gauge invariance}
Gauge invariance is a main ingredient not only of QCD but of the
Standard Model as a whole. We start with the Lagrangian for a single 
free Dirac fermion:
\begin{equation}
\cL_0= \overline{\psi}(x)\,i\,\slashed{\partial} \, \psi(x) - m\,
\overline{\psi}(x)\, \psi(x)~, \hspace*{2cm} \slashed{a}:=
\gamma^\mu a_\mu ~. 
\label{eq:Dirac}
\end{equation}
This Lagrangian and the resulting Dirac equation are invariant under 
a phase transformation (global $U(1)$)
\begin{equation} 
\psi(x) \lra \psi^\prime(x)=e^{-\displaystyle{iQ\,\ve}}\psi(x)
\label{eq:gt}
\end{equation} 
with $Q \ve$ an arbitrary real constant. One may now pose the
question whether the phase in the
transformation law (\ref{eq:gt}) must really be the same here and
``behind the moon'', as is the case in (\ref{eq:gt}) with a space-time
independent phase $Q \ve$. Instead of
experimenting behind the moon, we replace the constant $\ve$ with an
arbitrary real function $\ve(x)$ and see what happens. As is easily
checked, the mass term in (\ref{eq:Dirac}) remains invariant but not the
kinetic term because
\begin{equation}
\partial_\mu \psi(x) \lra e^{-\displaystyle{iQ\ve(x)}}\left(\partial_\mu
- i Q \partial_\mu \ve(x)  \right)\psi(x)~.
\label{eq:obn}
\end{equation}
The conclusion is that the phase must indeed be the
same here and behind the moon for the theory to be invariant under
transformations of the form (\ref{eq:gt}).

However, there is a well-known procedure for enforcing local
invariance, i.e. invariance for a completely arbitrary space-time
dependent phase $\ve(x)$. We enlarge the theory by introducing a
spin-1 vector field $A_\mu$ that has precisely the right
transformation property to cancel the obnoxious piece in
(\ref{eq:obn})  with
$\partial_\mu \ve(x)$. The idea is to replace the ordinary derivative
$\partial_\mu$ by a covariant derivative $D_\mu$:  
\begin{eqnarray} 
D_\mu \psi(x) &=& \left(\partial_\mu + iQ A_\mu(x) \right) \psi(x)~,
\end{eqnarray}
with
\begin{eqnarray}
A_\mu(x) \lra A^\prime_\mu(x) &=& A_\mu(x) + \partial_\mu \ve(x)~.
\end{eqnarray} 
It is easy to check that $D_\mu \psi$ transforms covariantly, 
\begin{equation} 
D_\mu \psi(x) \lra (D_\mu \psi)^\prime(x) =
e^{-\displaystyle{iQ\ve(x)}} D_\mu \psi(x) ~,
\end{equation}
so that the enlarged Lagrangian
\begin{eqnarray} 
\cL = \overline{\psi}(x)\left(i\,\slashed{D} -m\right) \psi(x)=
\cL_0 - Q A_\mu(x) \overline{\psi}(x)\,\gamma^\mu \psi(x)
\end{eqnarray} 
is invariant under local $U(1)$ transformations (gauge invariance).

The most important feature of this exercise is that the requirement of
gauge invariance has generated an interaction between the fermion
field $\psi$ and the gauge field $\displaystyle{A_\mu}$. Introducing a 
kinetic term for the gauge field to promote it to a propagating 
quantum field, the full Lagrangian 
\begin{equation}
\cL = \overline{\psi}\left(i\,\slashed{D} -m\right) \psi -
\displaystyle\frac{1}{4} F_{\mu\nu} F^{\mu\nu} 
\end{equation}
is still gauge invariant because the field strength tensor 
$F_{\mu\nu}=\partial_\mu A_\nu -\partial_\nu A_\mu$ is
automatically gauge invariant. Setting $Q=-e$ for the electron field,
we have ``deduced'' QED from the free electron theory and the
requirement of gauge invariance. For completeness, we take note that a
mass term of the form $M_\gamma^2 A_\mu A^\mu$ is forbidden by gauge
invariance implying massless photons.

\subsection{$SU(3)_c$ and the QCD Lagrangian}
As we have seen, quarks come in three colours. Suppressing all
space-time dependence, the free quark
Lagrangian for a single flavour has the form 
\begin{equation} 
\cL_0= \displaystyle\sum_{i=1}^3
\overline{q_i}\left(i\,\slashed{\partial}  - m_q \right) q_i ~. 
\end{equation}
Assuming the three quarks with different colours to have the same mass
$\displaystyle{m_q}$ (different flavours have different masses, of 
course), we can ask for the global invariances of
$\displaystyle{\cL_0}$. All transformations that 
leave $\cL_0$ invariant are of the form
\begin{eqnarray} 
q_i \lra q_i^\prime = U_{ij} q_j ~, \qquad
U\,U^\dg=U^\dg\,U=\mathbbm{1} 
\end{eqnarray} 
with arbitrary unitary matrices $U_{ij}$. Splitting off a common
phase transformation $\displaystyle{q_i \to e^{-\displaystyle{i\ve}} 
q_i}$ treated
previously and generating electromagnetic interactions that we know 
to be colour blind, we are
left with the special unitary group $SU(3)$ comprising all
three-dimensional unitary matrices with unit determinant.

In contrast to the $U(1)$ case treated before, we now have eight
independent transformations in $SU(3)$ (an 8-parameter Lie
group). Therefore, continuing in the same spirit as before, it is not
just a question of demanding gauge invariance but we also have to
find out which part of $SU(3)$ should be gauged. With hindsight, the
following two criteria lead to a unique solution.
\begin{enumerate} 
\item[i.] The three colours are not like three arbitrary electric
  charges but are instead intimately connected through gauge
  transformations. This requires the quarks to be in an
  irreducible three-dimensional representation leaving only two
  possibilities: either all of $SU(3)$ or one of the $SU(2)$ subgroups
  must be gauged.
\item[ii.] Quarks and antiquarks transform differently under gauge
  transformations ($\underline{3} \neq \underline{3}^*$). This closes
  the case and implies that all $SU(3)$ transformations must be
  gauged.
\end{enumerate} 
The above requirements guarantee that both $\overline{q}q$ and $qqq$ 
contain colour singlets $=$ hadrons,
\begin{eqnarray}
 \underline{3}^* \otimes  \underline{3} = \underline{1} \oplus
 \underline{8}, \qquad
\underline{3} \otimes  \underline{3} \otimes  \underline{3} =
\underline{1} \oplus \underline{8} \oplus \underline{8} \oplus
 \underline{10}~, 
\end{eqnarray}
but neither $qq$ nor $\overline{q}qqq$ or other exotic combinations. 

Every three-dimensional unitary matrix with $\det U=1$ can be written
as 
\begin{equation} 
U(\ve_a) = \exp\{-i \displaystyle\sum_{a=1}^8 \ve_a
  \displaystyle\frac{\lambda_a}{2}\}
\end{equation}
with eight parameters $\ve_a$ and 
with eight traceless hermitian Gell-Mann matrices
$\displaystyle{\lambda_a}$.  Their
commutation relations define the Lie algebra of $SU(3)$:  
\begin{equation} 
\left[\lambda_a,\lambda_b\right] = 2 i\,f_{abc} \lambda_c~,
\end{equation} 
with real, totally antisymmetric structure constants $f_{abc}$.

The gauge principle demands invariance of the theory for arbitrary
space-time dependent functions $\ve_a(x)$. Instead of a single gauge
field $A_\mu$, we now need eight vector fields
$\displaystyle{G_a^\mu(x)}$  entering the covariant derivative
\begin{eqnarray} 
(D^\mu q)_i = \left(\partial^\mu \delta_{ij}+ i\,g_s 
\displaystyle\sum_{a=1}^8 G_a^\mu
  \displaystyle\frac{\lambda_{a,ij}}{2}\right) q_j =:
\{(\partial^\mu + i g_s G^\mu) q\}_i~.
\end{eqnarray}
The real coupling constant $\displaystyle{g_s}$ measures the strength 
of the quark
gluon interaction just as the charge $Q$ is a measure of the
electromagnetic interaction. Using the convenient matrix notation
(summation convention implied)
\begin{equation}
G^\mu_{ij} := G_a^\mu \displaystyle\frac{\lambda_{a,ij}}{2}~,
\end{equation}
the covariant derivative now transforms as
\begin{eqnarray}
G_\mu \lra G_\mu^\prime = U(\ve) G_\mu U^\dg(\ve) +
\displaystyle\frac{i}{g_s} \left(\partial_\mu U(\ve)\right) U^\dg(\ve)
\label{eq:covd} 
\end{eqnarray}
in order for $(D_\mu q)_i$ to transform like the $q_i$ themselves
(covariance requirement).

Because of the nonabelian character of $SU(3)$, the transformation
laws are more complicated than in the electromagnetic case. The 
differences can already be seen in the infinitesimal transformations of 
the gluon fields $\displaystyle{G_a^\mu}$ following from (\ref{eq:covd}):
\begin{eqnarray}
G_a^\mu \lra G_a^{\mu\prime} = G_a^\mu +
\displaystyle\frac{1}{g_s}\partial^\mu \ve_a + f_{abc} \ve_b G_c^\mu 
+ O(\ve^2)~.
\end{eqnarray}

In order to have propagating gluon fields, we need an analogue of the
electromagnetic field strength tensor $F_{\mu\nu}$. The simplest
approach is to 
calculate the commutator of two covariant derivatives:
\begin{eqnarray}
\left[D_\mu,D_\nu \right] = \left[\partial_\mu + i\,g_s G_\mu,
\partial_\nu + i\,g_s G_\nu\right] =: i\,g_s G_{\mu\nu}~.
\end{eqnarray}
The nonabelian field strength tensor $\displaystyle{G^{\mu\nu}=
G_a^{\mu\nu} \displaystyle\frac{\lambda_a}{2}}$ has the explicit form
\begin{eqnarray}
G^{\mu\nu} &=& \partial^\mu G^\nu - \partial^\nu G^\mu + i g_s \left[
G^\mu,G^\nu\right] \\
G^{\mu\nu}_a &=& \partial^\mu G^\nu_a - \partial^\nu G^\mu_a - 
g_s f_{abc} G^\mu_b G^\nu_c \no
\end{eqnarray}
and it transforms covariantly under $SU(3)$ gauge transformations:
\begin{eqnarray}
G_{\mu\nu} \lra G_{\mu\nu}^\prime = U(\ve) G_{\mu\nu} U^\dg(\ve) ~.
\end{eqnarray}
The gauge invariant colour trace
\begin{eqnarray}
{\rm tr}(G_{\mu\nu} G^{\mu\nu}) = \displaystyle\frac{1}{2} G^{\mu\nu}_a
G_{\mu\nu}^a  
\end{eqnarray}
has the right structure for a gluon kinetic term leading immediately to
the $\displaystyle{SU(3)_c}$ invariant QCD Lagrangian for
$f=1,\dots,N_F$ quark flavours: 
\begin{eqnarray}
\cL_{\rm QCD} = 
- \displaystyle\frac{1}{2} {\rm tr}
(G_{\mu\nu} G^{\mu\nu}) + \displaystyle\sum_{f=1}^{N_F}
\overline{q}_f \left(i \slashed{D} - m_f \mathbbm{1}_c \right) q_f~. 
\label{eq:LQCD}
\end{eqnarray}
As in the $U(1)$ case, gauge invariance requires massless gluons.
Writing out the Lagrangian (\ref{eq:LQCD}) in detail, one finds
three types of vertices instead of a single one for QED: \\[.1cm] 
\hspace*{4.3cm} $\overline{q} q G$ \hspace*{2.1cm}  $G G G$ 
\hspace*{2.0cm} $GGGG$
\begin{center} 
\begin{figure}[H] 
\leavevmode 
\includegraphics[width=8cm]{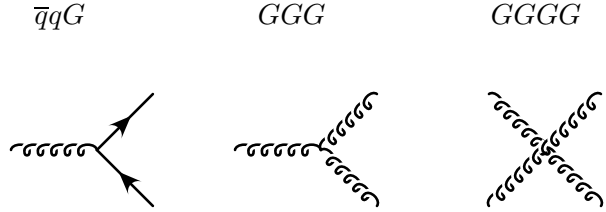}
\caption{Basic vertices of QCD.}
\label{fig:QCD}
\end{figure}
\end{center} 
In addition to the quark masses, QCD has a single parameter describing
the strength of the strong interactions, the strong coupling constant
$g_s$ ($\alpha_s=g_s^2/4\pi$).
\begin{center} 
\begin{fmpage}{4.5cm}
\vspace*{.1cm} 
\begin{center}
Experimental group theory
\end{center}
\vspace*{.1cm}  
\end{fmpage} 
\end{center} 
Can experimentalists determine more than a single coupling strength 
$\displaystyle{\alpha_s}$ in a strong process? All the information is 
contained in the vertices: in addition to $\displaystyle{g_s}$, the 
vertices also contain the two matrices 
\begin{eqnarray}
(t_a^F)_{ij} = \displaystyle\frac{1}{2} (\lambda_a)_{ij}~, &\qquad&
(t_a^A)_{bc} = -i f_{abc}~, 
\end{eqnarray} 
defining the fundamental and adjoint representations
of (the Lie algebra of) $SU(3)$:
\begin{equation}
[t_a,t_b]=i\,f_{abc} t_c~.
\end{equation} 
Let us pretend for a moment that we don't know that there are three
colours and eight gluons. For a general (compact Lie) group of
symmetry transformations, the vertices are again determined by quark 
and gluon representation matrices $\displaystyle{t_a^F, t_a^A}$. The 
combinations that actually appear in measurable quantities are the 
following traces and sums:
\begin{eqnarray}
{\rm tr} (t_a^R t_b^R)= T_R \delta_{ab}, \qquad 
\displaystyle\sum_a (t_a^R)_{ij} (t_a^R)_{jk} = C_R \delta_{ik} \qquad
(R=F,A)~, 
\label{eq:trs}
\end{eqnarray} 
with
\begin{center} 
\begin{tabular}{lcl}
$T_R$: & \mbox{     }\hspace*{.5cm}  & Dynkin index
  for the representation R; \\
$C_R$: & & (quadratic) Casimir for R.
\end{tabular}  
\end{center}  
For a $d_R$-dimensional representation, one derives from the
definitions (\ref{eq:trs}) the general relation
\begin{equation}
d_R\,C_R = n_G\,T_R
\end{equation}
where $n_G$ is the number of independent parameters of
$G$. Restricting the discussion to $SU(n)$, the two cases of interest
are \\[.1cm] 
$R=A$ (adjoint representation): \\
\hspace*{1cm} $d_A=n_G \qquad \lra \qquad C_A=T_A=n$ ~~for $SU(n)$;
\\[.1cm] 
$R=F$ (fundamental representation of $SU(n)$): \\
\hspace*{1cm} $d_F=n, ~n_G=n^2 -1, ~T_F=1/2 \qquad \lra \qquad 
C_F = \displaystyle\frac{n^2-1}{2n}$ \\[.1cm]
and for the special case of $SU(3)$: 
\hspace*{1cm} $C_F=\displaystyle\frac{4}{3},~C_A=T_A=N_c=3$~.  \\[.2cm] 
The independent quantities that can be measured are $C_F$ and
$C_A$. A combined jet analysis in $e^+ e^-$ annihilation at LEP found 
\cite{Kluth:2003yz}
\begin{eqnarray}
C_F = 1.30 \pm 0.01({\rm stat}) \pm 0.09({\rm sys}), & ~&
C_A = 2.89 \pm 0.03({\rm stat}) \pm 0.21({\rm sys})
\end{eqnarray}
in manifest agreement with $SU(3)$.

Feynman diagrams are constructed with the vertices and propagators of 
the QCD Lagrangian (\ref{eq:LQCD}). The problem here is the same as in 
QED: due to the gauge invariance of
(\ref{eq:LQCD}), the gluon propagator does not exist. At least for
perturbation theory, the inescapable consequence is that gauge
invariance must be broken in the Lagrangian! Or, in a more euphemistic
manner of speaking, the gauge must be fixed.  In the simplest and
widely used version (covariant gauge with real parameter $\xi$) the
Lagrangian (\ref{eq:LQCD}) is replaced by
\begin{eqnarray}
\cL_{\rm QCD} \qquad \lra \qquad \cL_{\rm QCD} -
\displaystyle\frac{\xi}{2} \left(\partial_\mu G^\mu_a \right)^2 +
\cL_{\rm ghost} ~.
\end{eqnarray}
The gluon propagator now exists ($\xi=1$: Feynman gauge):
\begin{eqnarray} 
\Delta_{ab}^{\mu\nu}(k) = \delta_{ab}
\displaystyle\frac{-i}{k^2+i\epsilon} 
\left(g^{\mu\nu} + (\xi^{-1}-1)\frac{k^\mu k^\nu}{k^2}\right) 
\st{\xi=1}{=}
\delta_{ab} \displaystyle\frac{-i\,g^{\mu\nu}}{k^2+i\epsilon}~.
\end{eqnarray} 
The additional ghost Lagrangian $\cL_{\rm ghost}$ repairs the damage
done by gauge fixing: although Green functions are now gauge dependent,
observable S-matrix elements are still gauge invariant and therefore
independent of $\xi$ (Feynman, Faddeev, Popov, BRST, \dots).

\section{Perturbative QCD}

\subsection{QCD at tree level}
\label{sec:QCDtree}
The calculation of tree amplitudes in QCD is straightforward but
\begin{itemize} 
\item[$\bullet$] to compare theory with experiment, we must have hadrons
  rather than quarks and gluons in the initial and final states;
\item[$\bullet$] amplitudes and cross sections are in general infrared
  divergent for massless gluons.
\end{itemize} 

\noindent
The general recipe is to consider infrared safe quantities, the
more inclusive the better. A good example is once more $e^+ e^- \to$
hadrons.
\begin{center} 
\begin{figure}[H] 
\leavevmode 
\includegraphics[width=10cm]{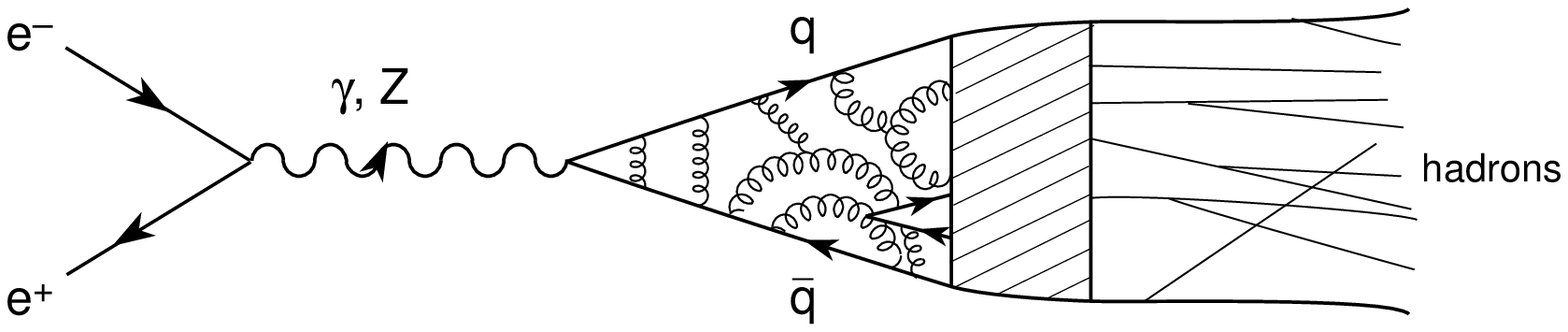}
\caption{~$e^+ e^- \to$ hadrons.}
\label{fig:eehad}
\end{figure}
\end{center} 
The sum over all hadronic final states can be expressed in terms of the
imaginary (absorptive) part of the two-point
function of electromagnetic currents (photonic case), the hadronic 
vacuum polarization:
\begin{eqnarray}
\Pi_{\rm em}^{\mu\nu}(q)= i \displaystyle\int d^4x \, 
e^{\displaystyle  i q\cdot x}
\langle 0|\,T J^\mu_{\rm em}(x) J^\nu_{\rm em}(0) |0\rangle =
\left(-g^{\mu\nu} q^2 + q^\mu q^\nu \right) \Pi_{\rm em}(q^2)~.
\end{eqnarray}
\begin{center} 
\begin{figure}[H] 
\leavevmode 
\includegraphics[width=10cm]{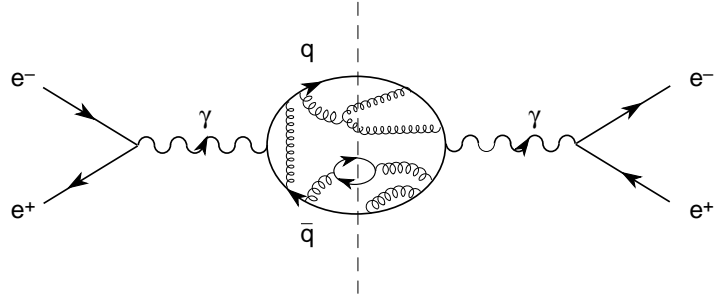}
\caption{Hadronic vacuum polarization.}
\label{fig:vacpol}
\end{figure}
\end{center} 
The sum over all intermediate states can be performed either with
quarks and gluons or with hadrons. Since there are no
massless hadrons, the hadronic vacuum polarization is infrared safe. 

\begin{center} 
\begin{figure}[H] 
\leavevmode 
\includegraphics[width=10cm]{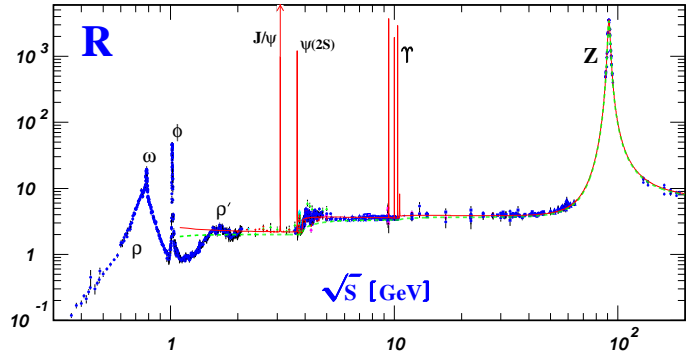}
\caption{Experimental data for $R_{e^+ e^-}$ taken from
  Ref.~\cite{Ezhela:2003pp}.} 
\label{fig:Ree}
\end{figure}
\end{center}

To lowest order in QCD, the amplitude for
\begin{equation}
e^+ e^- ~\to ~\gamma^*(Z^*) ~\to ~\overline{q} q
\end{equation}
is in fact independent of the strong coupling constant
$\displaystyle{g_s}$.  Except
for the charges, masses and multiplicities of quarks, it is 
the same calculation as for $\displaystyle{e^+ e^- \to \mu^+ \mu^-}$ in
QED. Therefore, for quarks with given flavour $f$ and colour $i$ the
amplitude is (neglecting $\displaystyle{m_e}$, $\displaystyle{m_\mu}$, 
$\displaystyle{m_q}$)
\begin{equation}
A(e^+ e^- \to \overline{q}^i_f q^i_f) = \displaystyle\frac{Q_f}{e}
A(e^+ e^- \to \mu^+ \mu^-)~.
\end{equation}
Quarks and antiquarks with different colour and flavour are in principle
distinguishable so that the total hadronic cross section, normalized
to $\sigma(e^+ e^- \to \mu^+ \mu^-)$, is
\begin{eqnarray}
R_{e^+ e^-} = \displaystyle\frac{\sigma(e^+ e^- \to {\rm ~hadrons})}
{\sigma(e^+ e^- \to \mu^+ \mu^-)}= \displaystyle\sum_{i,f} Q_f^2/e^2=
N_c \displaystyle\sum_f Q_f^2/e^2 ~. 
\end{eqnarray}
As shown in Fig.~\ref{fig:Ree}, this is a good approximation to the
experimental data between quark thresholds. 
A similar result is obtained for the hadronic width of the $Z$:
\begin{eqnarray} 
R_Z = \Gamma(Z \to {\rm~hadrons})/\Gamma(Z \to e^+
e^-) = N_c (1+\delta_{\rm EW})
\displaystyle\sum_f (v_f^2 + a_f^2)/(v_e^2 + a_e^2)~,
\end{eqnarray} 
where $v_F, a_F$ are the (axial-)vector couplings for $Z \to
\overline{F} F$. 

\begin{center} 
\begin{fmpage}{2.5cm}
\begin{center}
$e^+ \,e^- \to$ jets
\end{center} 
\end{fmpage} 
\end{center}
At high energies, the two-jet structure from $\displaystyle{e^+ e^- 
\to \overline{q} q}$ dominates, being the only process at
$\displaystyle{O(\alpha_s^0)}$. At
$O(\alpha_s)$ and omitting $Z$ exchange, we have in addition gluon
bremsstrahlung off quarks giving rise to a three-jet structure:
\begin{equation}
e^+(q_1) e^-(q_2) \to 
q (p_1) \overline{q}(p_2) G(p_3)~.
\end{equation}
\begin{center} 
\begin{figure}[H] 
\leavevmode 
\includegraphics[width=10cm]{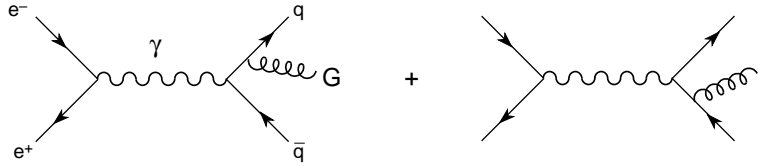}
\caption{Leading-order diagrams for three-jet production.}
\label{fig:3jet}
\end{figure}
\end{center} 
The calculation is again identical to QED bremsstrahlung except for a
factor (sum over all final states in the rate)  
\begin{eqnarray}
\displaystyle\sum_{a} {\rm tr} (t_a^F t_a^F)= 
T_F \displaystyle\sum_{a} \delta_{aa} =T_F \,n_{SU(3)} = d_F C_F =
3 \,C_F = 4 ~.
\end{eqnarray}
With the kinematics specified by
\begin{eqnarray} 
s=(q_1+q_2)^2,\quad (p_i+p_j)^2=(q_1+q_2-p_k)^2=:s(1-x_k) \\[.1cm] 
x_1+x_2+x_3=2,\hspace*{2cm}  {\rm CMS}: x_i=2 E_i/\sqrt{s}~,\no
\end{eqnarray}
the double differential cross section (for massless quarks) is found
to be
\begin{eqnarray} 
\displaystyle\frac{d^2
  \sigma}{dx_1\,dx_2} =\displaystyle\frac{2 \alpha_s \sigma_0}{3\pi}
  \displaystyle\frac{x_1^2+x_2^2}{(1-x_1)(1-x_2)} \qquad {\rm with}
  \qquad
\sigma_0=\displaystyle\frac{4\pi\alpha^2}{s}\displaystyle\sum_f
  \left(Q_f/e \right)^2~. 
\label{eq:ddcs} 
\end{eqnarray} 
The problem with this cross section is that it diverges for
$\displaystyle{x_i} \to
1$ ($i=1,2$). This infrared divergence is due the singular behaviour
of the quark propagator and it happens even for massive quarks:
\begin{eqnarray} 
(p_2 + p_3)^2 - m_q^2 =2 p_2\cdot p_3 = s (1 - x_1)~. 
\end{eqnarray}   
$m_q > 0: \qquad x_1 \to 1$ only possible for $p_3 \to 
0$ ~(soft gluon singularity); \\[.1cm] 
$m_q = 0: \qquad x_1= 1$ also possible for $p_3 \,||\, p_2$ 
~(collinear singularity)~. \\[.2cm]
To understand the origin of infrared divergences, we first take the
viewpoint of an experimentalist measuring three-jet events where the
jets stand for the quarks and the gluon in the final state.
\begin{itemize}
\item[$\bullet$] Depending on the detector resolution, a quark and a
  soft gluon cannot be distinguished from a single quark. In that
  case, the event will be counted as a two-jet event.
\item[$\bullet$] Two collinear massless particles can never be
  resolved: they always stay together.
\end{itemize} 
From the viewpoint of a theorist, we recall that perturbation theory
is built on the assumption that particles do not interact when they are
sufficiently far apart. This assumption is not really satisfied for 
massless quanta like photons or gluons that give rise to long-range
forces. In other words, an electron (a quark) can never be separated 
from its cloud of soft photons (gluons).\\[.2cm]
The practitioner's solution of the infrared problem is well understood:
\begin{itemize}  
\item[$\bullet$] One must define criteria to distinguish between (in
  the present case) two- and three-jet events (jet algorithms).
\item[$\bullet$] Virtual gluon (loop) corrections for the process 
$\displaystyle{e^+ e^- \to \overline{q} q}$ must be included.
\end{itemize}  

\subsection{Higher-order corrections and renormalization}
The loop corrections of $O(\alpha_s)$ for $e^+ e^- \to \overline{q} q$
are calculated from the Feynman diagrams below.
\begin{center} 
\begin{figure}[H] 
\leavevmode 
\includegraphics[width=12cm]{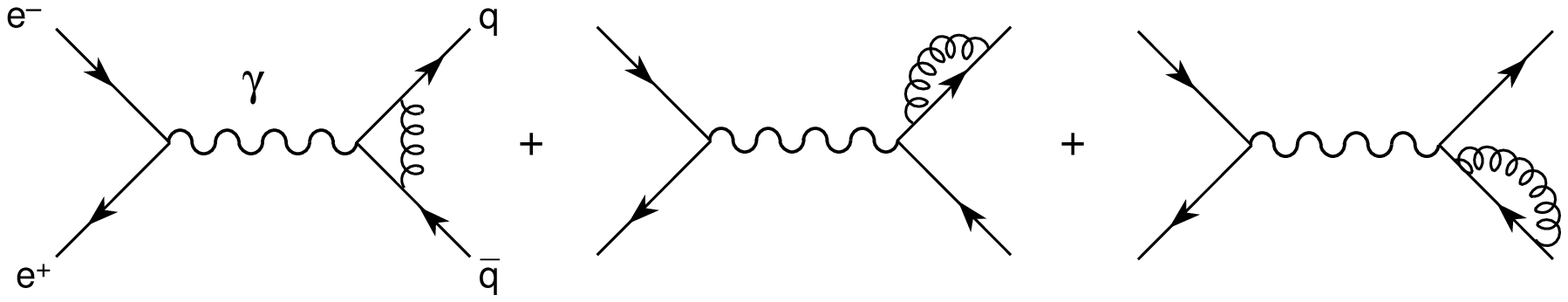}
\caption{One-loop diagrams for $e^+ e^- \to \overline{q} q$.}
\label{fig:eeqqloop}
\end{figure}
\end{center} 
The resulting amplitudes are both infrared and ultraviolet divergent.

In contrast to infrared divergences, ultraviolet divergences are
due to the high-momentum components of the particles in loops. Common
sense tells us that those components cannot influence physics at
low energies. If this were the case we would
have to give up all hopes of being able to make predictions at
presently accessible energies. 

The recipe to handle ultraviolet divergences is also well
understood. One first has to choose a method to cut off the
high-momentum components. There are infinitely many ways to do that so
the question is legitimate whether the final amplitudes will depend on
that procedure rendering the result completely arbitrary. The answer
is that the cutoff procedure (regularization) must always be
accompanied by renormalization. Before choosing a suitable
regularization procedure let us therefore try to understand the idea
of renormalization, using the most naive regularization method.

To simplify matters as much as possible, we consider the elastic
scattering of two particles in massless scalar
$\displaystyle{\phi^4}$ theory ($\cL_{\rm int} \sim \lambda \phi^4$):
\begin{equation}
\phi \phi \to \phi \phi~,
\end{equation}
with scattering amplitude  $A(s,t)$ in terms of the usual Mandelstam
variables. We now define what we mean by the physical (renormalized)
coupling constant. The definition should be applicable at every
order of perturbation theory and it should coincide with the
constant $\lambda$ in the Lagrangian at tree level. A possible
definition in scalar $\phi^4$ theory is
\begin{equation}
\lambda_r(\mu) := A(s=-t=\mu^2)
\end{equation}
with an arbitrary renormalization scale $\mu$. At tree level,
$A(s,t)$ is momentum independent and with the proper normalization we
have indeed $\displaystyle{\lambda_r(\mu)=\lambda}$.

\begin{center} 
\begin{figure}[H] 
\leavevmode 
\includegraphics[width=12cm]{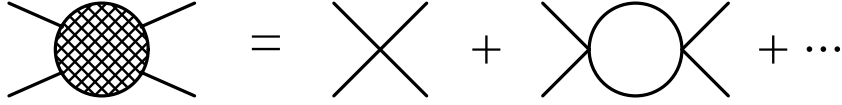}
\caption{Scattering amplitude for $\phi \phi \to \phi \phi$ to
one-loop order.} 
\label{fig:phiphi}
\end{figure}
\end{center}

Beyond tree level, the amplitude has an ultraviolet divergence that we
regularize with a simple momentum cutoff $\Lambda$ here. The relevant
diagrams up to one loop are shown in Fig.~\ref{fig:phiphi}.
Setting $s=-t=\mu^2$, one finds ($\beta_0$ is a constant)
\begin{eqnarray} 
\lambda_r(\mu)=A(s=-t=\mu^2) &=& \lambda + \beta_0 \lambda^2 
\log{\Lambda/\mu} \\
&+& \mu{\rm -independent~terms~of}~O(\lambda^2) + O(\lambda^3)~.\no
\end{eqnarray} 
Since $\lambda_r(\mu)$ is finite, being equal to the physical
scattering amplitude at some fixed point in phase space, the bare
coupling $\lambda$ diverges as the cutoff $\Lambda \to \infty$. 
However, the bare
coupling is not related to any physical quantity. Therefore, we are
free to ``sweep the infinities under the rug'' as long as this is done
in a transparent and controllable way.

To do this, we change the renormalization scale by a small amount
$\delta\mu$: 
\begin{eqnarray} 
\lambda_r(\mu+ \delta\mu) - \lambda_r(\mu) =  ~~\beta_0 \lambda^2 
\log{\left(\displaystyle\frac{\Lambda}{\mu + \delta\mu} 
\displaystyle\frac{\mu}{\Lambda}\right)} + O(\lambda^3) \nn
 \hspace*{0cm} = ~~\beta_0 \lambda_r^2 \log{\displaystyle\frac{\mu}{\mu +
    \delta\mu}} + O(\lambda_r^3) 
= - \beta_0 \lambda_r^2 \displaystyle\frac{\delta\mu}{\mu} +
O[(\delta\mu)^2]  + O(\lambda_r^3)~.
\end{eqnarray}  
The bare coupling and the cutoff have disappeared in the last
equation. Expanding $\displaystyle{\lambda_r}(\mu+ \delta\mu)$ around 
$\mu$ and letting $\delta\mu \to 0$, we recover the $\beta$ function
of $\displaystyle{\phi^4}$ theory to one-loop order:
\begin{eqnarray}
\mu \displaystyle\frac{d \lambda_r(\mu)}
{d\mu}  = - \beta_0  \lambda_r^2(\mu) + O(\lambda_r^3) =
\beta(\lambda_r(\mu))~. 
\end{eqnarray}
Unlike in Yang-Mills theories, $\beta_0 < 0$ so that $\phi^4$
theory is ultraviolet unstable like all quantum field theories except 
nonabelian
gauge theories \cite{Zee:1973gn,Coleman:1973sx}. However, for  
understanding the essence of renormalization the important observation
is that physical quantities do not depend on the bare coupling constant
$\lambda$ nor on the cutoff
$\Lambda$  (for $\Lambda \to \infty$) but only on the renormalized
coupling  $\displaystyle{\lambda_r}(\mu)$. For the purpose of
comparing theory with experiment at present energies, we will never 
notice the stuff that was swept under the rug.

We now turn to the choice of a regularization scheme. Although there
are infinitely many possibilities, some choices are clearly better than
others. The main criteria are:
\begin{dinglist}{42}
\item The regularization method should respect symmetries of the
theory as much as possible. In this respect, the
previously employed momentum cutoff is as bad as it gets violating 
Poincar\'e symmetry, gauge invariance, etc.
\item The scheme should violate only those symmetries that are
necessarily violated by quantum effects (anomalies).
\item The method should be simple to handle in practice. 
\end{dinglist}  
From the practitioner's point of view, dimensional regularization is
the almost unique choice fulfilling these criteria. Let us demonstrate
the method with a simple example, electronic vacuum polarization
(setting $m_e=0$).
 
\parbox[c]{0.45\textwidth}{
\begin{flushleft} 
\begin{figure}[H] 
\leavevmode 
\includegraphics[width=5cm]{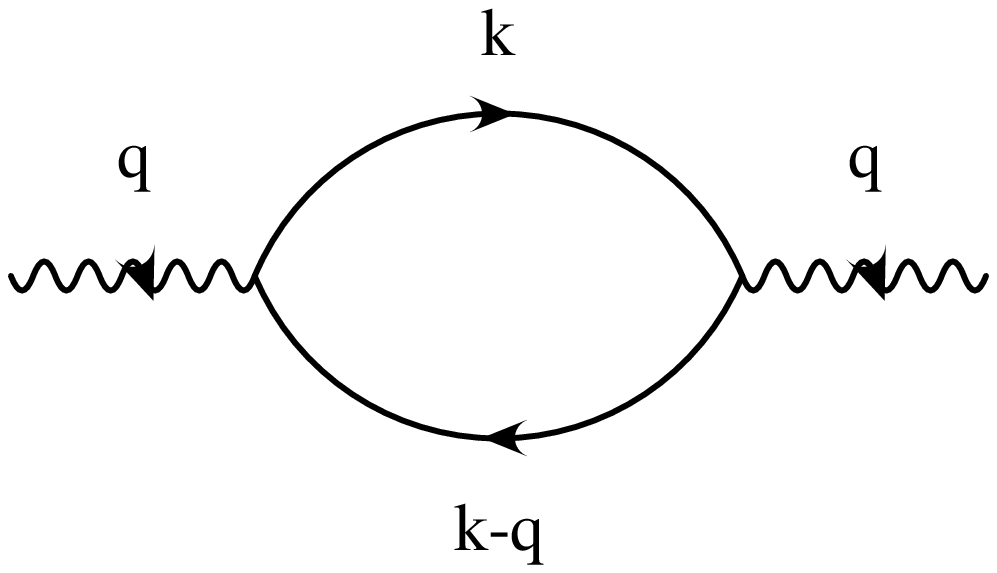}
\caption{Vacuum polarization at one loop.}
\label{fig:evacpol}
\end{figure}
\end{flushleft} 
}

\vspace*{-3.5cm}
\hspace*{5.5cm} 
\parbox[c]{0.55\textwidth}{
Gauge invariance, guaranteed by dim. regularization, \\
implies the same structure as in the hadronic case:
\begin{eqnarray} 
\Pi^{\mu\nu}(q) &=& 
\left(-g^{\mu\nu} q^2 + q^\mu q^\nu \right) \Pi(q^2)\\
\Pi(q^2) &=& \displaystyle\frac{8 e^2 \Gamma(\ve)}{(4\pi)^{2-\ve}}
\displaystyle\int_0^1 \displaystyle\frac{dx\,x(1-x)}{[-q^2
x(1-x)]^\ve} \nn
\Gamma(x) &=& 1/x - \gamma +O(x),\quad 2\ve=4-d~.  \no 
\end{eqnarray} 
}

\noindent 
Since dimensional regularization works in $d$ dimensions, there is a
small problem here: there is no scale for $\log{(-q^2)}$ that will
appear in the explicit form of $\Pi(q^2)$. To solve the
problem, we insert unity (the scale $\mu$ and the constant $c$ are
completely arbitrary) in the expression
and expand the second factor in $\ve$:
\begin{eqnarray}
1 = (c\mu)^{-2\ve} (c\mu)^{2\ve} = (c\mu)^{-2\ve} \left[1 + \ve
\log{\mu^2}+ 2 \ve \log{c} + O(\ve^2) \right]~. 
\end{eqnarray}  
Various schemes on the market differ by the constant $c$:\\[.1cm] 
\begin{tabular}{clcl}
\mbox{} \hspace*{2cm} & MS & \mbox{} \hspace*{2cm} & $c=1$ \\[.1cm]      
& $\overline{\rm MS}$ & & $\log{c}=
(\gamma - \log{4\pi})/2$~.
\end{tabular} 

\noindent 
Using the most popular scheme ($\overline{\rm MS}$), the final result
is
\begin{eqnarray} 
\Pi(q^2) &=& \displaystyle\frac{e^2}{12\pi^{2}}
\left\{\displaystyle\frac{(c\mu)^{-2\ve}}{\ve} - \log{(-q^2/\mu^2)} +
\displaystyle\frac{5}{3} \right\} + O(\ve) \nn
&=& \Pi_{\rm div}^{\overline{\rm MS}}(\ve,\mu) 
- \displaystyle\frac{e^2}{12\pi^{2}}
\left\{\log{(-q^2/\mu^2)} - \displaystyle\frac{5}{3} \right\} ~.
\end{eqnarray} 
The divergent part $\Pi_{\rm div}^{\overline{\rm MS}}(\ve,\mu)$ has been
isolated and it will be absorbed by wave function renormalization of
the photon field contributing to charge renormalization. We also
notice that the coefficients of $1/\ve$ and $-\log{(-q^2/\mu^2)}$ are
identical: the $\beta$ function can be extracted from the divergent
part. The stuff under the rug is useful after all.

Back to the loop corrections of Fig.~\ref{fig:eeqqloop}, we observe that
the diagrams give rise to an amplitude proportional to $g_s^2$, whereas 
$A(e^+ e^- \to \overline{q} q G) \sim g_s$. How can the infrared
divergences cancel among amplitudes of different order in
$\displaystyle{g_s}$? The
answer is that they cannot cancel on the level of amplitudes because
the final states are different. Not the amplitudes but the rates must
be added. Interference with the tree amplitude produces an
$O(\alpha_s)$  term in $\sigma(e^+ e^- \to \overline{q} q)$ that can
and will cancel the infrared divergence in $\sigma(e^+
e^- \to \overline{q} q G)$. For details of the calculation I refer to
the monograph \cite{esw}, an excellent source for applications of
perturbative QCD in general.

The easier part are the loop corrections for $\sigma(e^+ e^- \to
\overline{q} q)$. With dimensional regularization to regularize the
infrared divergences, one finds
\begin{eqnarray} 
\sigma_{\overline{q} q}^{\rm interference}=\sigma_0 C_F
\displaystyle\frac{\alpha_s}{4\pi}H(\ve)\left\{-
\displaystyle\frac{4}{\ve^2} - \displaystyle\frac{6}{\ve} - 16 + O(\ve)  
 \right\}~, \hspace*{2cm} H(0)=1 ~.  
\end{eqnarray} 
The less familiar part is the three-body phase space integration in
$d$ dimensions   giving rise to
\begin{eqnarray} 
\sigma_{\overline{q} q G}=\sigma_0 C_F
\displaystyle\frac{\alpha_s}{4\pi}H(\ve)\left\{
\displaystyle\frac{4}{\ve^2} + \displaystyle\frac{6}{\ve} + 19 + O(\ve)  
 \right\} ~.  
\end{eqnarray} 
The infrared divergences cancel as expected.

Adding the lowest-order cross section, one obtains finally
\begin{eqnarray}
\sigma(e^+ e^- \to {\rm ~hadrons})=\sigma_0 \left(1 + 3 C_F
\displaystyle\frac{\alpha_s}{4\pi} + O(\alpha_s^2) \right)=
\sigma_0 \left(1 + \displaystyle\frac{\alpha_s}{\pi} + O(\alpha_s^2) 
\right) ~,
\end{eqnarray}   
with $\sigma_0$ defined in Eq.~(\ref{eq:ddcs}). The cross section
$\sigma(e^+ e^- \to {\rm ~hadrons})$ is nowadays known up to
$O(\alpha_s^3)$. Replacing $\displaystyle{\alpha_s}$ by the 
renormalization group improved running coupling 
$\displaystyle{\alpha_s}(\sqrt{s})$, the general result for
$R_{e^+e^-}$ can be written
\begin{eqnarray}
R_{e^+e^-}(s) &=& N_c \displaystyle\sum_f Q_f^2/e^2 \left\{1 +
\displaystyle\sum_{n\ge 1} C_n \left(
\displaystyle\frac{\alpha_s(\sqrt{s})}{\pi}\right)^n  \right\} \\
&=& R_{e^+e^-}^{(0)} \left\{1 + C_1
\displaystyle\frac{\alpha_s(\mu)}{\pi} 
+ \left[C_2 - C_1
\displaystyle\frac{\beta_0}{4}\log{(s/\mu^2)}    \right]\left(
\displaystyle\frac{\alpha_s(\mu)}{\pi}\right)^2 + \dots \right\}~. \no
\end{eqnarray} 
The normalization is such that $C_1 = 1$, the coefficients $C_2,C_3$
being also known. 

In principle, $R_{e^+e^-}(s)$ is independent of the arbitrary scale
$\mu$ by construction. In reality, the unavoidable truncation of the
perturbative series introduces a scale dependence. Although there is
no unique prescription for the optimal choice of $\mu$, the obvious
choice here is $\displaystyle{\mu^2}=s$ to avoid large logarithms. If
the perturbative expansion is to make sense, we expect higher orders to 
mitigate the scale dependence. This is nicely demonstrated in
Fig.~\ref{fig:Resw} taken from Ref.~\cite{esw} where
the deviation (in percent) of $R_{e^+e^-}(\sqrt{s}=33~{\rm GeV})$ from
$R_{e^+e^-}^{(0)}$ is plotted as a function of $\mu$. As expected,
$\mu^2=s$ is indeed a very reasonable choice already at $O(\alpha_s)$
(L). 
\begin{center}
\vspace*{-.7cm}  
\begin{figure}[H] 
\leavevmode 
\includegraphics[width=16cm]{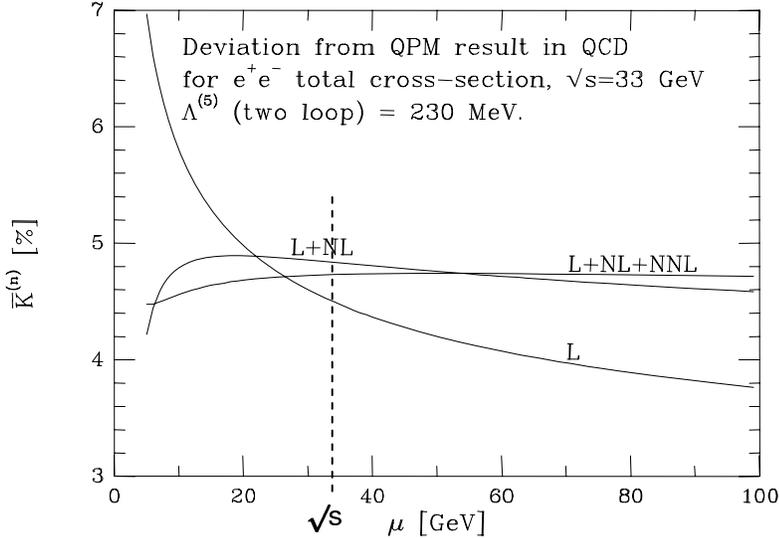}
\vspace*{-2cm} 
\caption{Improvement of the scale dependence in higher orders of
perturbation theory for $R_{e^+e^-}(\sqrt{s}=33~{\rm GeV})$ (taken from 
the book of Ellis, Stirling and Webber \cite{esw}).}
\label{fig:Resw}
\end{figure}
\end{center}

\subsection{Measurements of $\alpha_s$}
How should one characterize the coupling strength of QCD? 
After all,  the scale $\mu$ is arbitrary and, in addition,
$\alpha_s(\mu)$  is in general scheme dependent. For
qualitative purposes, one may introduce a scale $\Lambda_{\rm QCD}$
that is independent of the renormalization 
scale $\mu$. The drawback is that this quantity is scheme independent
only at leading (one-loop) order:
\begin{equation}
\alpha_s(E) = \displaystyle\frac{4\pi}{\beta_0 \log{(E^2/\Lambda_{\rm
QCD}^2)}}~.
\end{equation}
The coefficient $\beta_0$ is defined by rewriting the $\beta$ function 
for $\displaystyle{\alpha_s}$ (instead for $g_s$ as in Sec.~\ref{sec:hist}):
\begin{eqnarray} 
\mu \displaystyle\frac{d\alpha_s(\mu)}{d\mu}=2 \beta(\alpha_s)=
- \displaystyle\frac{\beta_0}{2\pi} \alpha_s^2 -
\displaystyle\frac{\beta_1}{4 \pi^2} \alpha_s^3 + \dots 
\label{eq:beta_alpha}  
\end{eqnarray} 
The $\beta$ function is known up to four loops (coefficient
$\displaystyle{\beta_{\,3}}$)
but only the first two coefficients
\begin{eqnarray} 
\beta_0=11 - 2 N_F/3~, \qquad & & \qquad \beta_1 =51 - 19 N_F/3
\end{eqnarray} 
are scheme and gauge independent.

Since the scheme dependence is unavoidable, the coupling strength is
nowadays usually given in the form of
$\alpha_s^{\overline{\rm MS}}(M_Z)$. Of course, this fixes
$\alpha_s^{\overline{\rm MS}}(\mu)$ at any scale via the integral
\begin{equation}
\log{(\mu_2^2/\mu_1^2)} =
\displaystyle\int_{\alpha_s(\mu_1)}^{\alpha_s(\mu_2)}
\displaystyle\frac{dx}{\beta(x)} ~.
\end{equation}
To get a first rough estimate of $\alpha_s$, consider the
leading-order prediction
\begin{equation}
R_{e^+e^-}(M_Z)=R_{e^+e^-}^{(0)}(M_Z)\left(1 +
\displaystyle\frac{\alpha_s(M_Z)}{\pi}  \right)~.
\end{equation}
Comparing the combined LEP result \cite{PDG04} $R_{e^+e^-}(M_Z)=20.767
\pm 0.025$ with the tree-level prediction
$R_{e^+e^-}^{(0)}(M_Z)=19.984$,
one obtains
\begin{equation}
\alpha_s(M_Z)=0.123 \pm 0.004~,
\end{equation}
close to the full three-loop result and not bad at all for a first 
estimate. A compilation of results can be found in the Review of
Particle Properties \cite{PDG04}. In Figs.~11 and \ref{fig:alpha_s},
the most recent data compiled by Bethke \cite{Bethke:2004uy} are shown.

\parbox[l]{0.5\textwidth}{
\begin{flushleft}
\begin{figure}[H] 
\leavevmode 
\includegraphics[width=4.5cm]{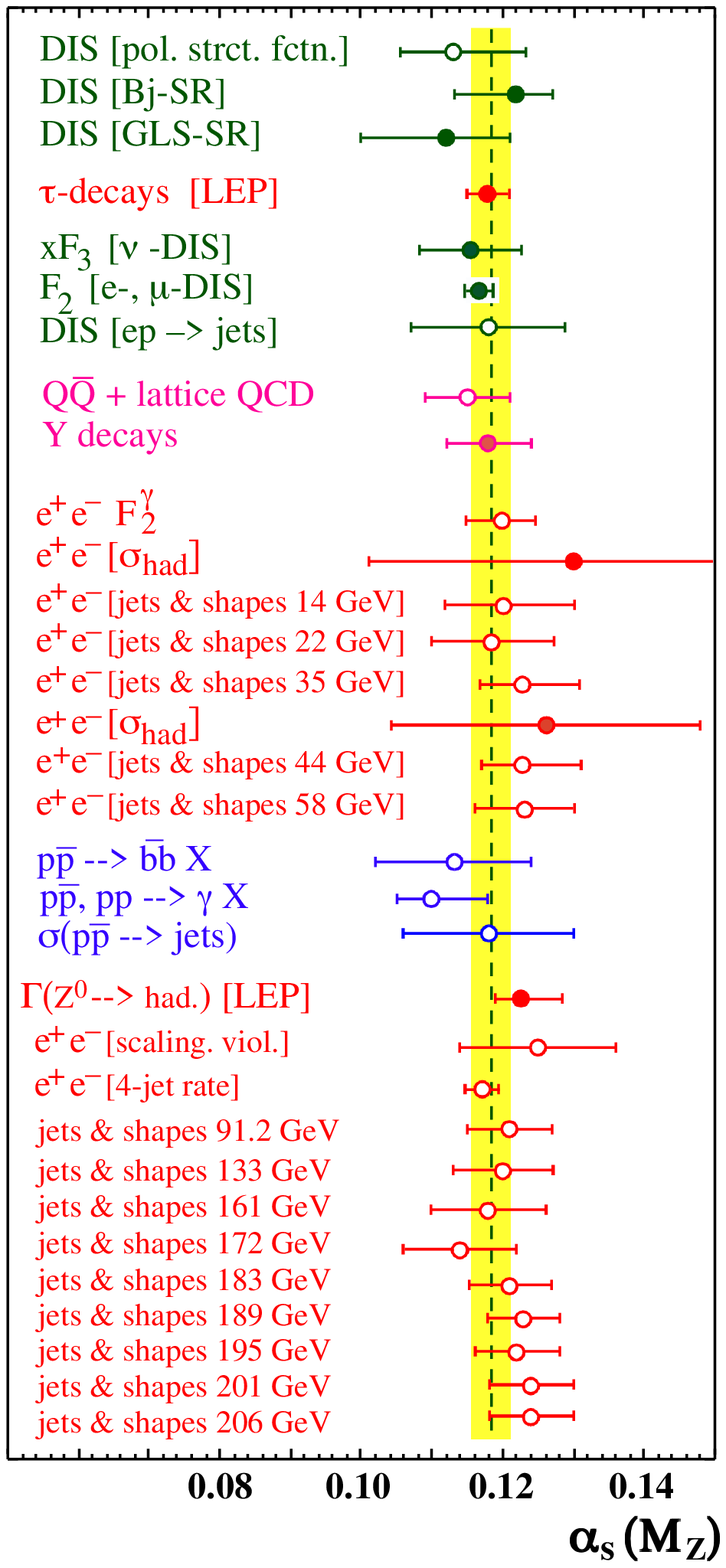}
\label{fig:compalpha}
\end{figure} 
\end{flushleft}  
}

\addtocounter{figure}{1}
\vspace*{-10.2cm}
\hspace*{6.5cm}
\parbox[r]{0.5\textwidth}{
{\bf Fig.~11:} ~Compilation of data for the extraction of
$\alpha_s^{\overline{\rm MS}}(M_Z)$  by Bethke \cite{Bethke:2004uy}.

\vspace*{.3cm} 
\begin{center} 
\begin{figure}[H] 
\leavevmode 
\includegraphics[width=5cm]{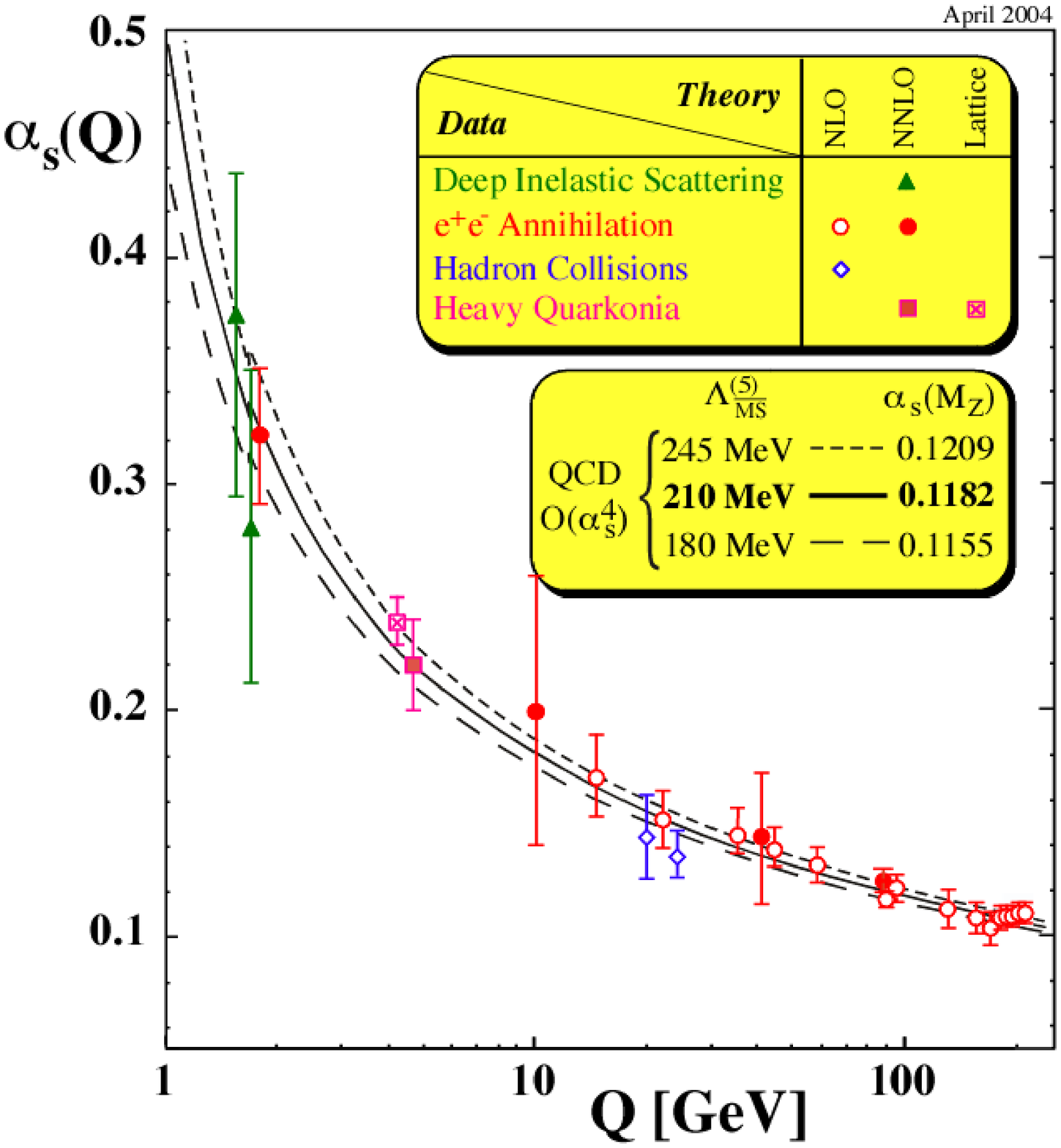}

\vspace*{.4cm} 
\caption{Energy dependence of the running coupling constant
$\alpha_s^{\overline{\rm MS}}(Q)$ \cite{Bethke:2004uy}.}
\label{fig:alpha_s}
\end{figure}
\end{center} 
}

\noindent 
The first impression is the remarkable agreement among experiments and
with theory. However, for determining the best value of
$\alpha_s(M_Z)$, the following two problems must be kept in mind:
\begin{itemize} 
\item  Different observables are known with different theoretical
accuracy: next-to-leading order (NLO) vs. next-to-next-to-leading
order (NNLO). Not only the scale dependence but also different scheme
dependences must be taken into account.
\item Theoretical errors are not normally distributed.
\end{itemize}
Using only NNLO results, Bethke found \cite{Bethke:2004uy}
\begin{equation}
\alpha_s(M_Z)= 0.1182 \pm 0.0027~,
\end{equation}
very similar to the PDG average \cite{PDG04} (using a different
procedure)
\begin{equation} 
\alpha_s(M_Z)= 0.1187 \pm 0.0020~.
\end{equation}
All values in this paragraph refer to $\alpha_s^{\overline{\rm
MS}}(M_Z)$.

\subsection{Hadronic $\tau$ decays}
\label{sec:hadtau}
A remarkably precise value for $\alpha_s^{\overline{\rm MS}}(M_Z)$
comes from hadronic $\tau$ decays. At first sight, this is quite
surprising because at the natural scale $\displaystyle{\mu=m_\tau}$ 
one has
approximately $\alpha_s(m_\tau) \simeq 0.35$. Can one expect
reasonable convergence of the perturbative series for such a large
coupling and how big are the nonperturbative corrections?

The first systematic investigation of $\displaystyle{R_\tau} 
= \Gamma(\tau^- \to 
\nu_\tau + {\rm hadrons})/\Gamma(\tau^- \to \nu_\tau e^- 
\overline{\nu_e})$ was performed by Braaten, Narison and Pich
\cite{bnp}. The analysis is similar to the one for
$\displaystyle{R_{\, e^+e^-}}$, with 
obvious modifications: the electromagnetic current (coupling to 
$\displaystyle{e^{\hspace*{1pt}+} e^{\hspace*{1pt}-}}$) must be 
replaced by the 
charged weak current (coupling to $\displaystyle{\tau \nu_\tau}$).

We start again with the two-point function (of weak currents 
$\displaystyle{L^{\, \mu} = \overline{u} \gamma^\mu 
(1-\gamma_5) d_\theta}$):
\begin{eqnarray} 
\Pi_L^{\mu\nu}(q) &=& i \displaystyle\int d^4x \, 
e^{\displaystyle  i q\cdot x}
\langle 0|\,T L^\mu(x) L^\nu(0)^\dg |0\rangle  \nn
&=&
\left(-g^{\mu\nu} q^2 + q^\mu q^\nu \right) \Pi_L^{(1)}(q^2) +
q^\mu q^\nu  \Pi_L^{(0)}(q^2)~,
\end{eqnarray} 
with $d_\theta$ the Cabibbo rotated $d$-quark field. One major
difference to the electromagnetic case is that one has to integrate
over the neutrino energy or, equivalently, over the hadronic
invariant mass $s$:
\begin{eqnarray} 
R_\tau = 12\pi \displaystyle\int_0^{m_\tau^2}\displaystyle\frac{ds}
{m_\tau^2} \left(1 - \displaystyle\frac{s}{m_\tau^2} \right)^2 
\left\{\left(1 + 2 \displaystyle\frac{s}{m_\tau^2}\right) 
\IM{\Pi_L^{(1)}(s)} + \IM{\Pi_L^{(0)}(s)} \right\}~. 
\end{eqnarray}
The problem is that the integration extends all the 
way down to $s=0$ (for $\displaystyle{m_u=m_d=0}$) where perturbation 
theory is certainly not applicable. 

\begin{center} 
\begin{figure}[ht] 
\leavevmode 
\centering\includegraphics[width=8cm]{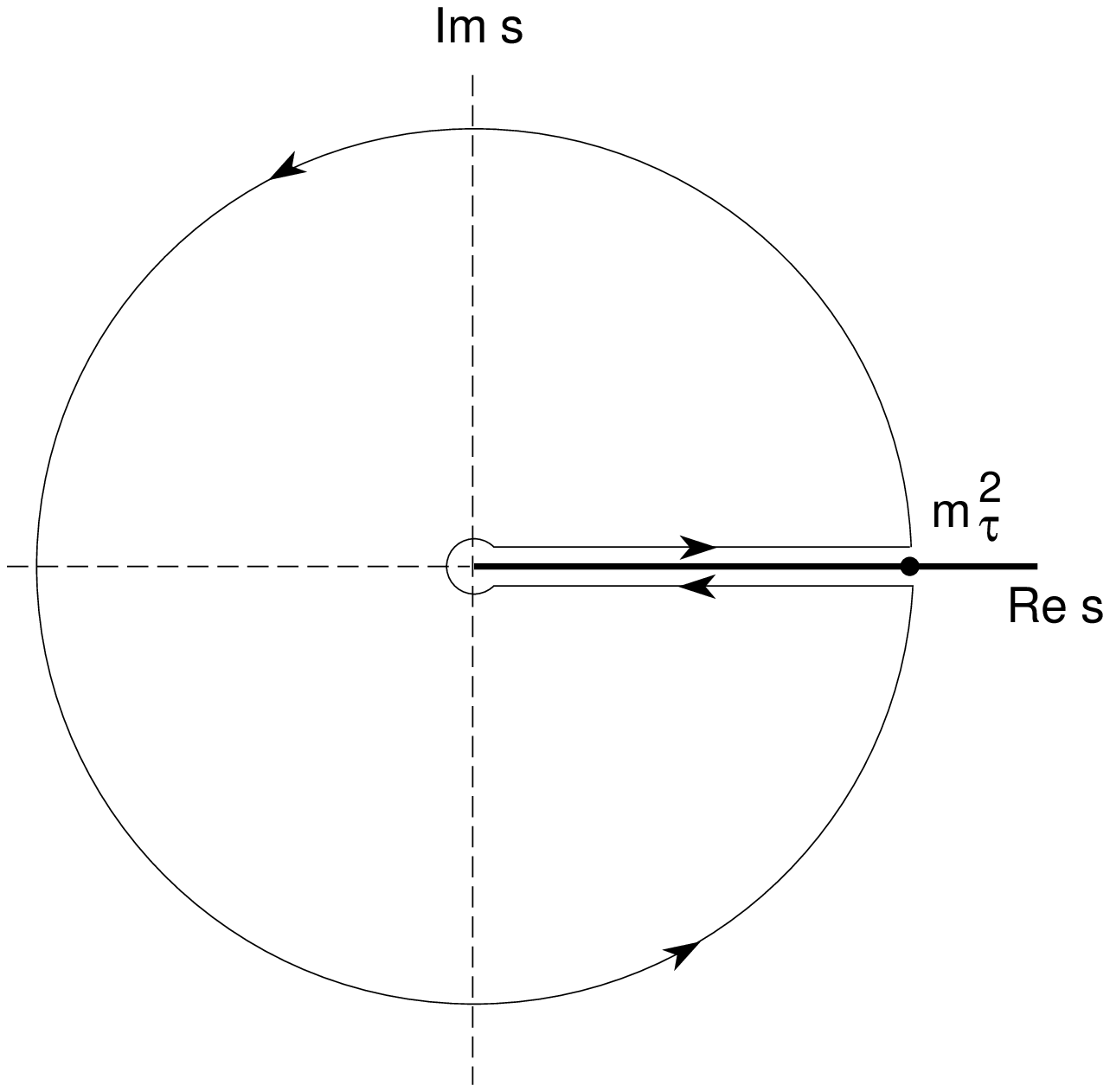}
\caption{Contour in the complex $s$-plane for the two-point functions
$\Pi_L^{(0,1)}(s)$.}
\label{fig:contour}
\end{figure} 
\end{center}

However, QFT provides information about the analytic structure of
two-point functions that can be used in a standard manner to
circumvent the problem. The invariant functions
$\displaystyle{\Pi_{\hspace*{1pt}L}^{\,(0,1)}}(s)$ are
known to be analytic in the complex $s$-plane with a cut on 
the positive real axis. Therefore, Cauchy's theorem tells us that 
the contour integral in Fig.~\ref{fig:contour} vanishes.

One can now trade the integral along the cut of 
\begin{equation}
\IM{\Pi_L^{(0,1)}(s)}=\displaystyle\frac{1}{2i}\left[\Pi_L^{(0,1)}(s+i\ve)-
\Pi_L^{(0,1)}(s-i\ve)\right]
\end{equation}
for an integral along the circle $|s|=m_\tau^2$ in the complex
$s$-plane. It turns out that the nonperturbative corrections are
now manageable, being suppressed as $(\Lambda_{\rm
QCD}/m_\tau)^6$. Very helpful in this respect is the factor $\left(1 - 
\displaystyle\frac{s}{m_\tau^2} \right)^2$ in the integrand that
suppresses potentially big contributions near the endpoint of the
cut. 

\vspace*{-.8cm}
\noindent  
The final result can be written in the form \cite{bnp}
\begin{equation}
R_\tau= 3 \left(|V_{ud}|^2 + |V_{us}|^2 \right) S_{\rm EW} \left\{1 +
\delta^\prime_{\rm EW} + \delta_{\rm pert} + \delta_{\rm nonpert} 
\right\}
\end{equation}
with leading and nonleading electroweak corrections 
$\displaystyle{S_{\hspace*{1pt}\rm EW}}=1.0194$ and 
$\delta^\prime_{\rm EW}= 0.0010$, respectively. The 
perturbative QCD corrections of interest for the extraction of
$\alpha_{s}$ are contained in
\begin{eqnarray} 
\delta_{\rm pert} &=& \displaystyle\frac{\alpha_s(m_\tau)}{\pi} +
\left(C_2 + \displaystyle\frac{19}{48}\beta_0   \right) \left( 
\displaystyle\frac{\alpha_s(m_\tau)}{\pi} \right)^2 + \dots \\
&=& \displaystyle\frac{\alpha_s(m_\tau)}{\pi} + 5.2 \left( 
\displaystyle\frac{\alpha_s(m_\tau)}{\pi} \right)^2 + 26.4 \left(
\displaystyle\frac{\alpha_s(m_\tau)}{\pi} \right)^3 +
O(\alpha_s(m_\tau)^4) ~.\no
\end{eqnarray} 
Finally, the best estimates of nonperturbative contributions, using
QCD sum rules and experimental input, yield
\begin{equation}
\delta_{\rm nonpert} = - 0.014 \pm 0.005~.
\end{equation}
From the PDG fit for $R_\tau$ one then obtains $\alpha_s(m_\tau)=0.35
\pm 0.03$. More interestingly, running this value down to $M_Z$ with
the help of the four-loop $\beta$ function, one finds
\begin{equation}
\alpha_s(M_Z) = 0.121 \pm 0.0007({\rm exp}) \pm 0.003({\rm th}),
\end{equation}
not only compatible but in fact very much competitive with other
high-precision determinations.

\subsection{Deep inelastic scattering}
From the conception of QCD till today, deep inelastic scattering of
leptons on hadrons has had an enormous impact on the field. It is also a
classic example for the factorization between long- and short-distance
contributions. 

Let us start with the kinematics of (in)elastic electron-proton 
scattering $\displaystyle{e^{\hspace*{1pt}-}(k) + p(p)} \to 
e^-(k^\prime) + X(p_X)$ 
shown in Fig.~\ref{fig:DIS}. 
In the case of elastic scattering ($X=p$), we have
\begin{eqnarray} 
W^2=m^2~,\qquad Q^2=2m\nu~, \qquad x=1
\end{eqnarray} 
and the usual two variables are $s=(p+k)^2, Q^2$ with the differential
cross section $d \sigma(s,Q^2)/d Q^2$.

For the inclusive scattering, there is a third independent variable: a
convenient choice is $s,x,y$ with $0 \le x \le 1,~~0 \le y \le 1$. In
general, we distinguish different types of deep inelastic scattering:
\begin{center} 
\begin{tabular}{lcl}
type & \hspace*{2cm} & exchange \\[.1cm]
\hline
& & \\[-.1cm]   
neutral current (NC) DIS & & $\gamma, ~Z, ~\gamma Z$-interference
\\[.1cm]
charged current (CC) DIS & & $W^\pm$ 
\end{tabular}
\end{center}  
For these lectures, I will restrict the discussion to photon
exchange and to unpolarized (spin-averaged) DIS. 
\\[-.2cm] 
\parbox[l]{0.5\textwidth}{
\begin{flushleft} 
\begin{figure}[H] 
\leavevmode 
\includegraphics[width=6.5cm]{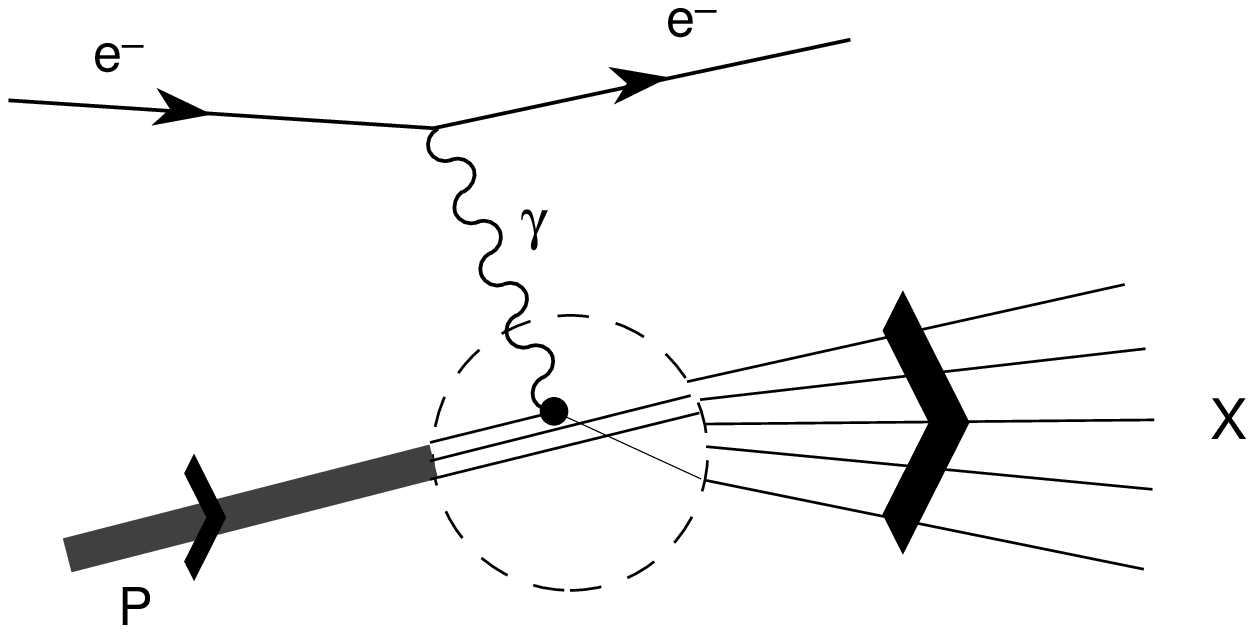}
\caption{Deep inelastic scattering.}
\label{fig:DIS}
\end{figure}
\end{flushleft} 
}

\vspace*{-4.5cm}
\noindent 
\hspace*{7.5cm} $q=k-k^\prime,~~Q^2=-q^2>0,~~p^2=m^2$ \\[.2cm] 
\hspace*{7.5cm} $\nu=p\cdot q/m = E - E^\prime$~~(target rest frame)
\\[.2cm]
\hspace*{7.5cm} $x=\displaystyle\frac{Q^2}{2 m \nu},
~~y=\displaystyle\frac{p\cdot q}{p\cdot k}=1 - E^\prime/E$ \\[.2cm] 
\hspace*{7.5cm} $W^2=p_X^2=(p+q)^2=m^2 +2 m \nu -Q^2 \ge m^2~.$ 
\\[1.4cm]
The matrix element for the diagram in Fig.~\ref{fig:DIS} has the
structure $e\,l({\rm epton})_\mu ~
\displaystyle\frac{g^{\mu\nu}}{Q^2}~ e\,h({\rm adron})_\nu$, with 
leptonic and hadronic current matrix elements $l({\rm epton})_\mu$ and
$h({\rm adron})_\nu$, respectively. The
resulting double differential cross section is of the form
\begin{eqnarray}
\displaystyle\frac{d^2 \sigma}{dx dy} &=& x (s-m^2)
\displaystyle\frac{d^2 \sigma}{dx dQ^2} = \displaystyle\frac{2\pi y
\alpha^2}{Q^4} L_{\mu\nu}\,H^{\mu\nu} \\[.1cm]
L_{\mu\nu} &=& 2(k_\mu k^\prime_\nu +  k^\prime_\mu k_\nu - k\cdot
k^\prime g_{\mu\nu}) \nn
H^{\mu\nu}(p,q) &=& \displaystyle\frac{1}{4\pi} \displaystyle\int d^4 z
e^{\displaystyle i q\cdot z} \langle p,s|\left[J_{\rm elm}^\mu(z),
J_{\rm elm}^\nu(0) \right]|p,s \rangle ~.\no  
\end{eqnarray}
One now performs a Lorentz decomposition of the hadronic tensor
$H^{\hspace*{.5pt}\mu\nu}$ and contracts it with the 
leptonic tensor 
$L_{\mu\nu}$ (setting $m_e=0$). The differential cross section then 
depends on two invariant structure functions $F_1, F_2$ (in the
photon case), which are themselves functions of the scalars $p.q$,~$q^2$ 
or $\nu, Q^2$ or $x, Q^2$:
\begin{eqnarray}
\displaystyle\frac{d^2 \sigma}{dx dy} = \displaystyle\frac{Q^2}{y}
\displaystyle\frac{d^2 \sigma}{dx dQ^2} = \displaystyle\frac{4\pi 
\alpha^2}{x y Q^2} \left\{\left(1-y - \displaystyle\frac{x^2 y^2
m^2}{Q^2}\right) F_2(x,Q^2) + y^2 x F_1(x,Q^2)   \right\}~. 
\label{eq:sigxy}
\end{eqnarray}
Deep inelastic scattering corresponds to $\displaystyle{Q^2} \gg m^2$ 
and $W^2 \gg m^2$. While the cross section shows a rather complicated 
behaviour at low and intermediate momentum transfer, the structure 
functions exhibit an originally unexpected simple behaviour in the 
so-called Bjorken limit
\begin{eqnarray}
 Q^2 \gg m^2,\quad \nu \gg m \quad {\rm with} \quad
x=\displaystyle\frac{Q^2}{2 m \nu} ~~{\rm fixed}~.
\end{eqnarray}
As shown for $F_2$ in Fig.~\ref{fig:scaling}, 
in the Bjorken limit the structure functions seem to depend on the
variable $x$ only:
\begin{equation}
F_i(x,Q^2) \lra F_i(x)~. 
\end{equation}
This scaling behaviour suggested that the photon scatters off
point-like constituents (no scale) giving rise to the quark parton
model (QPM).

\vspace*{.3cm} 
\begin{center} 
\vspace*{.3cm} 
\begin{figure}[H] 
\leavevmode 
\includegraphics[width=5cm,angle=-90]{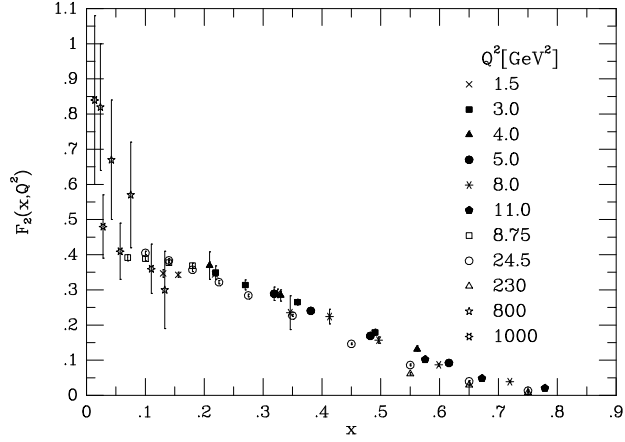}
\vspace*{.2cm} 
\caption{Evidence for Bjorken scaling taken from Ref.~\cite{esw}.}
\label{fig:scaling}
\end{figure}
\end{center} 
\begin{center}
\begin{fmpage}{4cm}
\begin{center}
\vspace*{.1cm}
QPM in the Breit frame
\vspace*{.1cm}
\end{center} 
\end{fmpage}
\end{center} 
The characteristics of the QPM can best be visualized in the so-called
Breit frame where the proton and the virtual photon collide head-on.\\
\parbox[l]{0.5\textwidth}{
\begin{center}  
\begin{figure}[H] 
\leavevmode 
\includegraphics[width=6cm]{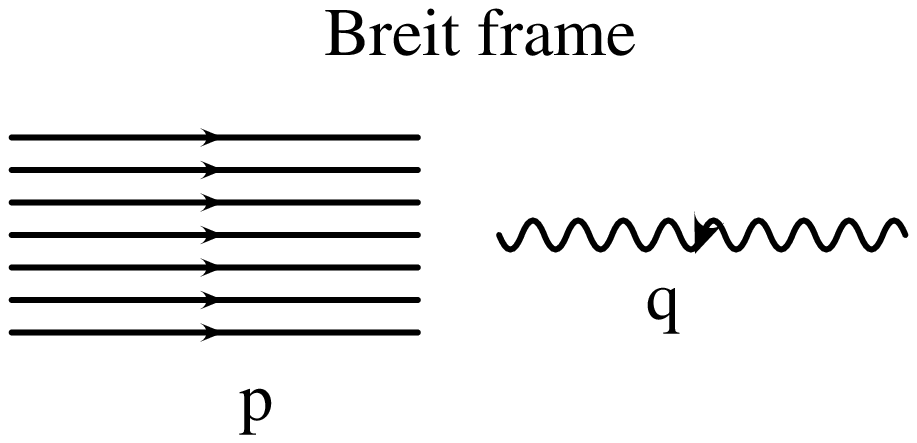}
\caption{DIS in the Breit frame.}
\label{fig:breit}
\end{figure}
\end{center}  
}
   
\vspace*{-4cm}
\hspace*{7cm}
\parbox[l]{0.5\textwidth}{
\begin{eqnarray}
p &\simeq & (P,P,0,0) ~~{\rm with} ~~P \gg m \no \\[.2cm] 
q &=& (0,-\sqrt{Q^2},0,0) \no
\end{eqnarray} 
}  

\vspace*{1.3cm}
\noindent 
The nucleon is pictured as a bunch of partons with negligible
transverse momenta. Each parton carries a fractional momentum 
$\xi \,p$. Since the scattered quark is massless (compared to $P$), 
we have
\begin{eqnarray} 
(q+\xi p)^2 \simeq - Q^2 +2 \xi p.q = 0
\end{eqnarray} 
and therefore
\begin{eqnarray}
\xi=x~,\qquad P=\displaystyle\frac{\sqrt{Q^2}}{2 x}~, \qquad 
q + x p =(xP,-\sqrt{Q^2}/2,0,0)~.
\end{eqnarray} 
The struck parton scatters with momentum $q + x p$ backwards, i.e. in 
the direction of the virtual photon, justifying a major assumption of
the QPM: the virtual  photon scatters incoherently on the partons.

The fundamental process of the QPM is elastic electron-quark
scattering 
\begin{equation}
e^-(k) + q(\xi p) \to e^-(k^\prime) + q(\xi p + q)~.
\end{equation}
Since there are now only two independent variables, the double
differential cross section in Eq.~(\ref{eq:sigxy}) contains a
$\delta$-function setting $x$ equal to $\xi$:
\begin{eqnarray} 
\displaystyle\frac{d^2 \sigma_{(q)}}{dx dy} = \displaystyle\frac{4\pi 
\alpha^2}{y Q^2}\left[1 + (1-y)^2 \right] \displaystyle\frac{Q_q^2}{2}
\delta(x-\xi)~.   
\end{eqnarray} 
In the notation of Eq.~(\ref{eq:sigxy}),
\begin{eqnarray}
F_{2(q)} = x Q_q^2 \delta(x-\xi) = 2 x F_{1(q)}~.
\end{eqnarray}  
The incoherent sum of partonic cross sections amounts to an integral
over quark distribution functions $q(\xi), \overline{q}(\xi) $:
\begin{eqnarray} 
F_2(x) = \displaystyle\sum_{q,\overline{q}} \displaystyle\int_0^1 d\xi
\,q(\xi) \,x \,Q_q^2 \,\delta(x-\xi) = \displaystyle\sum_{q,\overline{q}}
Q_q^2 \,x \,q(x)~, 
\end{eqnarray} 
implying the Callan-Gross relation \cite{Callan:1969uq}
\begin{equation}
F_2(x) = 2x F_1(x)
\end{equation}
that is due to the spin-1/2 nature of quarks. The so-called
longitudinal structure function $F_{\hspace*{1pt}L}=F_2 - 2 x F_1$ 
therefore vanishes in the QPM.

\begin{center} 
\begin{figure}[ht] 
\leavevmode 
\centering\includegraphics[width=8cm]{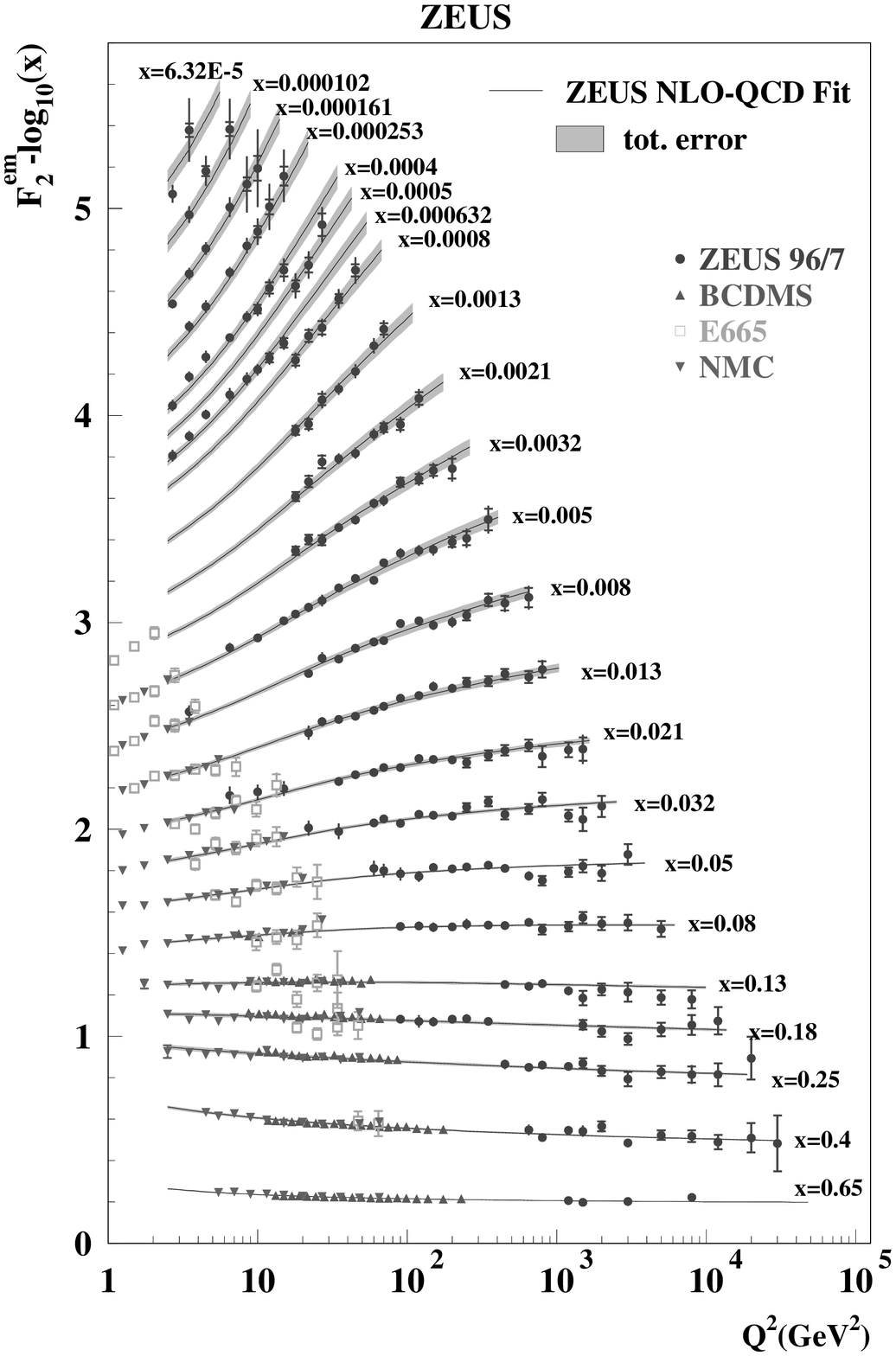}
\vspace*{.2cm} 
\caption{Experimental evidence for the violation of Bjorken scaling
taken from Ref.~\cite{Kappes:2002dz}.}
\label{fig:scaleviol}
\end{figure}
\end{center}

It was already known at the beginning of the seventies, before the
advent of QCD, that exact scaling in the sense of the QPM was
incompatible with a nontrivial QFT. QCD must therefore account for the
systematic deviation from scaling that is clearly seen in the data
(e.g., in Fig.~\ref{fig:scaleviol}): with increasing $Q^2$, the
structure function $F_2$ increases (decreases) at small (large) $x$.
Qualitatively, scaling violation is due to the radiation of (hard)
gluons generating transverse momenta for the quarks. More gluons are
radiated off when $Q^2$ increases, leading to 
logarithmic scaling
violations in the structure functions and to scale dependent parton 
distribution functions (pdf) $q_i(x,\mu^2),~g(x,\mu^2)$ as we will now
discuss in more detail.
\begin{center} 
\begin{fmpage}{2.3cm}
\vspace*{.1cm}
\begin{center}
DIS in QCD
\end{center}
\vspace*{.1cm}
\end{fmpage}
\end{center}
In QCD at leading order in $g_s$, the same diagrams need to be
considered as the ones in Figs.~\ref{fig:3jet}, \ref{fig:eeqqloop} 
relevant for $e^+ e^- \to$ hadrons except for crossing. 

Previously, the sum of real and virtual gluon emission was infrared 
finite because a sum over all final state quarks and gluons was
involved. In DIS the situation is different because the initial state 
contains a quark. Since different incoming quark momenta are in principle
distinguishable, gluons collinear with the incoming quark generate in
fact an infrared divergence. How to get rid of this divergence will be
discussed later but for now we regulate the infrared divergence with a
cutoff on the transverse quark momentum $\displaystyle{k_\perp^2} \ge
\kappa^2$. Adding the contribution to $F_{2(q)}$ from real gluon
emission ($\xi=1$ for simplicity) one finds
\begin{equation}
F_{2(q)}(x,Q^2)=Q_q^2 \,x \left[\delta(1-x) +
\displaystyle\frac{\alpha_s}{2\pi} \left(P_{qq}(x)
\log{\displaystyle\frac{Q^2}{\kappa^2} } + C_q(x) \right)  \right]~.
\label{eq:F2q}
\end{equation}
Introduction of the cutoff $\kappa$ has produced a logarithmic
dependence on $Q^2$. This dependence is governed by the 
so-called
quark-quark splitting function $P_{qq}(x)$. This function is
universal, in contrast to the non-logarithmic coefficient function
$C_q(x)$ that is scheme dependent. 

The origin of the scheme-independent $\log{Q^2}$ can 
be understood as
follows. The struck quark acquires a transverse momentum $k_\perp$ 
with probability $\alpha_s
\displaystyle\frac{dk_\perp^2}{k_\perp^2}$. Since $k_\perp^2$ cannot
be bigger than $Q^2$, integrating over all $k_\perp$ produces the term 
$\alpha_s \log{Q^2/\kappa^2} $. Virtual gluon contributions
(diagrams in Fig.~\ref{fig:eeqqloop}) must still be added and they are
ultraviolet finite as before. Since this is a contribution from elastic
scattering it must be proportional to $\delta(1-x)$. Altogether, the
quark-quark splitting function at leading order is 
\begin{equation}
P_{qq}(x)=\displaystyle\frac{4}{3}\left(
\displaystyle\frac{1+x^2}{[1-x]_+}\right) + 2\,\delta(1-x) ~.
\label{eq:splfunc}
\end{equation}
Also the first part of $P_{qq}(x)$ is actually a distribution. The
distribution $\left[F(x) \right]_{+}$ is defined in such a way that for
every sufficiently regular (test) function $f(x)$ one has
\begin{equation}
\displaystyle\int_0^1 dx\,f(x)\left[F(x) \right]_+ =
\displaystyle\int_0^1 dx\,\left(f(x)-f(1) \right) F(x)~.
\end{equation}
It is then straightforward to show that $P_{qq}(x)$ in
Eq.~(\ref{eq:splfunc}) can also be written as
\begin{equation}
P_{qq}(x)=\displaystyle\frac{4}{3}  
\left[\displaystyle\frac{1+x^2}{(1-x)}\right]_+~.
\label{eq:spf}
\end{equation}

The problem remains how to interpret (or rather get rid of) the
infrared cutoff $\kappa$. Up to now, we have only considered the quark
structure function $F_{2(q)}(x,Q^2)$. To get 
$F_2(x,Q^2)$ for the nucleon, we convolute
$F_{2(q)}(\displaystyle\frac{x}{\xi},Q^2)$ with a (bare) pdf
$q_0(\xi)$:
\begin{equation}
F_2(x,Q^2) = x \displaystyle\sum_{q,\overline{q}} Q_q^2 \left[q_0(x) +
\displaystyle\frac{\alpha_s}{2\pi} \displaystyle\int_x^1
\displaystyle\frac{dy}{y} q_0(y) \left\{P_{qq}(x/y)
\log{\displaystyle\frac{Q^2}{\kappa^2} } + C_q(x/y) \right\}
\right]~.
\end{equation}
One now absorbs the collinear singularity $\sim \log{\kappa^2}$ into 
$q_0(x)$ at a factorization scale $\mu$ to define a
renormalized pdf  $q(x,\mu^2)$:
\begin{equation}
q(x,\mu^2)=q_0(x) + \displaystyle\frac{\alpha_s}{2\pi} \displaystyle\int_x^1
\displaystyle\frac{dy}{y} q_0(y) \left\{P_{qq}(x/y)
\log{\displaystyle\frac{\mu^2}{\kappa^2}} + C_q^\prime(x/y) \right\}~.
\end{equation}
The interpretation of $q(x,\mu^2)$ is straightforward: the soft part 
$k_\perp^{2} \le \mu^2$ is now included in the pdf. Since the scale
$\mu$ is arbitrary, the renormalized pdf is necessarily scale
dependent, in complete analogy with the renormalization of ultraviolet
divergences. As a small aside, we take note that the coefficient
function $C_q^\prime$ need not be the same as $C_q$ in
Eq.~(\ref{eq:F2q}), both being scheme dependent. The final form for
the nucleon structure function $F_2$ in the $\overline{\rm MS}$ 
scheme (except for a contribution from the gluon pdf) is then
\begin{eqnarray} 
F_2(x,Q^2) & = & x \displaystyle\sum_{q,\overline{q}} Q_q^2
\displaystyle\int_x^1 \displaystyle\frac{dy}{y} q(y,\mu^2) 
\left[\delta(1-x/y) + \displaystyle\frac{\alpha_s}{2\pi} 
 \left\{P_{qq}(x/y) \log{\displaystyle\frac{Q^2}{\mu^2}} + 
C_q^{\overline{\rm MS}}(x/y) \right\} \right] \no  \\[.1cm] 
&=& x \displaystyle\sum_{q,\overline{q}} Q_q^2
\displaystyle\int_x^1 \displaystyle\frac{dy}{y} q(y,Q^2) 
\left[\delta(1-x/y) + \displaystyle\frac{\alpha_s}{2\pi}
C_q^{\overline{\rm MS}}(x/y) \right]~.
\label{eq:fac}
\end{eqnarray}
This factorization formula can be proven to all orders in
$\alpha_{s}$, separating the calculable 
hard part from the soft part contained in the scale dependent
pdfs. The pdfs $\displaystyle{q(x,\mu^{2})}$, 
$\displaystyle{\overline{q}(x,\mu^{2})}$, $\displaystyle{g(x,\mu^{2})}$ 
describe the composition of nucleons and are, of course, not calculable in
perturbation theory. They can be extracted from experimental data with
the help of appropriate parametrizations (cf., e.g., Ref.~\cite{esw})
but the question remains what the factorization result (\ref{eq:fac})
is  actually good for. 

The answer is that even though the functional dependence of the pdfs
cannot be calculated their scale dependence is calculable in QCD
perturbation theory. The derivation of the so-called DGLAP evolution
equations \cite{DGLAP} is very similar to the derivation of the
$\beta$ function in Sec.~\ref{sec:hist}, starting from the observation
that the measurable structure function must be scale independent:
\begin{eqnarray}
\mu^2 \displaystyle\frac{d F_2(x,Q^2)}{d\mu^2} = 0 \quad & \lra &
\quad \mu^2 \displaystyle\frac{d q(x,\mu^2)}{d\mu^2} =
\displaystyle\frac{\alpha_s(\mu)}{2\pi} \displaystyle\int_x^1
\displaystyle\frac{dy}{y} P_{qq}(x/y,\alpha_s(\mu)) q(y,\mu^2) 
\end{eqnarray}
\vspace*{-.4cm} 
\begin{eqnarray} 
P_{qq}(x,\alpha_s(\mu)) &=& P_{qq}^{(0)}(x) +
\displaystyle\frac{\alpha_s(\mu)}{2\pi} P_{qq}^{(1)}(x) + \dots 
\end{eqnarray} 
With $P_{qq}^{(0)}(x)$ given by Eq.~(\ref{eq:spf}), the evolution
equation at leading order takes the explicit form
\begin{eqnarray} 
\mu^2 \displaystyle\frac{d q(x,\mu^2)}{d\mu^2} &=&
\displaystyle\frac{2\alpha_s(\mu)}{3\pi} \displaystyle\int_x^1
\displaystyle\frac{dz}{z}q(x/z,\mu^2) \displaystyle\frac{1+z^2}{1-z}
- \displaystyle\frac{2\alpha_s(\mu)}{3\pi} q(x,\mu^2)
\displaystyle\int_0^1 dz \displaystyle\frac{1+z^2}{1-z} ~.
\end{eqnarray} 
Due to soft gluons, both terms on the right-hand side are divergent: 
the first term with positive sign is due to quarks with momentum
fraction larger than $x$ radiating off gluons whereas the second term
leads to a decrease from quarks with given $x$ that radiate
gluons. The overall result is finite.

At $O(\alpha_s)$ also the gluon pdf enters via $\gamma^{*} + g \to q + 
\overline{q}$. At any order, the DGLAP equations
are in general $(2 N_F +1)$-dimensional matrix equations for 
$\displaystyle{q_{i}(x,\mu^2)}$, $
\displaystyle{\overline{q_{i}}(x,\mu^2) ~(i=1,\dots,N_F)}$,
$\displaystyle{g(x,\mu^2)}$ with splitting functions $P_{qq}(x)$, 
$P_{qg}(x)$, $P_{gq}(x)$ and $P_{gg}(x)$. The analytic calculation of
these splitting functions to next-to-next-to-leading order (three
loops) has just been completed \cite{Vermaseren:2005qc} allowing for 
precise tests of scaling violations. Finally, we 
note that the longitudinal structure function
$F_{L}(x,Q^2)$  that was
zero in the QPM is generated in QCD already at  $O(\alpha_{s})$.

For more applications of perturbative QCD I refer
once again to the book of Ellis, Stirling and Webber \cite{esw}: 
jets in $e^+ e^-$ and hadroproduction, vector boson
production (Drell-Yan), heavy quark production and
decays, Higgs production at the LHC, \dots

\section{Heavy and light quarks}

\subsection{Effective field theories}
\label{sec:EFT}
Unlike QED, QCD is valid down to shortest distances because of
asymptotic freedom. However, at long distances where quarks and gluons
are practically invisible, perturbative QCD is not applicable and a
nonperturbative approach is needed. Many models can be found in the
literature that are more or less inspired by QCD. Qualitative insights
into the structure of the strong interactions have been found from
model studies but quantitative predictions require methods that can be
related directly to QCD. There are essentially only two approaches
that satisfy this criterion.
\begin{dinglist}{42}
\item Lattice QCD has already scored impressive results and may in the
long run be the most predictive method. At present, the range of
applicability is still limited.
\item Effective field theories (EFTs) are the quantum field
theoretical formulation of the ``quantum ladder'': the relevant
degrees of freedom depend on the typical energy of the problem. EFTs
become practical tools for phenomenology when the characteristic
energy scales are well separated.  
\end{dinglist}   
Let a given step of the quantum ladder be characterized by an
energy scale $\Lambda$. The region $E > \Lambda$ is the short-distance
region where the fundamental theory is applicable. At long
distances ($E < \Lambda$), on the other hand, an effective QFT can and 
sometimes must be used. By definition, the notions ``fundamental'' and
``effective'' only make sense for a given energy scale $\Lambda$. As
we probe deeper into the physics at short distances, today's
fundamental theory will become an effective description of an even
more fundamental underlying theory.

To understand the different effective field theories that are being
used in particle physics, it is useful to classify them
as to the structure of the transition between the fundamental and
the effective level.
\begin{dinglist}{108} 
\item \underline{Complete decoupling} \\[.1cm]
The heavy degrees of freedom (heavy with respect to 
$\Lambda$) are integrated out, i.e., they disappear from the
spectrum of states that can be produced with energies $< \Lambda$. 
Correspondingly, the effective Lagrangian contains only light fields 
(once again, light stands for masses $< \Lambda$) and may be written
symbolically as
\begin{equation}
\cL_{\rm eff} = \cL_{d\le 4} + \displaystyle\sum_{d>4}
\displaystyle\frac{1}{\Lambda^{d-4}} 
\displaystyle\sum_{i_d} g_{i_d} O_{i_d} ~.
\label{eq:Leff}
\end{equation}        
The first part $\cL_{d\le 4}$ contains all the renormalizable
couplings for the given set of fields. The best-known example for
such a Lagrangian is the Standard Model itself. The second part of the
Lagrangian (\ref{eq:Leff}) contains the nonrenormalizable couplings
having operator dimension $d>4$. The best-known example here is the
Fermi theory of weak interactions with $d=6$ and $\Lambda =
M_{W}$. For the Standard Model, on the other hand, we do 
not know the scale where new physics will appear. Present
experimental evidence implies $\Lambda >$ 100 GeV but there are good
reasons to expect new physics around $\Lambda \sim 1$ TeV.
\item \underline{Partial decoupling} \\[.1cm]
In this case, heavy fields do not disappear completely in the EFT. Via
so-called field redefinitions, only the high-momentum modes are
integrated out. The main application of this scenario in particle
physics is for heavy quark physics.
\item \underline{Spontaneous symmetry breaking (SSB)} \\[.1cm]
In the previous two classes, the transition from the
fundamental to the effective level was smooth. Some of the
fields or at least their high-energy modes just drop out and the
effective description involves the remaining fields only. In the
present case, the transition is more dramatic and involves a phase
transition: SSB generates new degrees of freedom, the
(pseudo-)Goldstone bosons associated with spontaneously broken
symmetries (to be discussed in more detail in Sec.~\ref{sec:SCSB}). 
The prefix pseudo accounts for the frequent case that the
symmetry in question is not only spontaneously but also explicitly
broken. Goldstone bosons in the strict sense are massless and the
associated SSB relates processes with different numbers of Goldstone
bosons. As a consequence, the distinction in the Lagrangian 
(\ref{eq:Leff}) between renormalizable ($d \le 4$) and
nonrenormalizable ($d > 4$) terms becomes meaningless. Therefore, EFTs
in this category are generically nonrenormalizable. An important but 
maybe too simple exception is the Higgs model for electroweak SSB. 

The generic EFT Lagrangian is organized in the number of derivatives of 
Goldstone fields and in the number of terms with explicit symmetry
breaking. An important concept is universality: it turns out that
EFTs describing different physical situations have very similar
structure. In QCD, the symmetry in question is chiral symmetry that
becomes exact in the limit of massless quarks. In the real world, SSB
of chiral symmetry generates pseudo-Goldstone bosons that are
identified with the pseudoscalar mesons $\pi, K, \eta$.    
\end{dinglist} 
We are used to derive quantitative predictions from renormalizable
QFTs in the framework of perturbation theory but how should we treat
nonrenormalizable EFTs? The clue to the answer is Goldstone's theorem
\cite{Gold} that makes two crucial predictions. The first prediction
is well known: SSB implies the existence of massless Goldstone
bosons. The second consequence of
Goldstone's theorem is not that well known but very important as well: 
Goldstone bosons decouple when their energies tend to
zero. In other words, independently of the strength of the underlying
interaction (the strong interaction in our case!), Goldstone bosons
interact only weakly at low energies. This important feature allows
for a systematic expansion of strong amplitudes even in the
confinement regime, which is precisely the low-energy regime of QCD.

However, in contrast to the decoupling case (e.g., in heavy quark
physics), the coupling constants of the low-energy EFT cannot be
obtained by perturbative matching with the underlying theory of
QCD. Other methods have to be used to get access to the low-energy
couplings.

\subsection{Heavy quarks}
\label{sec:hqm}
Quarks cannot be put on a balance, or more realistically,
their energies and momenta cannot be measured directly. How do we then 
determine their masses? Two methods have been used.
\begin{dinglist}{42}
\item The first approach ignores confinement and calculates the pole
of the quark propagator just as we determine at least in theory the
mass of the electron. This looks rather artificial because the full
quark propagator should have no pole because of
confinement. Going ahead nevertheless, one expects those pole masses
to be very much affected by nonperturbative infrared effects. In
practice, this method is only used for the top quark with \cite{PDG04}
\begin{equation}
m_t = 174.3 \pm 5.1 ~{\rm GeV}~.
\end{equation}
\item Quark masses are parameters of the QCD Lagrangian just like the
strong coupling constant $g_{s}$. One therefore studies 
the influence of these parameters on measurable quantities and
extracts specific values
for the masses by comparison with experimental measurements. As for
the strong coupling constant, renormalization is crucial and
introduces a scale dependence also for quark masses. The scale
dependence is governed by a differential equation very similar to the
renormalization group equations (\ref{eq:beta_gs}) or 
(\ref{eq:beta_alpha}) for the strong coupling:
\begin{equation}
\mu \displaystyle\frac{d m_q(\mu)}{d \mu} = -
\gamma(\alpha_s(\mu)) m_q(\mu)
\end{equation}
where the anomalous dimension $\gamma$ is nowadays known up to
four-loop accuracy:
\begin{equation}
\gamma(\alpha_s) = \displaystyle\sum_{n=1}^4 \gamma_n
\left(\displaystyle\frac{\alpha_s}{\pi} \right)^n~.
\end{equation}
The solution of this renormalization group equation for 
$m_{q}(\mu)$ is
flavour independent (in the $\overline{\rm MS}$ scheme):
\begin{equation}
m_q(\mu_2) = m_q(\mu_1) \exp\left\{ - 
\displaystyle\int_{\alpha_s(\mu_1)}^{\alpha_s(\mu_2)}
dx\,\displaystyle\frac{\gamma(x)}{2\beta(x)}
\right\}~.  
\end{equation}
Since $\gamma(\alpha_s)$ is positive, quark masses decrease with
increasing scale  $\mu$, e.g., 
\begin{equation} 
\displaystyle\frac{m_q(1~{\rm GeV})}{m_q(M_Z)} = 2.30 \pm 0.05 ~.
\end{equation} 
\end{dinglist}  

\begin{center} 
\begin{figure}[ht] 
\leavevmode 
\centering\includegraphics[width=12.5cm]{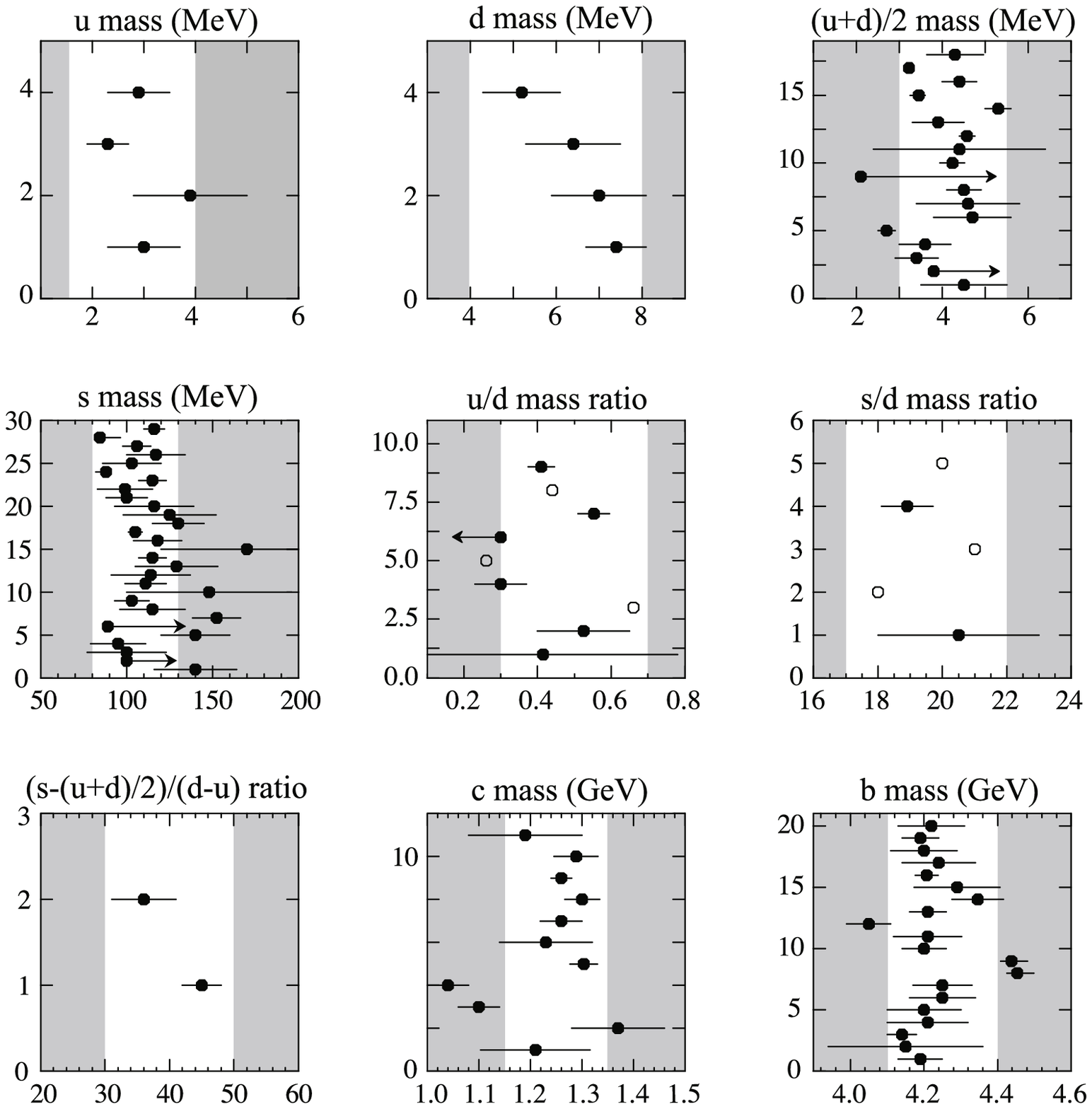}
\caption{The values of quark masses taken from the 2004 Review of
Particle Properties \cite{MS_PDG04}. The most recent data points are
at the top of each plot.}
\label{fig:qm}
\end{figure}
\end{center}

Different methods have been used to determine the quark masses: 
H(eavy) Q(uark) E(ffective) T(heory) (see below), QCD sum rules
(section \ref{sec:QCDSR}), lattice QCD, \dots The current state is
summarized in Fig.~\ref{fig:qm} taken from Ref.~\cite{MS_PDG04}. All
values correspond to the $\overline{\rm MS}$ scheme: light quarks are
given at the scale 2 GeV whereas the heavy quarks $\displaystyle{m_c,
m_b}$  are listed as $\displaystyle{m_q(m_q)}$.
  
\begin{center} 
\begin{fmpage}{6cm}
\vspace*{.1cm}
\begin{center}
Heavy quark effective theory (HQET)
\end{center}
\vspace*{.1cm}
\end{fmpage}
\end{center}
Why should one use an EFT for $b$-quark physics? After all, QCD is
still accessible in perturbation theory for $\mu=m_b$. The main
arguments in favour of HQET are the following.
\begin{dinglist}{42} 
\item Although the hard effects are calculable in QCD perturbation
theory, there are inevitably incalculable soft effects because hadrons
rather than quarks and gluons appear in the final states of $B$
decays. The necessary separation between perturbative and
nonperturbative contributions is much easier to achieve in an EFT
description. The keyword is the same as in deep inelastic scattering:
factorization. 
\item Approximate symmetries that are hidden in full QCD become
manifest in an expansion in $1/m_{Q}$.
\item Explicit calculations simplify in general, in particular the
resummation of large logs via renormalization group equations.
\end{dinglist} 
The physics behind HQET can be understood by an analogy with atomic
physics. The spectrum of the hydrogen atom is to a good approximation
insensitive to the proton mass. In fact, the atomic spectra of
hydrogen and deuterium are practically identical. The implementation
of this analogy is most straightforward for hadrons with a single
heavy quark ($b$ or $c$). In the hadron rest frame the heavy quark 
``just sits there'' acting as a colour source just as the proton acts 
as a Coulomb source in the hydrogen atom.

We decompose the momentum of the heavy quark as
\begin{equation}
p^\mu = m_Q v^\mu + k^\mu~,
\end{equation} 
where $v$ is the hadron velocity normalized to $v^{2}=1$ 
($v=(1,0,0,0)$ in the hadron rest frame). The residual
quark momentum $k$ is then expected to be of $O(\Lambda_{\rm
QCD})$ only. Starting from the QCD Lagrangian for a heavy quark $Q$,
\begin{equation}
\cL_Q = \overline{Q} (i \slashed{D} - m_Q) Q~,
\end{equation}
we decompose the quark field $Q(x)$ into two fields $h_{v}(x)$ and
$H_{v}(x)$ by using energy projectors $P_v^\pm=(1 \pm \slashed{v})/2$ and
applying (shift) factors $e^{\,i m_Q v\cdot x}$:
\begin{eqnarray}
\label{redef}
Q(x) = e^{-i m_Q v\cdot x} \left(h_v(x) + H_v(x) \right) & &\\
h_v(x) = e^{i m_Q v\cdot x} P_v^+ Q(x)~, & & H_v(x) = 
e^{i m_Q v\cdot x} P_v^- Q(x)~. \no
\end{eqnarray}
It is easy to check that in the hadron rest frame the fields 
$h_{v}$ and $H_{v}$ are just the upper (big) 
and lower (small) components of the spinor field $Q$, respectively. 
Expressing $\displaystyle{\cL_Q}$ in terms of $\displaystyle{h_v}$
and $H_v$, one finds
\begin{eqnarray}  
\cL_Q &=& \overline{Q} (i \slashed{D} - m_Q) Q \nn
&=& \overline{h_v} \,i v \cdot D \,h_v - \overline{H_v} ( i v\cdot D + 2
m_Q ) H_v + {\rm mixed~~terms}~.
\end{eqnarray} 
For the purpose of illustration, we use the field equation 
$(i \slashed{D}  - m_Q) Q=0$ to eliminate $H_{v}$ in this Lagrangian:
\begin{equation}
\cL_Q = \overline{h_v} \,i v \cdot D \,h_v +
\overline{h_v} \,i \slashed{D}_\perp \frac{1}{i v \cdot D + 2 m_Q -
  i\epsilon} i \slashed{D}_\perp h_v \quad {\rm with} \quad
D^\mu_\perp=(g^{\mu\nu}-v^\mu v^\nu) D_\nu ~.
\end{equation}
The heavy quark mass $m_{Q}$ has disappeared from the kinetic term of
the shifted field $h_{v}$ and has moved to the
denominator of a nonlocal Lagrangian that is in fact the starting 
point for a systematic expansion in $1/m_{Q}$.

The Lagrangian for $b$ and $c$ quarks to leading-order in $1/m_{Q}$ is
therefore 
\begin{equation} 
\cL_{b,c}= \overline{b_v} \,i v \cdot D \,b_v + 
\overline{c_v} \,i v \cdot D \,c_v ~.
\label{eq:LHQET}
\end{equation} 
This Lagrangian exhibits two important symmetries. The symmetries
are only approximate because the Lagrangian (\ref{eq:LHQET}) is not
full QCD but the first approximation in an expansion in 
$1/m_{Q}$. The
symmetries are manifest in (\ref{eq:LHQET}) but they are hidden in
full QCD.
\begin{dinglist}{42}
\item A heavy-flavour symmetry  $SU(2)$ relates $b$
and $c$ quarks  moving with the same velocity.
\item Because there is no Dirac matrix in the Lagrangian
(\ref{eq:LHQET}), both spin degrees of freedom couple to gluons in the
same way. Together with the flavour symmetry, this leads to an overall 
spin-flavour symmetry $SU(4)$.
\end{dinglist} 
The simplest spin-symmetry multiplet $\cM$ consists of a pseudoscalar
$M$ and a vector meson $M^{*}$. One of the first
important applications
of spin-flavour symmetry was for the semileptonic decays 
$B \to D^{(*)} l \nu_l$. In general, there are several 
form factors governing the two matrix elements (for $D$ and 
$D^{*}$). To leading order in $1/m_{Q}$, 
all those form factors are given up to
Clebsch-Gordan  coefficients by a single function $\xi(v\cdot
v^\prime)$ called Isgur-Wise function:
\begin{eqnarray} 
\langle \cM(v^\prime) | \overline{h_{v^\prime}} \Gamma h_v |\cM(v)
\rangle \sim \xi(v\cdot v^\prime)~.
\end{eqnarray}
Moreover, $\overline{h_{v}} \gamma^\mu h_{v}$ is the conserved current of 
heavy-flavour symmetry. Similar to electromagnetic form factors that
are normalized at $q^{2}=0$ due to charge conservation, the Isgur-Wise 
function is fixed in the no-recoil limit $v=v^{\prime}$ to be
\begin{equation} 
\xi(v\cdot v^\prime =1)=1~.
\end{equation} 
Of course, there are corrections to this result valid only in the
symmetry limit, both of $O(\alpha_{s})$ and in general of 
$O(1/m_{Q})$. For the decay $B \to D^{*} l \nu_l$, the 
leading mass corrections turn out to be of $O(1/m_{Q}^2)$
only. HQET provides the standard method for the
determination of the CKM  matrix element $V_{cb}$ (see Ref.~\cite{HQET}
for reviews).

HQET has been extended in several directions.
\begin{dinglist}{108}  
\item Soft collinear effective theory (SCET) \\[.1cm]
HQET cannot be applied to decays like $B \to X_s \gamma$ or 
$B \to \pi \pi$ where the light particles in the final state can have 
momenta of $O(m_{Q})$. SCET accounts for those energetic light states
but it is more complicated than HQET. Because of the presence of
several scales, several effective fields must be introduced by
successive field transformations. A major achievement is again the
proof of factorization that is for instance crucial for a
reliable extraction of the CKM  matrix element $V_{ub}$ 
from experiment.
\item Nonrelativistic QCD (NRQCD) \\[.1cm]
This extension of HQET includes quartic interactions 
to treat heavy quarkonia $\overline{b}\,b$ 
and $\overline{c}\,c$. In this case, three widely 
separate scales are involved: the heavy mass $m_{Q}$, the 
bound-state momentum $p \sim m_{Q} v$ ($v \ll 1$) and the kinetic 
energy $E \sim m_{Q} v^2$. Applications
include the analysis of the $q\,\overline{q}$ potential and the
production and decay of quarkonia \cite{Brambilla:2004wf}.      
\end{dinglist} 

\subsection{QCD sum rules}
\label{sec:QCDSR}
The general idea of QCD sum rules is to use the analyticity properties
of current correlation functions to relate low-energy hadronic
quantities to calculable QCD contributions at high energies. We recall
the example of the two-point functions $\Pi_{L}^{(0,1)}(q^2)$ in
hadronic $\tau$ decays discussed in Sec.~\ref{sec:hadtau}.

In general, the QCD contribution consists itself of two different
parts,
\begin{itemize} 
\item a purely perturbative part and
\item a partly nonperturbative part that is important at intermediate
energies and makes use of the operator product expansion (OPE, another
case of factorization).
\end{itemize} 
Altogether, a typical two-point function (QCD sum rules are not
restricted only to two-point functions, however) has the form
\begin{equation}
\Pi(q^2) = \Pi_{\rm pert}(q^2) + \displaystyle\sum_d C_d(q^2) \langle
0|O_d|0\rangle ~.
\label{eq:Wilson}
\end{equation}
The so-called Wilson coefficients $C_{d}(q^2)$ are calculable 
perturbatively and they depend on the two-point function under
consideration. Up to logs, they decrease for large 
$|q^{2}|$ as $(q^{2})^{-n(d)}$  with some positive integer $n(d)$. The
vacuum condensates $\langle 0|O_d|0 \rangle$, on the other hand, are
universal and they absorb long-distance contributions with
characteristic  momenta $< \sqrt{|q^{2}|}$. 

\begin{center} 
\begin{figure}[ht] 
\leavevmode 
\centering\includegraphics[width=6cm]{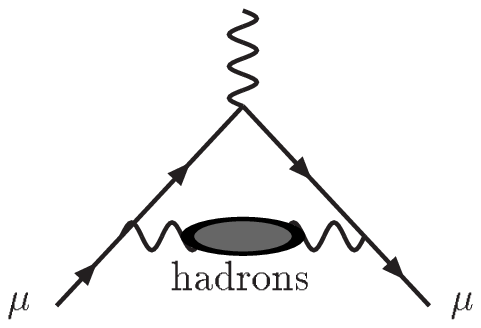}
\caption{Lowest-order hadronic vacuum polarization contribution to the 
anomalous magnetic moment of the muon.}
\label{fig:amu}
\end{figure}
\end{center}

Three main types of applications of QCD sum rules can be
distinguished.
\begin{itemize} 
\item[$\bullet$] Using experimental data as input for the low-energy 
hadronic part, one can extract universal QCD parameters: ~$\alpha_{s}$, 
quark masses, condensates, \dots
\item[$\bullet$]  With QCD parameters known, one can predict hadronic 
quantities: hadron masses, decay constants, amplitudes, \dots
\item[$\bullet$] The compatibility of low-energy data with QCD can be 
checked. I will discuss a recent example of topical interest, the 
spectral data relevant for the leading hadronic contribution to the 
anomalous magnetic moment of the muon.
\end{itemize} 
\begin{center} 
\begin{fmpage}{1.6cm}
\vspace*{.1cm}
\begin{center}
$\mathbf{(g-2)_\mu}$
\end{center}
\vspace*{.1cm}
\end{fmpage}
\end{center} 
The biggest source of uncertainty in the calculation of the anomalous
magnetic moment of the muon $a_\mu = (g_\mu -2)/2$ in the Standard
Model is at present the lowest-order hadronic vacuum polarization
\begin{eqnarray}
a_\mu^\mathrm{had,LO}=a_\mu^\mathrm{vac.pol.}=\displaystyle\int_{4
M_\pi^2}^{\infty} dt K(t) \sigma_0(e^+ e^- \to
~\mathrm{hadrons})(t)
\end{eqnarray}
depicted in Fig.~\ref{fig:amu}.

\vspace*{-.4cm}
\noindent   
The kernel $K(t)$ is a known function \cite{Gourdin:1969dm}. 
Although the integral extends
from threshold to infinity, about 73 \% of $a_\mu^{\mathrm{had,LO}}$ are
due to the $\pi\pi$ intermediate state in Fig.~\ref{fig:amu}, governed 
by the pion form factor. Moreover, 70 \% of 
$a_\mu^{\mathrm{had,LO}}$ come from the region $t \le 0.8$
GeV$^{2}$. Therefore, by far the most important part is not 
calculable in QCD perturbation theory.

A few years ago, Alemany, Davier and H\"ocker \cite{ADH97} suggested to
use not only data from $e^{+} e^{-} \to \pi^{+} \pi^{-}$ to 
extract the pion form
factor but also from the decay $\tau^{-} \to \pi^{0} \pi^{-}
\nu_{\tau}$. In the isospin limit, it is straightforward to derive the
relation 
\begin{equation}
\sigma_0(e^+ e^- \to \pi^+
\pi^-)(t)= h(t)  
\displaystyle\frac{d \Gamma(\tau^-\to \pi^0 \pi^- \nu_\tau)}{dt}
\end{equation}
with a known function $h(t)$ where $t$ is the two-pion invariant mass
squared.

At the level of precision required for comparison with experiment
(better than 1\,\% for $a_{\mu}^{\mathrm{had,LO}}$),
isospin violating and
electromagnetic corrections are mandatory \cite{DEHZ03,CEN}. The status
until recently was summarized by H\"ocker at the High Energy
Conference in Beijing \cite{Hoe_beijing}.
\begin{dinglist}{42}
\item There was a significant discrepancy between the $\tau$ and 
$e^{+} e^{-}$ data, mainly above the $\rho$ resonance
region, as shown in Fig.~\ref{fig:eevstau}.
\begin{center} 
\begin{figure}[H] 
\leavevmode 
\centering\includegraphics[width=10cm]{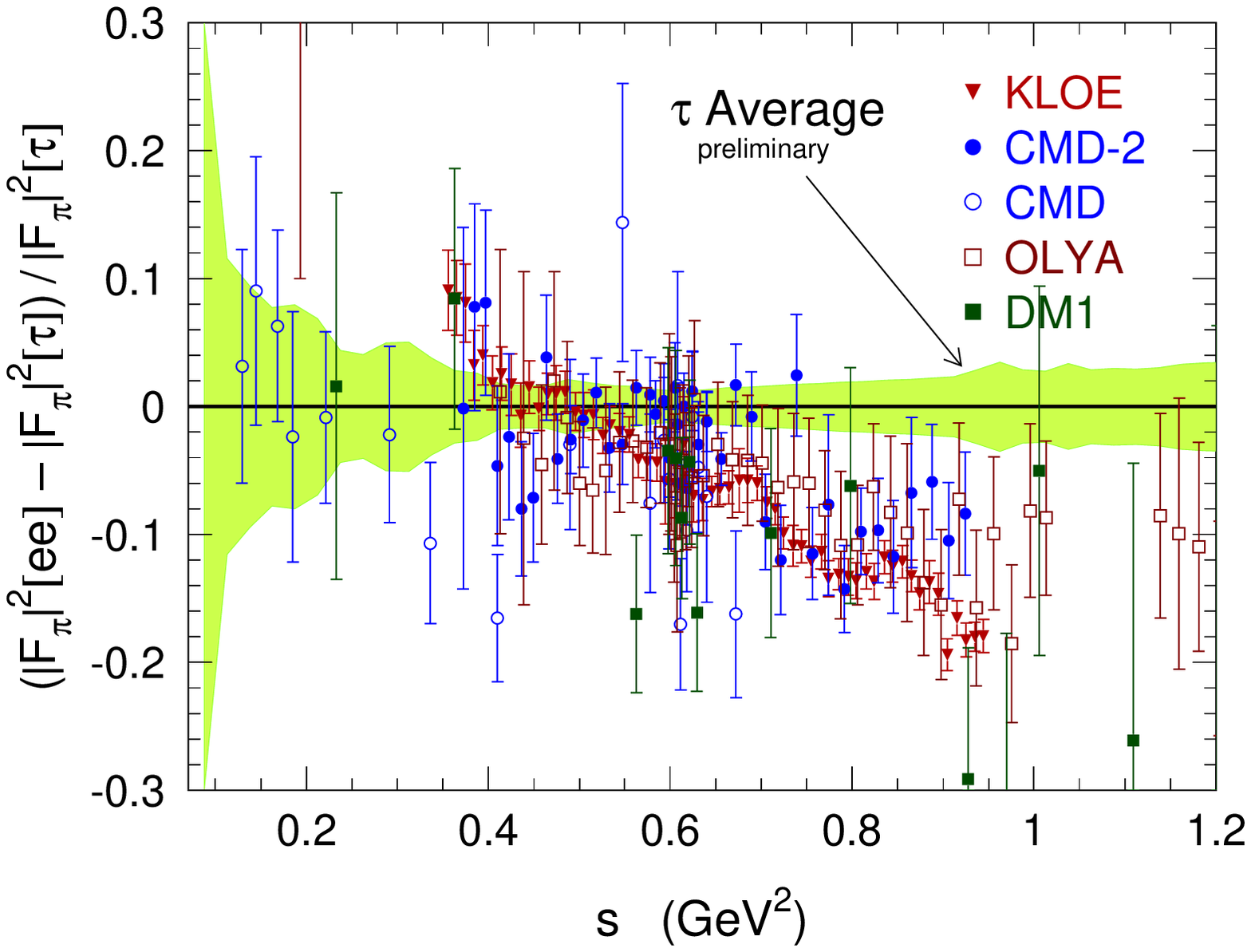}
\caption{Comparison between $e^{+} e^{-}$ and $\tau$ data for the pion
form factor from Ref.~\cite{Hoe_beijing}; plotted is the difference of
the form factors squared normalized to the $\tau$ data.}
\label{fig:eevstau}
\end{figure}
\end{center}  
\item Isospin violation cannot explain the difference.
\item The $e^{+} e^{-}$ data from the KLOE experiment confirm the previous
trend of the CMD-2 data although the agreement among the 
$e^{+} e^{-}$ data is not impressive.
\end{dinglist} 
The widely accepted recommendation at the Beijing Conference was to
ignore the $\tau$ data until the origin of the discrepancy is
understood \cite{Hoe_beijing}.

The situation has changed both on the theoretical and on the
experimental side. The theoretical clarification is due to Maltman
\cite{Maltman} who checked the consistency of experimental data with
QCD by investigating sum rule constraints for the two spectral
functions relevant for the $e^{+} e^{-}$ and the $\tau$ case,
respectively:
\begin{center} 
$\rho_{\rm em}(s)=\IM \,\Pi_{\rm elm}(s)$ \hspace*{.5cm} and
\hspace*{.5cm} $\rho_{V}^{I=1}(s)=\IM \,\Pi_{L,ud}(s)$ ~.
\end{center}
By using a contour integral in the complex $s$-plane as in
Fig.~\ref{fig:contour}, with $m_{\tau}^{2}$ replaced by an 
a priori arbitrary $s_0$, one derives a so-called 
F(inite)E(nergy)S(um)R(ule)  
\begin{equation}
\displaystyle\int_0^{s_0} w(s) \rho(s) ds = -\displaystyle\frac{1}{2\pi} 
\oint_{|s|=s_0} w(s) \Pi(s) ds  ~.
\label{eq:FESR}
\end{equation}
$\Pi(s), \rho(s)$ refer to the two cases considered (electromagnetic
currents in the $e^{+} e^{-}$ case, charged currents for 
$\tau$) and $w(s)$ is an analytic function (actually a polynomial)
that will be chosen conveniently. 

The right-hand side of the FESR can be estimated in QCD as exemplified
by Eq.~(\ref{eq:Wilson}). The purely perturbative part
is known up to $\alpha_{s}^{3}$, with
estimates of the $O(\alpha_{s}^{4})$ contribution available. For not too
small $s_0$ the purely perturbative part dominates the right-hand side
depending only on $\alpha_{s}(M_Z)$. The $d=2$ part is 
completely
negligible in the $\tau$ case and it depends only on the mass of the
strange quark in the electromagnetic case. The $d=4$ contributions
involve the quark condensates $\langle 0|m_{q} \overline{q} q|0 \rangle 
(q=u,d,s)$ (very well known from chiral perturbation theory,
cf. Sec.~\ref{sec:CHPT}) and the gluon condensate $\langle 0|\alpha_{s} 
G^a_{\mu\nu} G_a^{\mu\nu}|0\rangle $ (less well but sufficiently known
from charmonium sum rules). For $d \ge 6$, the relevant condensates
are practically unknown. However, by using the pinching trick
($w(s_{0})=0$) appropriately, Maltman eliminates the 
$d=6$ OPE contributions to 
$\Pi(s)$ altogether. The negligible effect of $d \ge 8$ contributions 
can be checked by varying $s_{0}$.

Turning now to the left-hand side of the FESR (\ref{eq:FESR}), Maltman
uses the most precise experimental data for the spectral functions:
ALEPH (compatible with CLEO and OPAL) for
$\rho_{V}^{I=1}(s)$, CMD-2 for $\rho_{\rm em}(s)$. As a 
first test, he
fits  $\alpha_{s}(M_{Z})$ (keeping the remaining OPE input fixed) from the
experimentally determined left-hand side of the FESR. The outcome is,
quite independently of the weight function $w(s)$ that is always chosen
positive and monotonically increasing for $0 \le s \le
s_{0}$, that the
fitted value of $\alpha_{s}(M_{Z})$ is systematically lower than the
high-energy determination dominated by LEP in the electromagnetic case
while there is perfect agreement in the $\tau$ case. This is a first
indication that the electromagnetic spectral density is too low. 

A second test, largely independent of the value of 
$\alpha_{s}(M_{Z})$, compares the slopes of the OPE parts 
and spectral integrals with respect to $s_{0}$. The
results are shown in Figs.~\ref{fig:malt_em6}, 
\ref{fig:malt_tau6}.

\noindent 
\parbox[l]{0.5\textwidth}{
\begin{center} 
\begin{figure}[H] 
\leavevmode 
\includegraphics[width=7cm]{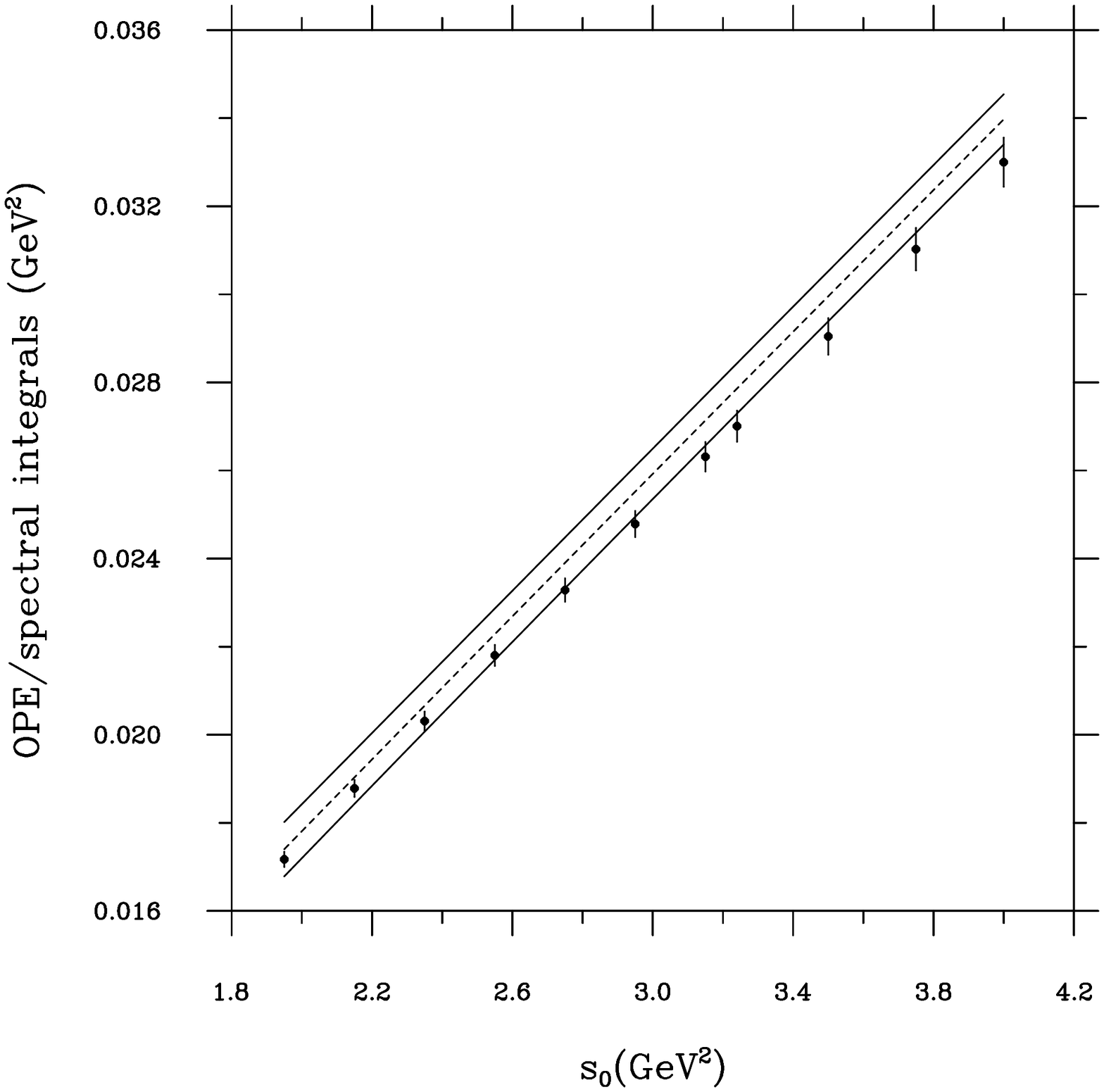}
\caption{Slopes with respect to $s_{0}$ in the $e^{+} e^{-}$ case for a
specific weight function $w_6(s)$. The dashed lines denote the central
values for the OPE input and the solid lines indicate the error
bands. The spectral integrals are shown for several points, error bars
included.} 
\label{fig:malt_em6}
\end{figure}
\end{center}
} 

\vspace*{-11.3cm}
\hspace*{7.5cm}
\parbox[r]{0.5\textwidth}{
\begin{center} 
\begin{figure}[H] 
\leavevmode 
\includegraphics[width=7cm]{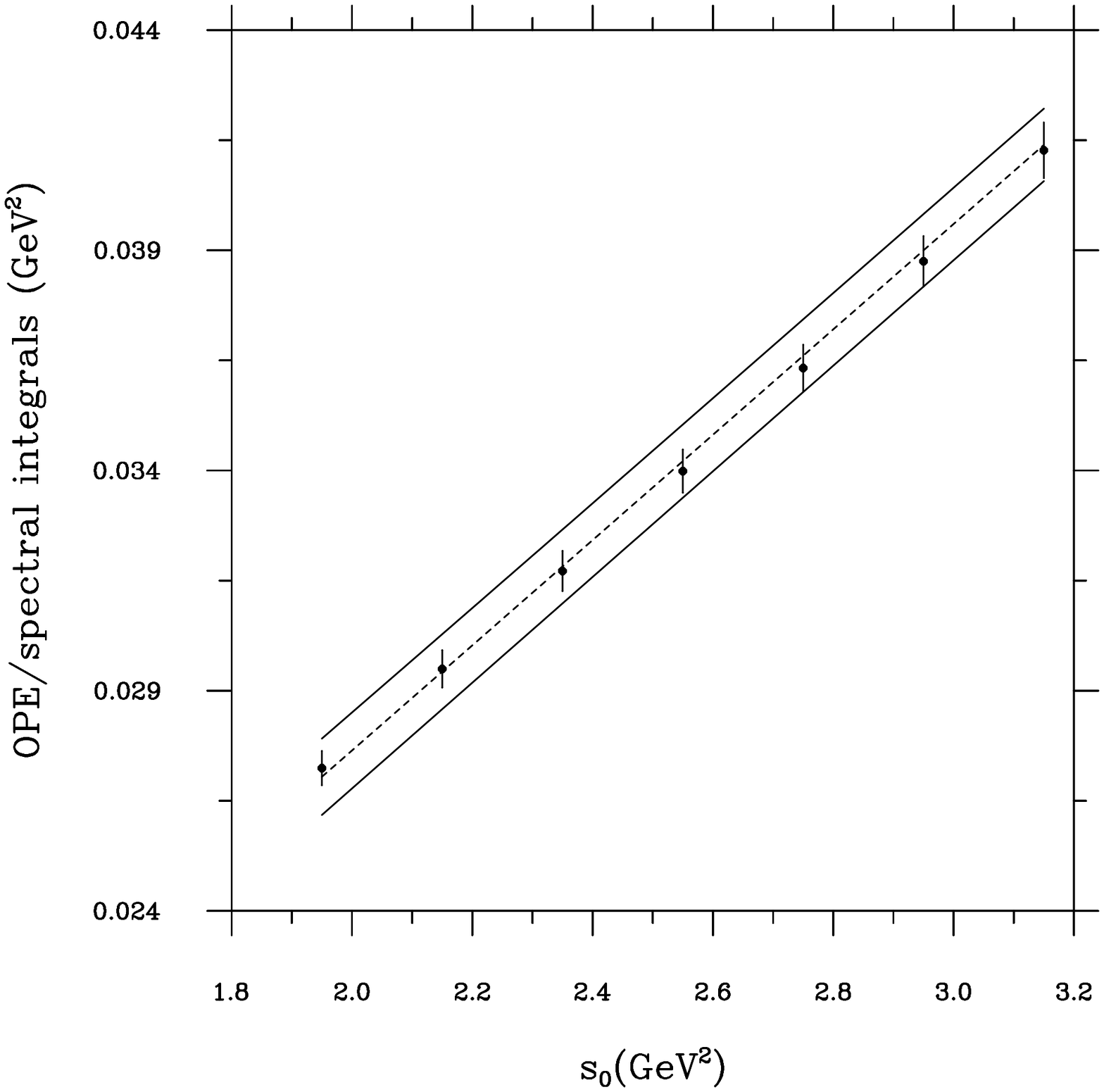}
\caption{Same as in the previous figure for the $\tau$ data. Both
  figures are taken from Ref.~\cite{Maltman}.}
\label{fig:malt_tau6}
\end{figure}
\end{center} 
}
   
\vspace*{2cm} 
\noindent 
The situation is similar as before. While the slopes differ between
data and QCD by $\gtrsim 2.5 \,\sigma$ in the electromagnetic case there
is perfect agreement in the $\tau$ case. The previous conclusion
is reinforced: the $e^{+} e^{-}$
spectral data are systematically too low whereas the $\tau$ data are
completely consistent with QCD, both in normalization and in the
slopes. The QCD sum rule tests clearly favour the $\tau$ over the 
$e^{+} e^{-}$ data for the pion form factor. 

The most recent development is again an experimental one. Only two
months before the School new data on $e^{+} e^{-} \to \pi^{+} \pi^{-}$
were 
released by the SND Collaboration from Novosibirsk. Their results
disagree with both CMD-2 and KLOE and they go in the right direction
towards reconciling the $e^{+} e^{-}$ with the $\tau$ data. 
The discrepancies between the three data sets in $e^{+}
e^{-}$ annihilation remain to be understood. Based on $\tau$ data for the
$2\pi$ and $4\pi$ channels in the hadronic vacuum polarization, the
calculation of the anomalous magnetic moment of the muon in the
Standard Model \cite{DM04} compares well with the measured value
\cite{amu_BNL}:
\begin{eqnarray}
a_\mu^\mathrm{exp} - a_\mu^\mathrm{SM} 
= (7.6 \pm 8.9) \cdot 10^{-10}~.
\end{eqnarray} 
There is at present no evidence for a discrepancy between the Standard 
Model and experiment.

\subsection{Chiral symmetry}
\label{sec:SCSB}
By construction, QCD is a gauge theory with gauge group
$SU(3)_{c}$. However, the QCD Lagrangian 
\begin{equation}
\cL_{\rm QCD} = 
- \displaystyle\frac{1}{2} {\rm tr}
(G_{\mu\nu} G^{\mu\nu}) + \displaystyle\sum_{f=1}^{N_F}
\overline{q}_f \left(i \slashed{D} - m_f \mathbbm{1}_c \right) q_f
\label{eq:LQCD2}
\end{equation}
possesses additional symmetries. As in QED, the theory is parity
invariant because of the absence of $\gamma_{5}$ 
(vector couplings only). Moreover, the coupling constant 
$g_{s}$ and the
quark masses $m_{f}$ are real so that QCD conserves also CP, ignoring
the so-called strong CP problem here.

Are there still additional symmetries in the QCD
Lagrangian (\ref{eq:LQCD2})? To answer this question, we first have a
look at the quark kinetic term only (with $N_{F}=6$ flavours):
\begin{eqnarray}
\cL_{\rm kin} = i \displaystyle\sum_{f=1}^{6}
\overline{q}_f \, \slashed{D} \, q_f = i \displaystyle\sum_{f=1}^{6}
\left\{ 
\overline{q}_{fL} \, \slashed{D} \, q_{fL} + \overline{q}_{fR} 
\, \slashed{D} \, q_{fR} \right\} ~,
\label{eq:Lkin}
\end{eqnarray}
with chiral components 
\begin{eqnarray}
q_L = \displaystyle\frac{1}{2} (1-\gamma_5) q, \qquad & \qquad 
q_R = \displaystyle\frac{1}{2} (1+\gamma_5) q ~.
\end{eqnarray}
Since the $q_{fL}$ and the  $q_{fR}$ do not talk to each other in
(\ref{eq:Lkin}), they can be rotated separately implying the maximal
global flavour symmetry $U(6) \times U(6)$. However, this is a
symmetry of the kinetic term only. In the full quark Lagrangian (colour
indices will be suppressed from now on) 
\begin{equation}
\cL_q = \displaystyle\sum_{f=1}^{6}
\left\{ 
\overline{q}_{fL} \, i \slashed{D} \, q_{fL} + \overline{q}_{fR} \, i
\slashed{D} \, q_{fR} - m_f (\overline{q}_{fR} q_{fL} +
\overline{q}_{fL} q_{fR}) \right\} 
\end{equation}
$q_{fL}$ and $q_{fR}$ can in general not be rotated separately any
longer because of the quark masses. The actual flavour symmetry
therefore  depends on the quark mass matrix
\begin{equation} 
\cM_q = {\rm diag}~(m_u,m_d,m_s,m_c,m_b,m_t)~.
\end{equation} 

In order to find all even only approximate symmetries of QCD, 
we distinguish several cases depending on the
specific values of the quark masses. 
\begin{itemize} 
\item[i.] In the real world, all quark masses are non-zero and they are
  all different from each other. In this case, the remaining flavour 
symmetry amounts to the phase transformations
$~q_{f\,L,R} \to e^{\displaystyle - i \ve_f} 
q_{f\,L,R}$  $(f=1,\dots,6)$ where
  the phase $\ve_f$ for a given flavour must be the same for $q_{fL}$
  and $q_{fR}$. The symmetry group reduces to the product
  $U(1) \times U(1) \times \dots \times U(1) = U(1)^6$ leading to the
  well-known conserved flavour quantum numbers 
  $N_{u}$, $N_{d}$, $N_{s}$, $N_{c}$, $N_{b}$ and  $N_{t}$. All these
  symmetries are broken by the weak interactions, except their
  sum (baryon number) 
\begin{equation} 
\quad B= \left(N_u + N_d + N_s + N_c + N_b + N_t \right)/3~.
\end{equation} 
\item[ii.] In some approximation, the quark masses are still non-zero
  but $n_{F}$ of them are equal 
  ($n_{F} < N_{F}=6$). In this
  case, the maximal symmetry group $U(6) \times U(6)$ reduces to 
\begin{equation} 
U(n_F) \times U(1)^{6-n_F} \simeq
  SU(n_F) \times U(1) \times U(1)^{6-n_F}~.
\end{equation}
The only realistic cases are $n_{F}=2$ or 3 and they lead to
well-known approximate symmetries:
\hspace*{1cm} $n_F=2$: \qquad $m_{u}=m_{d} \hspace*{1.8cm}  \lra \qquad$ 
isospin $SU(2)$ \\[.1cm]
\hspace*{1cm} $n_F=3$: \qquad $m_{u}=m_{d}=m_{s} \qquad \lra \qquad$
flavour $SU(3)$ 
\item[iii.] A much more radical approximation consists in setting some
  of the quark masses to zero:~~$m_{f}=0$
($f=1,\dots,n_{F}$). In this case, $n_{F}$ of the $q_{fL}$ and
  $q_{fR}$ can again be rotated
  separately implying the chiral symmetry
\begin{eqnarray}  
SU(n_F)_L \times SU(n_F)_R \times
  U(1)_V \times U(1)_A \left[ \times U(1)^{6-n_F} \right]~.
\label{eq:chiralsymm}
\end{eqnarray} 
To set $n_{F}=2$ \,of the quark masses to zero is a reasonable 
approximation in view
  of $m_{u,d} \ll \Lambda_{\rm QCD}$, whereas $n_{F}=3$ (setting also
  $m_{s}=0$) is certainly more daring. $U(1)_{V}$ is again
  responsible for baryon number conservation. The factor $U(1)_{A}$ is
  actually not a symmetry of full QCD at the quantum level (abelian
anomaly). 
\end{itemize}
We are familiar with isospin and flavour $SU(3)$ that we see at least 
approximately realized in the hadron spectrum. But what are the
consequences of the approximate chiral symmetry of QCD? If chiral
symmetry would manifest itself in the hadron spectrum, each hadron
would have to have a partner of opposite parity of approximately the
same mass. That is obviously not the case: chiral symmetry appears to be 
more hidden than isospin, for example. In order to
understand the manifestations of chiral symmetry, we have to recall
the main features of   
\begin{center}
\begin{fmpage}{5.3cm}
\vspace*{.1cm} 
\begin{center}
Spontaneous symmetry breaking
\end{center}
\vspace*{.1cm}
\end{fmpage}
\end{center}  
There are many examples of SSB in
physics and the best-known example in particle physics is the
spontaneously broken electroweak symmetry (see the Lectures of
W. Buchm\"uller at this School). 

The mechanism was first realized in
condensed matter physics and a good example for our purpose is the
ferromagnet. The underlying theory of the ferromagnet (eventually QED)
is certainly rotationally invariant. The Hamiltonian (e.g., of the
Heisenberg model) does not single out any
direction in space. Nevertheless, in the ground state of the
ferromagnet the spins align in a certain direction. The direction is
arbitrary and there is no trace of it in the Hamiltonian. In this sense 
rotational symmetry is ``spontaneously'' broken. It is certainly
not manifest in the ground state but it has other important
consequences such as the existence of excitations (called magnons or
spin waves in this case) with a very special dispersion law. The
dispersion law is the dependence of the frequency $\omega$
on the wave length $\lambda$ or on the wave number
$k=2\pi/\lambda$. SSB in condensed matter
physics implies that for some excitations the frequency tends to zero 
for infinite wave length:
\begin{eqnarray}
\displaystyle\lim_{k\to 0} \omega(k) = \displaystyle\lim_{\lambda \to
\infty}  \omega(k) = 0~.
\label{eq:displaw}
\end{eqnarray} 

\begin{center} 
\begin{figure}[ht] 
\leavevmode 
\centering\includegraphics[width=4.5cm]{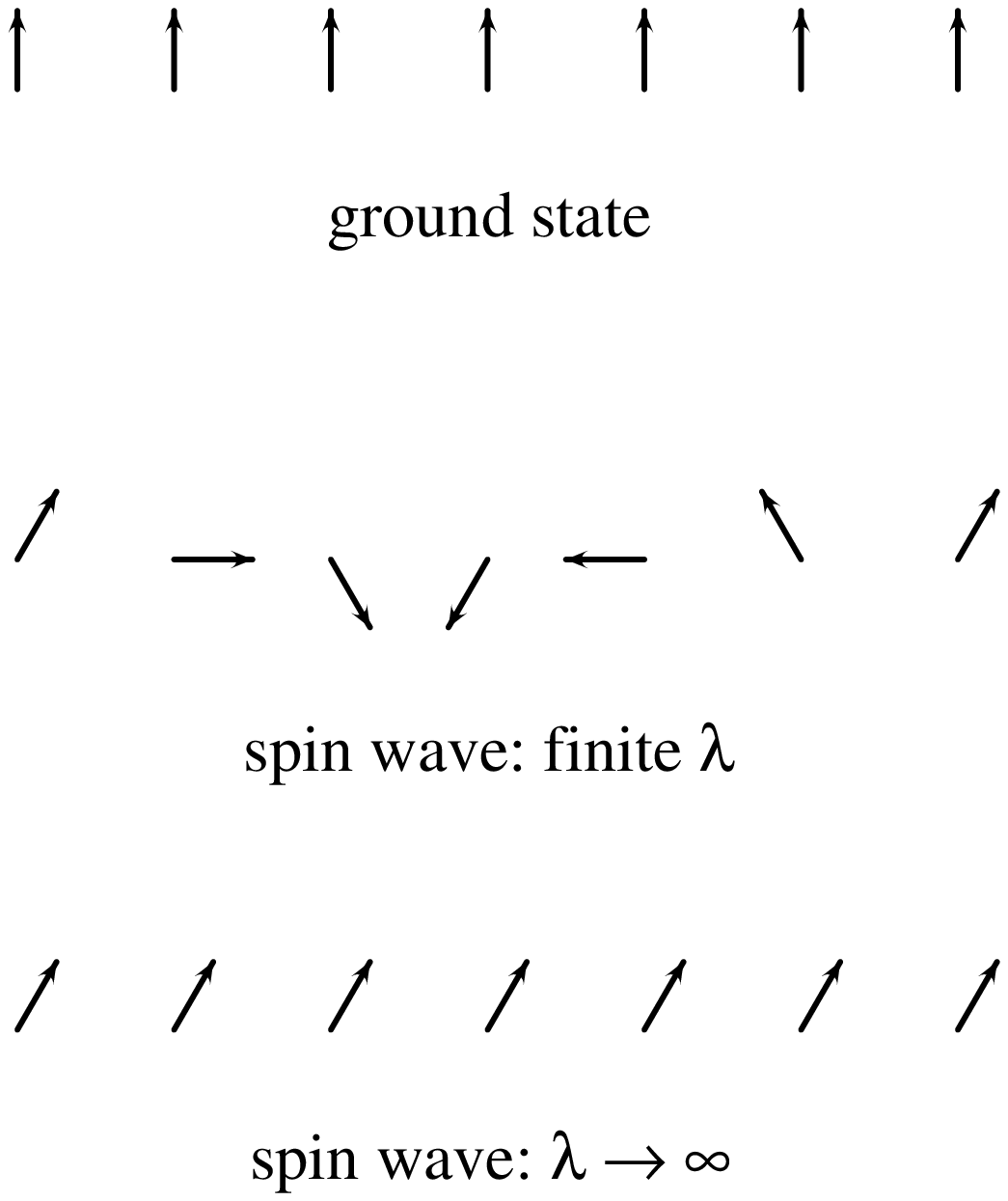}
\caption{Spin directions in the (one-dimensional) ferromagnet: ground
state, spin wave
  with finite wave length and a spin wave with infinite wave length.}
\label{fig:magnons}
\end{figure}
\end{center}

This property of magnons in particular is easy to visualize as shown
in Fig.~\ref{fig:magnons}.
In the ground state of the ferromagnet spins are
aligned. A typical spin wave is displayed in the second line:
the wave length is the distance 
between spins pointing in the same direction. In the limit $\lambda
\to \infty$, the spins become again aligned, albeit in a different 
direction in general. Since by the assumed rotational symmetry of the
theory  each direction is as good as any other, the configurations in
the first and in the last line must have the same energy as expressed 
by Eq.~(\ref{eq:displaw}). Put in another way, magnons do not have an
energy gap in their spectrum.

\vspace*{-1cm}
\noindent  
What is the analogy in a relativistic QFT like QCD? The frequency is
replaced by the energy of the particle, with the three-momentum
instead of the wave number. The dispersion law is just the
relativistic energy-momentum relation $E = \sqrt{p^2 + m^2}$. The
energy tends to zero for $p \to 0$ if and only if the particle is
massless: 
\begin{eqnarray} 
\hspace*{1.2cm} 
\displaystyle\lim_{p \to 0} E = \displaystyle\lim_{p \to 0} \sqrt{p^2
  + m^2} =0 \qquad & \Longleftrightarrow & \qquad m=0 ~.
\end{eqnarray} 
How can we distinguish if a symmetry is realized in the usual way
(like isospin) or if it is spontaneously broken? The crucial question
is what the conserved charge $Q = \int d^3x J^{0} (x)$ associated with a
symmetry current $J^\mu(x)$ ($\partial_\mu J^\mu =0$) does
when applied to the ground state (vacuum). In a relativistic quantum
field theory, there are only two options:
\begin{center}
\underline{Goldstone alternative} \\[.4cm] 
\begin{tabular}{cccc}
 & $Q |0\rangle = 0$ & \hspace*{.5cm} \mbox{ }
\hspace*{.5cm} & $||Q |0\rangle || = \infty$ \\*
& Wigner--Weyl & & Nambu--Goldstone \\*
& linear representation & & nonlinear realization \\*
& degenerate multiplets & & massless Goldstone bosons   \\*
& exact symmetry & &  spontaneously broken symmetry 
\end{tabular} 
\end{center}   
The left column describes the more familiar case (Wigner-Weyl) where
states are grouped in multiplets (irreducible representations of the 
symmetry group). The vacuum is annihilated by the charge and the
states in a given multiplet all have the same mass (degeneracy). In
the other possible realization (Nambu-Goldstone), applying the charge
to the vacuum is, strictly speaking, not defined. There are
no degenerate multiplets (therefore we don't see the symmetry in the
spectrum) but there must be massless particles in the theory
(Goldstone bosons). Although the charge cannot be applied to
the vacuum directly, the following matrix element called an order
parameter of SSB may be well defined:
\begin{eqnarray}
\langle 0|[Q,A]|0\rangle
\label{eq:op}
\end{eqnarray} 
where $A$ is some operator. If we can find an order parameter
that is different from zero, the symmetry associated with the charge
$Q$ is necessarily spontaneously broken. This is easy to understand:
because of the commutator in (\ref{eq:op}) the order
parameter vanishes if $Q$ annihilates the vacuum. In the familiar example
of electroweak symmetry breaking, scalar field operators
$\varphi_{i}$  take the
place of $A$ with $[Q,\varphi_{i}]=c_{ij} \varphi_{j}$. If 
$c_{ij} \langle 0|\varphi_{j}|0\rangle \neq 0$ (Higgs vacuum expectation 
value), the electroweak symmetry is spontaneously broken. This is also
a good example that the argument goes in one direction only. Even if
all scalars have zero vacuum expectation values or if there aren't any
scalar fields at all the symmetry may still be spontaneously broken.
The mechanism could involve some other operator $A$ in the order
parameter (\ref{eq:op}). 

For each spontaneously broken symmetry 
Goldstone's theorem implies the existence of a massless state 
$|{\rm G} \rangle$ with
\begin{equation}
\langle 0|J^{0}(0)|{\rm G} \rangle\langle {\rm G} |A|0\rangle \neq 0 ~.
\end{equation}
A necessary and sufficient condition for SSB is that the
\begin{center} 
\begin{fmpage}{7cm}
\vspace*{.1cm}
\begin{center} 
Goldstone matrix element $\langle 0|J^{0}(0)|{\rm G} \rangle \neq 0$  
\end{center}
\vspace*{.1cm} 
\end{fmpage}
\end{center} 
implying also that the Goldstone state $|{\rm G} \rangle$ has the
same quantum numbers as $J^{0}(0)|0\rangle$. The
following remarks are useful:
\begin{itemize} 
\item[$\bullet$]  The state $|{\rm G} \rangle$ need not correspond to 
a physical particle. This can only happen in the case of a spontaneously
broken gauge symmetry as in electroweak theory.
\item[$\bullet$]  $J^{0}(0)$ is usually a rotationally invariant
  bosonic operator and thus $|{\rm G} \rangle$ carries spin 0. 
\item[$\bullet$] Spontaneous breaking of discrete symmetries does not 
give rise to Goldstone bosons. 
\end{itemize} 
The main features of SSB can be 
discussed in the original Goldstone model. It has a single complex
scalar field $\phi(x)$ with the Lagrangian
\begin{equation}
\cL_{\rm Goldstone} = \partial_\mu \phi \partial^\mu
\phi^\dagger - \lambda ( \phi\phi^\dagger - 
\frac{v^2}{2})^2 \qquad 
(\lambda, ~v \,\,{\rm real\,and\,positive})~,
\end{equation} 
symmetric with respect to global $U(1)$ transformations $\phi(x) \to
e^{i\alpha} \phi(x)$. The minimum of the Mexican hat 
potential occurs
at $\phi\phi^\dagger =  \displaystyle\frac{v^2}{2}$. Decomposing the
complex field $\phi(x)$ into two hermitian fields $R(x), G(x)$ with
\begin{eqnarray}
\phi(x) &=& (R(x)+iG(x))/\sqrt{2} \\[.1cm]
\langle 0|R(x)|0\rangle = v, & \qquad & \langle 0|G(x)|0\rangle =
0~,\no  
\end{eqnarray}
the Lagrangian expressed in terms of the fields $R(x), G(x)$ displays
the following spectrum at tree level:
\begin{center}
\begin{tabular}{llll}
& Goldstone boson  field  \,$G(x)$  & \hspace*{1cm}   & $M_G=0$
  \\[.1cm] 
& massive field $H(x)=R(x)-v$ & &  $M_H=\sqrt{2\lambda}\,v$
\end{tabular}
\end{center}
Denoting the four-momenta of Goldstone particles generically as
$p_{G}$, one finds an unexpected behaviour for scattering 
amplitudes: they vanish for $p_{G} \to 0$, e.g.,
\begin{eqnarray}
A(GG \to GG)=O(p_G^4), \qquad A(GH \to GH)=O(p_G^2)
\end{eqnarray}
for arbitrary values of non-Goldstone momenta $p_{H}$. More generally,
Goldstone bosons decouple when their energies tend to zero.

This behaviour looks mysterious at first, but it can be understood by
choosing a different set of fields. Instead of the fields
$G(x)$ and $H(x)$, we choose another representation (polar
decomposition) that may be familiar from electroweak theory:
\begin{equation}
\phi(x)= \frac{1}{\sqrt{2}}[h(x)+v]e^{\displaystyle i g(x)/v}~.
\end{equation}
In terms of the hermitian fields $g(x),h(x)$ the Goldstone Lagrangian
takes the form
\begin{eqnarray}
\cL_{\rm Goldstone} &=& \frac{1}{2}(\partial_\mu g)^2+
\frac{1}{2v^2}(h^2+2vh)(\partial_\mu g)^2 \\[.1cm] 
&+ & \frac{1}{2}(\partial_\mu h)^2 - \lambda v^2 h^2
-\frac{\lambda}{4}(h^4+4vh^3)~.\no
\end{eqnarray}
A general theorem of QFT ensures that the fields $G,H$ on one side and 
$g,h$ on the other side
produce the same $S$-matrix elements although the Green functions are
in general different.
 
The main consequences are the following. \\[-.6cm] 
\begin{itemize} 
\item The particle spectrum is unchanged: 
\begin{center} 
\begin{tabular}{lll}
Goldstone  field  $g(x)$  & \hspace*{1cm} & $M_{g}=0$ \\
massive field $h(x)$ & & $M_{h}=\sqrt{2\lambda}\,v$
\end{tabular}
\end{center} 
\item The Goldstone field $g$ has only derivative  couplings implying
  for the scattering amplitudes considered previously:
\begin{equation} 
\displaystyle\lim_{p_{G}
\to 0} A(p_{G}) =0 ~. 
\end{equation} 
\end{itemize} 

The important lesson is very general and not restricted to the
Goldstone model. $S$-matrix elements with only Goldstone
states vanish for $p_{G} \to 0$. When other 
non-Goldstone
particles participate in the initial and final states, the statement
remains true for some matrix elements  like for elastic scattering
$G\,H \to G\,H$. In general, the interactions of Goldstone bosons
among themselves and with other matter become arbitrarily weak for
small momenta.

\subsection{Chiral perturbation theory}
\label{sec:CHPT}
We start from a theorist's paradise (copyright H. Leutwyler), QCD in 
the chiral limit where
$n_{F}=2$ [or 3] quarks $u,d~[,s]$ are massless:
\begin{equation}  
{\cal L}^0_{\rm QCD} = \overline{q_L} i \slashed{D} q_L + 
\overline{q_R} i \slashed{D} q_R + {\cal L}_{\mbox{\tiny heavy
    quarks}} +  {\cal L}_{\mbox{\tiny gauge}} 
\end{equation} 
with
\[
q^\top= ( u \,\, d \,\, [s]) ~.
\]
As explained in the previous section, this Lagrangian has a global
symmetry 
\begin{eqnarray}
SU(n_F)_L \times SU(n_F)_R \times U(1)_V \times U(1)_A 
\left[\times U(1)^{6-n_F}\right]~.
\end{eqnarray} 
The nonabelian factor $G=SU(n_{F})_L \times SU(n_{F})_R$ is called
the chiral group. 

Although not yet proven from QCD alone, there is strong evidence, both
from phenomenology and from theory, that chiral symmetry is
spontaneously broken. The spontaneous breaking does not affect all of
$G$ but, roughly speaking, half of it: $G \, \lra \, H  =
SU(n_F)_{V}$. The so-called vectorial subgroup $H$ of $G$ is nothing but
isospin (for $n_{F} = 2$) or flavour $SU(3)$ (for $n_{F} = 3$) and it is
realized in the familiar way \`a la Wigner-Weyl. Some arguments
in favour of this spontaneous breakdown are:   
\begin{dinglist}{108}  
\item As already emphasized before, there are no 
  parity doublets in the hadron spectrum.
\item There is no other convincing argument why the pseudoscalar
  mesons are the lightest hadrons. Spontaneous chiral symmetry
  breaking implies that they would be massless in the chiral limit 
 (pseudo-Goldstone bosons). 
\item The vector and axial-vector spectral functions are very
  different ($\rho$ vs. $a_{1}$).
\item The so-called anomaly matching conditions together with 
 confinement require that $G$  is spontaneously broken for $n_{F} \ge
 3$. 
\item Under very reasonable assumptions, $SU(n_F)_{V}$ is not
  spontaneously broken. It is of course expli\-citly broken by
  the differences between quark masses.
\end{dinglist}  
Even if it has not been possible so far to prove directly from QCD that
chiral symmetry is spontaneously broken, we can ask for possible order
parameters. It turns out (more details can be found in
Ref.~\cite{GE:beijing}, for instance) that the simplest such order
parameter involves the pseudoscalar operators $A_{b} =
\overline{q} \gamma_{5} \lambda_{b} q ~(a=1, \dots, 8)$ 
giving rise to the
quark condensate $\langle 0|\overline{q}\,q|0\rangle$. There is evidence
both from lattice QCD and from phenomenology 
that the quark condensate is non-zero
implying spontaneous chiral symmetry breaking. As will be discussed in
Sec.~\ref{sec:pipi}, the quark condensate is in fact the dominant
order parameter of spontaneous chiral symmetry breaking, in a sense to
be specified later.  

From Goldstone's theorem we know (still in the chiral limit) that
there are $n_{F}^{2} - 1$ massless Goldstone bosons:
\begin{center} 
\begin{tabular}{ccccl}
 $n_{F}$ & \mbox{} \hspace*{.5cm} & $n_{F}^{2}-1$ &
\mbox{} \hspace*{.5cm} & Goldstone bosons \\
\hline
 2 & & 3 & & $\pi$ \\
 3 & & 8 & & $\pi, K, \eta$ \\
\hline \\
\end{tabular}
\end{center} 
Although the real world is not a theorist's paradise, we still
expect low-energy amplitudes to be dominated by the exchange of
pseudoscalar mesons, which are the lightest hadrons also in the real 
world. In order to calculate such amplitudes, we construct
an effective field theory with only Goldstone fields. As already
explained in Sec.~\ref{sec:EFT}, the Lagrangian of Goldstone fields is
nonrenormalizable and it is in fact even nonpolynomial. The
underlying physical reason is that we can add any number of
sufficiently soft pions (still massless!) to a hadron state without
appreciably changing its energy. Therefore, we have degenerate states
with different numbers of particles that are related by chiral
symmetry transformations. For the Lagrangian this argument implies
that the symmetry transformations are nonlinear in the pion
fields. Starting from a Lagrangian with fixed powers in the Goldstone
fields, successive
nonlinear transformations generate any number of fields in the
Lagrangian. Since the Lagrangian is to be symmetric under such
transformations it must necessarily be nonpolynomial.

The basic building block of chiral Lagrangians is therefore a
nonpolynomial matrix function of the Goldstone fields, e.g.,
the exponential function (for $n_{F}=3$)
\begin{eqnarray}  
U(\phi)=\exp{(i \sqrt{2} \Phi/F)},  & \quad &
\Phi = \begin{pmatrix}
\dfrac{\pi^0}{\sqrt{2}} + \dfrac{\eta_8}{\sqrt{6}} & \pi^+ &  K^+
\\*[.1cm] 
\pi^- & -\dfrac{\pi^0}{\sqrt{2}} + \dfrac{\eta_8}{\sqrt{6}} &  K^0
\\*[.1cm] 
K^- & \overline{K^0} & - \dfrac{2 \eta_8}{\sqrt{6}} \end{pmatrix}
\end{eqnarray} 
where $F$ is the pion decay constant in the chiral limit
characterizing the size of the Goldstone matrix element 
$\langle 0|J^{0}(0)|{\rm G} \rangle $. 

Chiral Lagrangians are organized according to the number of 
derivatives of the
fields. The unique lowest-order Lagrangian of $O(p^{2})$ with two
derivatives is the so-called nonlinear $\sigma$ model:
\begin{eqnarray}
\cL_2^{(0)}  = \displaystyle\frac{F^2}{4} {\rm tr}_{n_F} \left(
\partial_\mu U  \partial^\mu U^\dagger  \right) =: 
\displaystyle\frac{F^2}{4} \langle \partial_\mu U 
\partial^\mu U^\dagger  \rangle 
= \partial_\mu \pi^+ \partial^\mu \pi^- + \displaystyle\frac{1}{2}
\partial_\mu \pi^0 \partial^\mu \pi^0 + O(\pi^4)~,  
\end{eqnarray} 
using a bracket notation for $n_{F}$-dimensional traces. 

So much for the paradise. Back to reality, we must admit that there is
no chiral symmetry in nature! In the Standard Model, it is explicitly
broken in two different ways.
\begin{itemize} 
\item[$\bullet$] Chiral symmetry is explicitly broken by nonvanishing 
quark masses. This should be a small modification for two, a more
  pronounced one for three flavours:
\begin{center} 
\begin{tabular}{cll} 
$m_u, m_d$ & $\ll \quad M_{\rho}
\quad$ \hspace*{1cm} & $n_F=2$ \\
$m_s$  & $< \quad M_{\rho}$ \hspace*{1cm} & $n_F=3$
\end{tabular} 
\end{center} 
\item[$\bullet$]  Also the electroweak interactions break chiral 
symmetry. If electroweak effects are to be included,  they can be
taken into account perturbatively in $\alpha, G_{F}$.
\end{itemize} 

The main assumption of chiral perturbation theory (CHPT) is that an
expansion around the chiral limit (the theorist's paradise) makes
sense. Therefore, even in the absence of electroweak interactions,
chiral Lagrangians are organized in a two-fold expansion.
\begin{itemize} 
\item[i.] Spontaneous chiral symmetry breaking gives rise to an
  expansion in derivatives of the fields leading to an expansion of
  amplitudes in the momenta of pseudo-Goldstone bosons.
\item[ii.] Explicit symmetry breaking suggests an expansion also in
the quark masses $m_{q}$.
\end{itemize} 
The two expansions can be related via the meson masses. As
will be discussed in the next subsection, the squares of the meson
masses start out linear in the quark masses:
\begin{equation}
M_M^2 \sim  m_q + O(m_q^2)~.
\end{equation}
The standard chiral counting therefore amounts to treating quark
masses like the second power of momenta:
\begin{eqnarray} 
m_q=O(M_M^2)=O(p^2)~.
\end{eqnarray} 
The effective Lagrangian (for pseudoscalar mesons) is therefore of the
form \cite{GL}
\begin{eqnarray}
\cL_{\rm eff}  &=& \cL_2 + \cL_4 + \cL_6 + \dots \no \\[.2cm] 
\cL_2  &=& \frac{F^2}{4} \langle \partial_\mu U \partial^\mu U^\dagger + 
\chi U^\dagger + \chi^\dagger  U \rangle 
\label{eq:L2}
\end{eqnarray} 
where $\chi$ represents the quark masses: $\chi = 2 B
\cM_{q} = 2 B\, {\rm diag}(m_u,m_d [,m_s])$. The lowest-order
Lagrangian contains only two parameters $F,B$ that are related to 
physical quantities as
\begin{eqnarray}
F_\pi = F \left[1+O(m_q)\right]~, & \hspace*{2cm}  &
\langle 0| \overline{u}u|0\rangle = -  F^2  B  \left[1+O(m_q)\right] ~.
\end{eqnarray} 

The lowest-order amplitudes of CHPT are of $O(p^{2})$ and
they correspond to the current algebra amplitudes of 40
years ago. The tree-level amplitudes can be read off directly from 
the Lagrangian (\ref{eq:L2}) depending only on $F_\pi$ and  $M_{M}^{2}$ 
~[$M_\pi^2= B (m_u+m_d)$, \dots]. For instance, the elastic $\pi\pi$
scattering amplitude of $O(p^{2})$ is given by
\begin{equation}
A_2(s,t,u) = \frac{s-M_\pi^2}{F_\pi^2} ~.
\label{eq:pipip2}
\end{equation}
Contrary to symmetries like isospin that relate different amplitudes,
the spontaneously broken chiral symmetry makes an absolute prediction
for this scattering amplitude. It was left as an exercise to the
audience to explain why that is possible.
 
The lowest-order results we have been discussing so far were already
known in the late sixties and early seventies of the last century
(current algebra, phenomenological Lagrangians). After an
influential paper of Weinberg \cite{Weinberg:1978kz}, but especially
with the work of Gasser and Leutwyler \cite{GL} the systematic treatment
of QCD at low energies became a respectable theory. The first step was
to construct the Lagrangian of next-to-leading order $\cL_{4}$ that
contains 10 (7) additional coupling constants (usually called LECs for
low-energy constants) for $SU(3)$ ($SU(2)$). With a hermitian
Lagrangian tree amplitudes are necessarily real but we know that
unitarity and analyticity require complex amplitudes. It is not
difficult to convince oneself that imaginary parts occur first at
$O(p^{4})$. The consequence is that a systematic low-energy  expansion
requires a loop expansion beyond lowest order
\cite{Weinberg:1978kz}. But loop amplitudes have a tendency to be
divergent. For a bona fide QFT we therefore need both regularization
and renormalization. As strange as it may sound, also
nonrenormalizable theories can and in fact must be renormalized to
qualify as respectable QFTs.

A nonrenormalizable QFT like CHPT has many common features with the
more standard renormalizable theories.
\begin{dinglist}{42}
\item Divergences are absorbed by the coupling constants in the
  higher-order Lagrangians $\cL_{4},
\cL_{6}, \dots$. Unlike in
  renormalizable theories, new LECs occur at every order of the chiral
  expansion.
\item The renormalized LECs are scale dependent just like the strong
  coupling constant $g_{s}(\mu)$. They can be interpreted 
  as describing the effect of all heavy hadronic states that are not
  represented by explicit fields in the Lagrangian.
\item Renormalization ensures that there is no dependence on some
  artificial cutoff.
\end{dinglist} 
For phenomenological applications, we have to know the values of the
various LECs. In principle, QCD fixes those constants but a matching
between QCD and CHPT is not possible in perturbation theory. This was
already discussed in Sec.~\ref{sec:EFT} for general EFTs with SSB
but it is also easy to understand in the present case: CHPT can only 
be applied for energies $E \ll M_{\rho}$ whereas
perturbative QCD only makes sense for $E \gtrsim $ 1.5 GeV. The most
successful approaches bridging this gap to get information on the LECs
are resonance saturation (based on the properties of QCD for large
$N_{c}$, i.e. for a fictitious world with many colours) 
and lattice QCD.

The chiral expansion is an expansion in $p^{2}/(4\pi
F)^{2}$ where $p$ is a characteristic momentum for the process in
question. Therefore, the chiral expansion should and does work better
for $SU(2)$ than for $SU(3)$: 
\begin{eqnarray} 
n_F=2 : \frac{p^2}{(4\pi F)^2}=0.014 \frac{p^2}{M_\pi^2}~, 
& \hspace*{1cm} &
n_F=3 : \frac{p^2}{(4\pi F)^2}=0.18 \frac{p^2}{M_K^2}~.
\label{eq:chexp}
\end{eqnarray} 

Most amplitudes and form factors for realistic processes have been 
calculated at least to next-to-leading order. There is an easy-to-use
Mathematica program to generate both strong and nonleptonic weak
transitions up to $O(p^{4})$ that is described in
Ref.~\cite{Unterdorfer:2005au} and is available for general use. The 
state of the art is next-to-next-to-leading order or
$O(p^{6})$.  A short 
introduction can be found in Ref.~\cite{GE:beijing}.

\subsection{Light quark masses}
\label{sec:qm}
In CHPT, the light quark masses always appear in the combination 
$B\,m_{q} \sim \, m_{q} \,\langle 0| \overline{u}\,u|0\rangle$. Since there
are no quarks or gluons in CHPT, only QCD scale invariant quantities 
can appear. As we discussed in Sec.~\ref{sec:hqm}, quark masses are
scale dependent whereas the product $m_{q} \,\langle 0|
\overline{u}\,u|0\rangle$ is not. The consequence is that CHPT can only
provide methods for extracting the ratios of quark masses. 

The lowest-order expressions for the meson masses in terms of quark
masses can be read off directly from the 
Lagrangian (\ref{eq:L2}):
\begin{eqnarray}
M^2_{\pi^+} =  2 \hat{m} B~, & \mbox{} \hspace*{.1cm} &
M^2_{\pi^0} =  2 \hat{m} B + O\left[(m_u - m_d)^2/(m_s -
  \hat{m})\right] \no \\[.1cm]
M^2_{K^+}  =  (m_u + m_s)B~, & & M^2_{\eta_8}  =  
\displaystyle\frac{2}{3}(\hat{m} + 2m_s)B +
O\left[(m_u - m_d)^2/(m_s - \hat{m})\right] \no \\[.1cm] 
M^2_{K^0}  =  (m_d + m_s)B~, & & \hat{m}  :=  \displaystyle\frac{1}{2}
(m_u + m_d)~. \label{eq:qmp2}
\end{eqnarray}
Several well-known relations follow from these expressions: \\[.2cm] 
\begin{tabular}{clcl} 
\mbox{}  \hspace*{.4cm}  & Gell-Mann--Oakes--Renner & \mbox{}  
\hspace*{.6cm} & 
$F_{\pi}^{2} M_\pi^2 = - 2 \hat{m} \langle 0|\bar u\,u|0\rangle$ \\[.1cm] 
 & GMOR, Weinberg & &
$B = \displaystyle\frac{M_\pi^2}{2 \hat{m}} = 
\displaystyle\frac{M_{K^+}^2}{m_s + m_u} =
\displaystyle\frac{M_{K^0}^2}{m_s + m_d}$ \\[.4cm] 
 & Gell-Mann--Okubo & &
$3M^2_{\eta_8} = 4 M_K^2 - M_\pi^2 \qquad$ (isospin limit)
\end{tabular} 

\begin{center} 
\begin{figure}[H] 
\leavevmode 
\includegraphics[width=7cm]{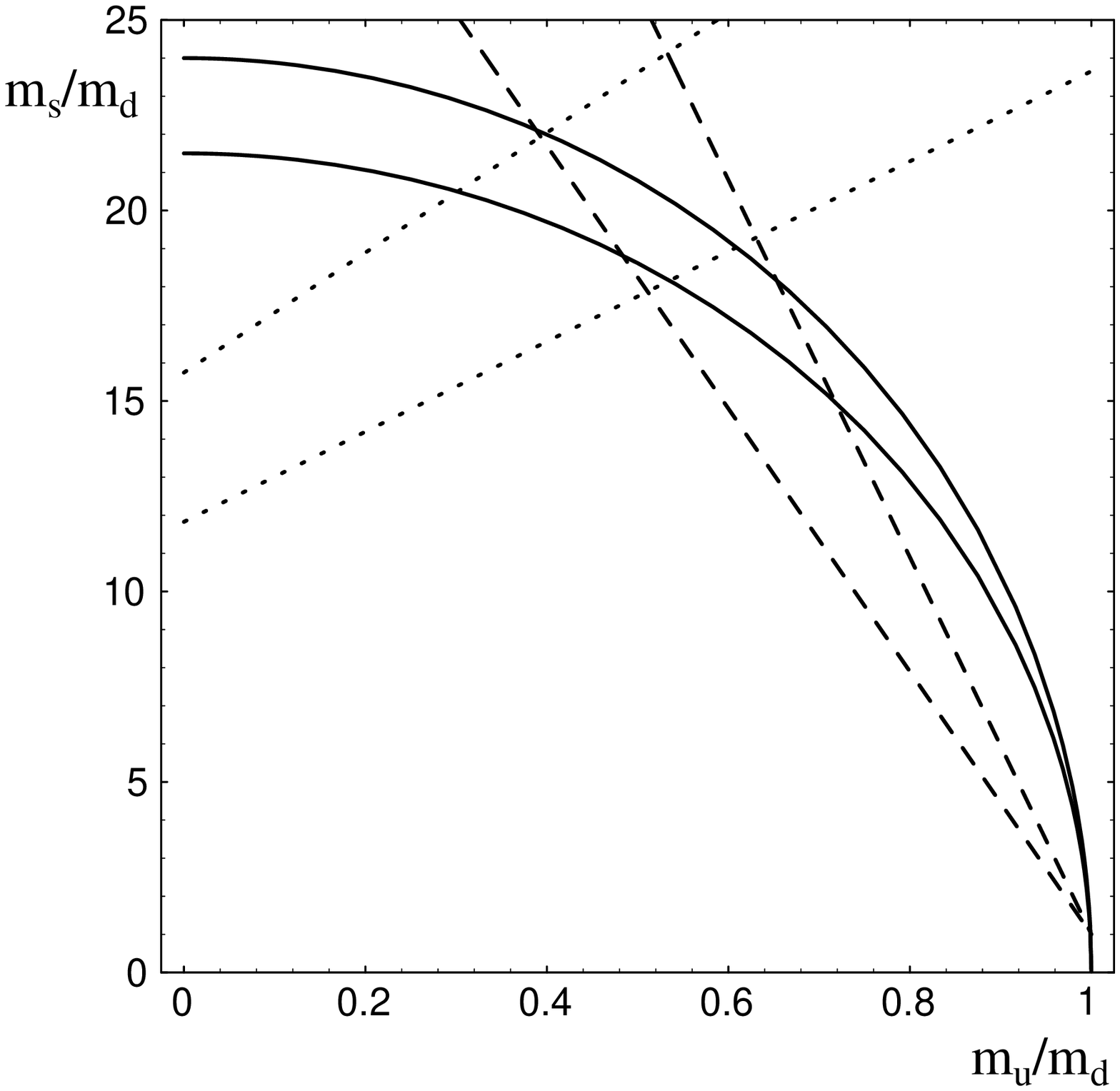}
\caption{Constraints on light quark mass ratios
  \cite{Leutwyler:1996qg}.}
\label{fig:hlellipse}
\end{figure}
\end{center}

\vspace*{.2cm} 
\noindent 
The relations (\ref{eq:qmp2}) are also the basis for the so-called
current algebra mass ratios
\begin{eqnarray} 
\hspace*{1cm} 
\displaystyle\frac{m_u}{m_d} = 0.55 ~, \quad & \quad 
\displaystyle\frac{m_s}{m_d} = 20.1 ~,  \quad  & \quad 
\displaystyle\frac{m_s}{\hat{m}} = 25.9 ~. 
\end{eqnarray} 
These ratios are subject to higher-order
corrections, most importantly of
$O(p^{4})=O(m_{q}^{2})$  and $O(e^2 m_{s})$. 
Because of an accidental symmetry  at
$O(p^{4})$, the ratios 
$m_s/m_d$, $m_u/m_d$ cannot be extracted separately from
S-matrix elements but only in the combination known as
Leutwyler's ellipse \cite{Leutwyler:1996qg} shown in 
Fig.~\ref{fig:hlellipse}.

\vspace*{.2cm} 
In addition to the full boundaries following directly from CHPT (the
difference is due to uncertainties in the electromagnetic corrections),
information is also available from $\eta - \eta^{\prime}$ 
mixing (dotted lines), baryon mass splittings and $\rho - \omega$ mixing 
(dashed lines). The overall conclusion is that the corrections of
$O(p^{4})$ are small for the ratios. The next-to-next-to-leading 
corrections of $O(p^{6})$ are also 
known \cite{Amoros:2001cp} but there are 
at the moment too many unknown LECs for quantitative conclusions. 
\begin{center} 
\renewcommand{\arraystretch}{1.1}
\begin{table}[H]
\begin{tabular}{|c|ccc|}  \hline
\mbox{   } &\mbox{   } $m_u/m_d$ \mbox{   }
 &\mbox{   } $m_s/m_d$ \mbox{   } & 
\mbox{   } $m_s/\hat m$ \mbox{   } \\[.1cm] 
\hline
\mbox{   } $O(p^{2})$ \mbox{   } & 0.55 & 20.1 &25.9  \\[.1cm] 
\mbox{   } $O(p^{4})$ \mbox{   } &\mbox{   } 0.55 $\pm$ 0.04 \mbox{   }
 &\mbox{   } 18.9 $\pm$ 0.8 \mbox{   }
&\mbox{   } 24.4 $\pm$ 1.5\mbox{   } \\
\hline
\end{tabular}
\caption{Quark mass ratios to $O(p^{4})$ \cite{Leutwyler:1996qg}.}
\label{tab:mqratios}
\end{table}
\end{center}
Absolute values of the quark masses are less well known than the
ratios. The main methods are QCD sum rules and lattice simulations,
most recently with full (unquenched) QCD. From the Review of Particle
Properties \cite{PDG04}, the combined result of lattice determinations
of the strange quark mass is $m_{s}(2~{\rm GeV}) = (100 \pm 25)$
MeV. More results are reproduced in Fig.~\ref{fig:qm}.

\subsection{Pion pion scattering}
\label{sec:pipi}
Pion pion scattering has a privileged status in CHPT. It is {\bf the}
fundamental scattering process of CHPT and it involves only pions. The
low-energy expansion can therefore be set up in the framework of chiral
$SU(2)$ and it can be expected to converge well. The scattering
amplitude is very sensitive to the mechanism of
spontaneous chiral symmetry breaking giving access to the quark
condensate in particular. 

The following review of recent developments will be restricted to the
isospin limit ($m_{u}=m_{d}$) in the absence of electromagnetic
corrections. In this case, the information for all possible scattering
channels is contained in a single amplitude $A(s,t,u)$ (with $s+t+u=4
M_{\pi}^{2}$). 

The lowest-order amplitude of $O(p^{2})$ was already
shown in Eq.~(\ref{eq:pipip2}):
\begin{eqnarray*} 
A_2(s,t,u) = \frac{s-M_\pi^2}{F_\pi^2} ~.
\end{eqnarray*} 
At the same order, the quark mass ratio 
$r=\displaystyle\frac{m_s}{\hat{m}} = \displaystyle\frac{2
  M_K^2}{M_\pi^2} -1 \simeq 26 $, as also shown in Table
\ref{tab:mqratios}. As the mass ratio $r$, also the S-waves are
very sensitive to the quark condensate. In a modified version 
of CHPT (Generalized CHPT \cite{{Stern:1997ri}}), one 
can tune the quark condensate. As an example, I show the
leading-order results for the $I=0$ S-wave scattering length
$a_{0}^{0}$ and for the quark mass ratio $r$ for both the 
standard and for a very small value of the quark condensate:
\begin{center} 
\begin{tabular}{ccc}
\hspace*{.3cm}  $a_{0}^{0}$ \hspace*{.3cm}  & \hspace*{.3cm}  $r$ 
\hspace*{.3cm}  & \mbox{    } $B(\nu=$ 1 GeV) \\ \hline
0.16 & 26 & 1.4 GeV ~(standard value)\\
0.26 & 10 &  $F_\pi$\\
\hline
\end{tabular}
\end{center} 
At next-to-leading order, the scattering amplitude was calculated in
1983 \cite{Gasser:1983kx}. It turns out that especially the S-wave
scattering lengths are quite sensitive to chiral corrections (chiral
logs). For instance, $a_{0}^{0}$ increases from 0.16 at 
$O(p^{2})$ to 0.20 at  $O(p^{4})$, an 
increase of 25 \% and thus quite a bit larger than
the natural estimate in Eq.~(\ref{eq:chexp}). Since the favoured
experimental value of  $a_{0}^{0}$ at that time was 0.26 
(with a 25 \% error), it seemed mandatory to perform one more step in 
the chiral expansion. From the amplitude to $O(p^{6})$ 
\cite{Bijnens:1995yn} it was clear that a value 
$a_{0}^{0}=0.26$ was not compatible with QCD. To narrow
down the uncertainties related to the LECs appearing in the $\pi\pi$
amplitude, the chiral amplitude was finally combined with dispersion
theory (Roy equations) \cite{Ananthanarayan:2000ht}.

CHPT together with dispersion theory predicts not only the S-wave
scattering lengths with amazing precision
\cite{Ananthanarayan:2000ht}, 
\begin{eqnarray}
a_0^0 = 0.220 \pm 0.005~,  & &
a_0^2 = - 0.0444 \pm 0.0010 ~,
\label{eq:a02Roy}
\end{eqnarray} 
but also the  S- and P-wave phase shifts. As a by-product, the pionium
lifetime is predicted to be $\tau=(2.9 \pm 0.1)\cdot 10^{-15}$. A
short history of the S-wave scattering lengths is shown in
Fig.~\ref{fig:CHPTRoy}.

There is a small caveat here. All the results have been derived in the
standard framework of CHPT that assumes a large quark
condensate. With recent experimental information on the pion-pion
phase shift difference $\delta^{0}_{0} - 
\delta^{1}_{1}$ from $K_{e4}$ decays,
even this last loophole could be closed. Using the correlation between
$a_{0}^{2}$ and $a_{0}^{0}$ implied by the Roy equations, the measured phase
shift difference \cite{Pislak:2001bf} can be used to determine $a_{0}^{0}$ 
as shown in Fig. \ref{fig:ke4}. The fitted value
\cite{Colangelo:2001sp} $a_{0}^{0} = 0.221 \pm 0.026$ is 
in perfect agreement with the prediction (\ref{eq:a02Roy}) from the 
combined analysis of CHPT and Roy equations.

The precise determination of the $\pi\pi$ scattering amplitude has
several important implications. One application concerns the chiral
expansion of the pion mass in terms of the light quark masses 
$m_{u}, m_{d}$: 
\begin{eqnarray} 
M_\pi^2 &=& M^2 - \displaystyle\frac{\bar{l}_3}{32 \pi^2 F^2} M^4
 + O(M^6) \\
M^2 &=& (m_u + m_d) |\langle 0|\overline{u}u|0\rangle| / F^2 ~.\no
\end{eqnarray} 
A by-product of the analysis of $\pi\pi$ scattering is a precise
determination of LECs like  $\bar{l}_{3}$, which implies 
in turn that more than 94 \% of $M_\pi$ are in fact due to the 
leading term  (the Gell-Mann--Oakes--Renner term) confirming the 
standard mechanism of spontaneous chiral symmetry breaking 
\cite{Colangelo:2001sp}. In other words, the quark
condensate is indeed the dominant order parameter of chiral symmetry
breaking. 

\vspace*{-1cm} 
\parbox[l]{0.47\textwidth}{  
\begin{flushleft} 
\begin{figure}[H]
\centering\includegraphics[width=6cm,angle=-90]{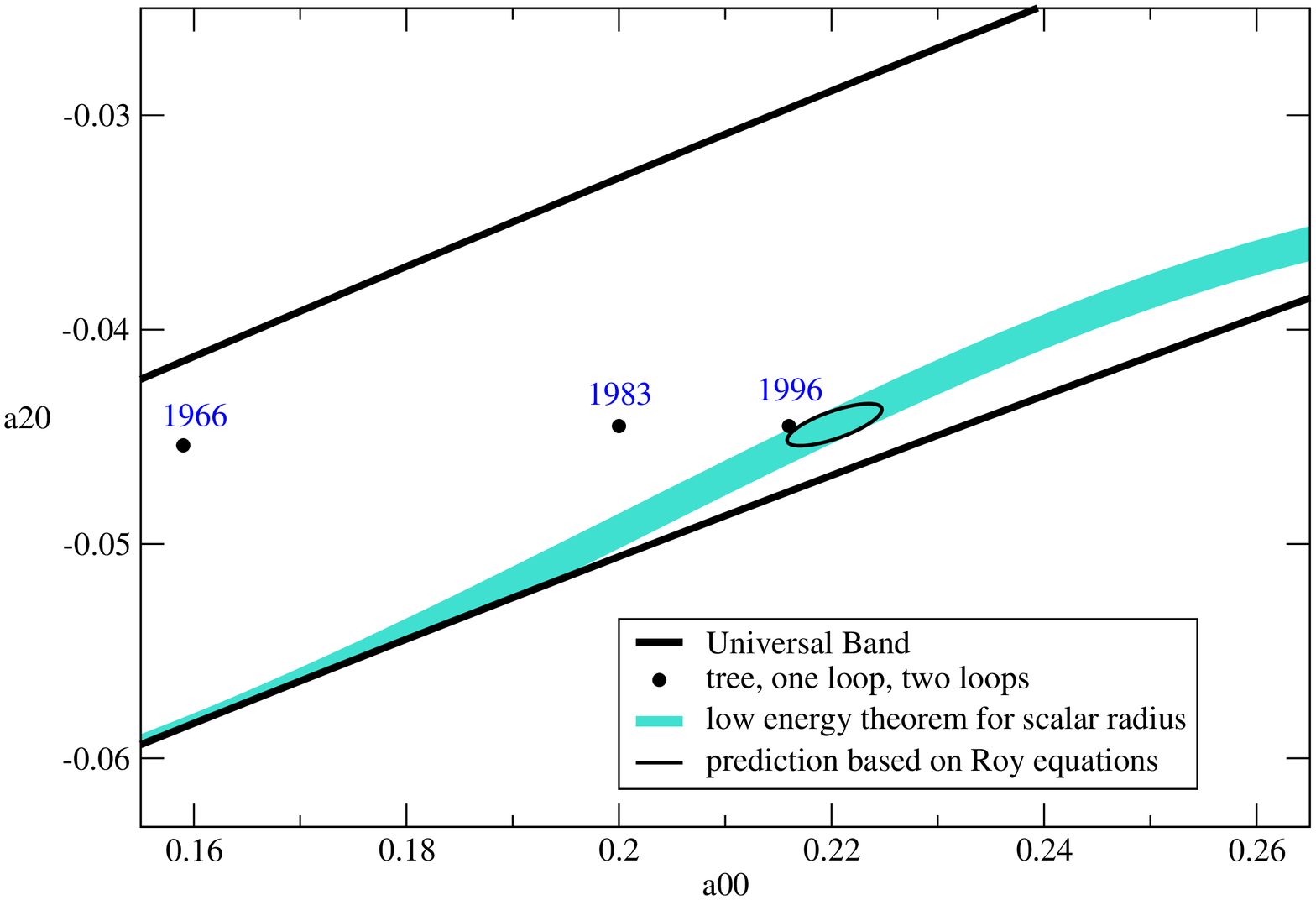}
\caption{Predictions for S-wave scattering lengths from current
  algebra till today, taken from Ref.~\cite{Ananthanarayan:2000ht}.} 
\label{fig:CHPTRoy}
\end{figure}
\end{flushleft} 
}

\vspace*{-7.6cm}
\hspace*{8.5cm}
\parbox[r]{0.45\textwidth}{
\begin{center} 
\begin{figure}[H] 
\leavevmode 
\includegraphics[width=6.5cm]{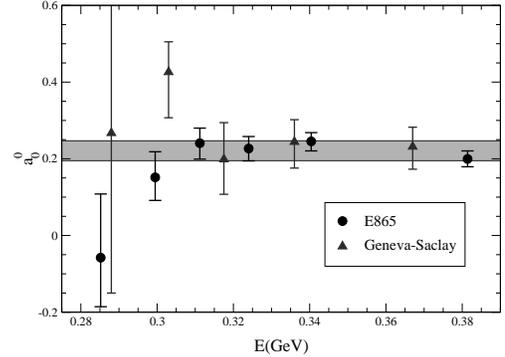}
\caption{$K_{e4}$ data translated into a determination of the $I=0$
$S$-wave scattering length $a_{0}^{0}$ \cite{Colangelo:2001sp}.}
\label{fig:ke4}
\end{figure}
\end{center}
}

\vspace*{.3cm} 
The precise knowledge of the $\pi\pi$ scattering amplitude from CHPT
and dispersion theory has more recently (after the School) produced
another important insight. The much discussed scalar isoscalar state
now called $f_0(600)$ by the Particle Data Group \cite{PDG04}, but
more commonly known as the $\sigma$ meson, was analysed by Caprini,
Colangelo and Leutwyler \cite{ccl} by extending the Roy equations to
complex values of the Mandelstam variable $s$. The result is an
astonishingly precise determination of mass and width of the lightest
hadronic resonance:
\begin{eqnarray}
M_\sigma = 441  \,\mbox{}_{- 8}^{+ 16}  {\rm ~MeV}~, & \qquad & 
\Gamma_\sigma = 544 \,\mbox{}_{- 18}^{+ 25} {\rm ~MeV}~.
\end{eqnarray}  
The $\sigma$ resonance has the quantum numbers of the vacuum and it
corresponds to an unambiguous pole on the second sheet of the scalar
isoscalar partial wave. Its real part is close to threshold but the
pole is quite far from the real axis. Although not as straightforward
as its position in the complex energy plane, the most appealing
interpretation of the $\sigma$ is a quasi-bound state of two pions,
quite different from a member of a standard $\overline{q}\,q$ meson
nonet \cite{Ananthanarayan:2000ht,ccl}.

\subsection{$K_{l3}$ decays and $V_{us}$}
\label{sec:kl3}
The Cabibbo--Kobayashi--Maskawa (CKM) matrix $V_{ij}$ determines the 
structure of the hadronic charged weak current. The matrix elements
are fundamental
parameters of the Standard Model. Together with the quark and lepton
masses, the CKM matrix and the corresponding lepton mixing matrix
contain information about the mechanism of mass
generation. Anticipating forthcoming LHC data to unveil the secrets of
mass generation (Higgs sector), both masses and mixing angles should
be measured as precisely as possible.

With three generations of quarks, the CKM matrix must be a unitary
matrix. For some time, a possible problem with CKM unitarity has been
discussed. With the PDG values of 2004 \cite{PDG04},
\begin{eqnarray} 
|V_{ud}|= 0.9738(5)~, & \qquad & |V_{us}|= 0.2200(26)~, 
\end{eqnarray} 
unitarity appeared to be violated at the 2.2 $\sigma$ level by the 
elements of the first row $V_{uj}~~(j=d,s,b)$:
\begin{equation}
\sum_{j=d,s,b} |V_{uj}|^2 -1 = - 0.0033(15)~.
\end{equation}
At this level of accuracy, the element $V_{ub}$ is still
negligible. On the other hand, the required precision for $V_{ud}$ and
$V_{us}$ calls for reliable isospin
violating and electromagnetic corrections.

\begin{center} 
\begin{figure}[H] 
\leavevmode 
\centering\includegraphics[width=8cm,angle=-90]{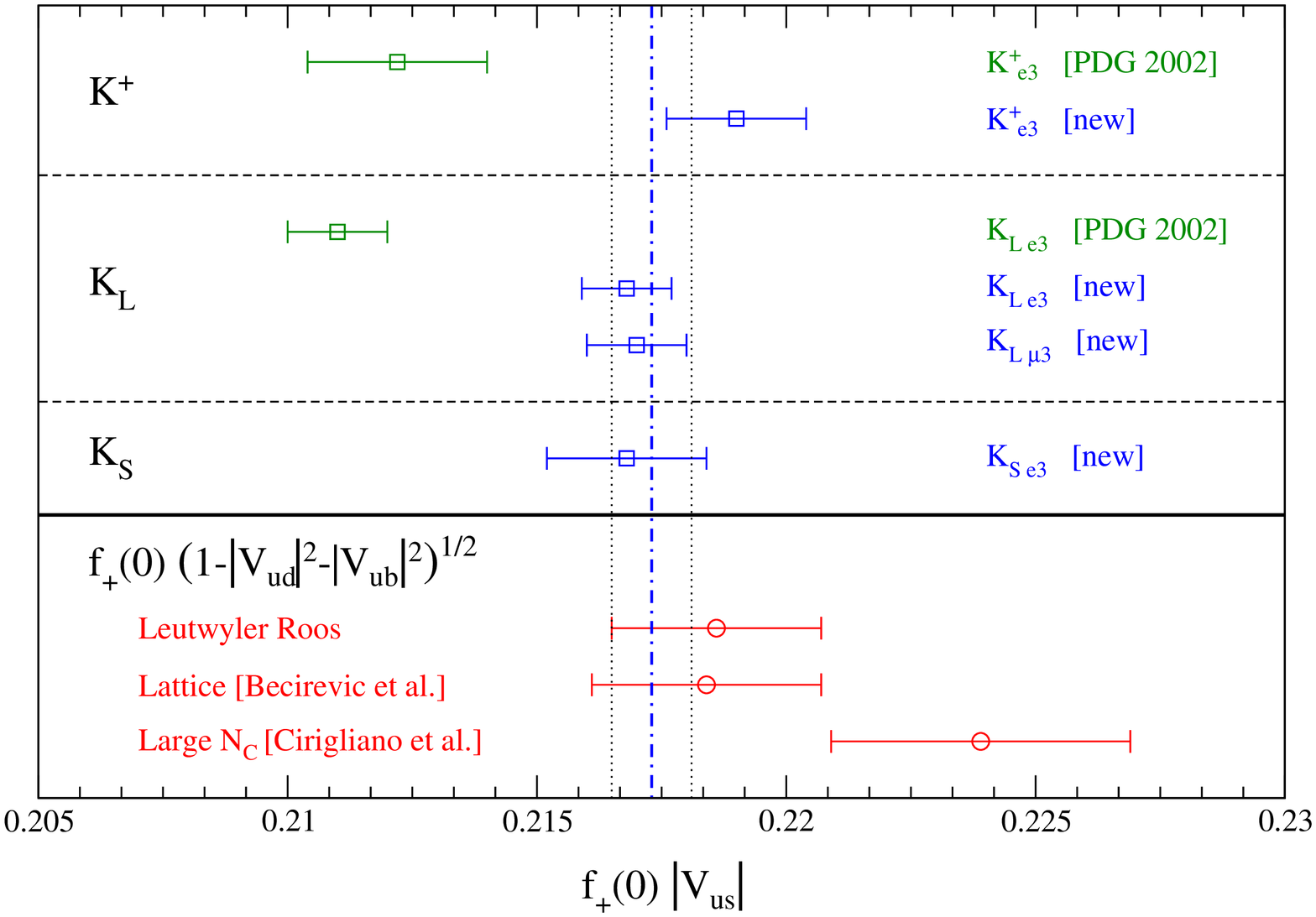}
\caption{Present experimental and theoretical status of
  $f_+(0)\,|V_{us}|$ taken from Ref.~\cite{Blucher:2005dc}.} 
\label{fig:Kl3data}
\end{figure}
\end{center}

The most accurate determination of $V_{us}$, both experimentally and
theoretically, comes from $K_{l3}$ decays that can be 
treated in the framework of
CHPT. The decay amplitude is governed by two form factors $f_{+}(t)$
and $f_{-}(t)$ with $t=(p_K - p_\pi)^2$:
\begin{equation}
\langle \pi^- (p_\pi) | \bar{s} \gamma_\mu u | K^0 (p_K) \rangle = 
f_{+}^{K^0 \pi^-} (t)  \, (p_K + p_\pi)_\mu  + 
f_{-}^{K^0 \pi^-} (t)  \, (p_K - p_\pi)_\mu ~.
\end{equation}
Both form factors are known to next-to-next-to-leading order in
CHPT. For the extraction of $V_{us}$, the crucial 
quantity is $f_{+}(0)$. The chiral expansion is of the form
\begin{equation}
f_{+}^{K^0 \pi^-} (0) = 1 + f_{p^4} + f_{e^2\,p^2} + f_{p^6} + O[(m_u
- m_d)p^4,e^2\,p^4]~.
\label{eq:fplusexp}
\end{equation}
The status of the various contributions is as follows:
\begin{center} 
\begin{tabular}{lclcl} 
$f_{p^4}$ & \mbox{  }  & $- 0.0227$ 
(no uncertainty) & \mbox{  }  & Gasser, Leutwyler \cite{Gasser:1984ux} \\
$f_{e^2\,p^2}$ & &  radiative corrections & & Cirigliano, Neufeld, Pichl
\cite{Cirigliano:2004pv} \\
$f_{p^6}$ & & loop contributions & & Bijnens, Talavera
\cite{Bijnens:2003uy}; Post, Schilcher \cite{Post:2001si}\\
& & tree contributions  & & LECs $L_5^{2}$, ~$C_{12}+C_{34}$
\end{tabular}
\end{center} 
As a first test, we look at the ratio
\begin{eqnarray} 
r_{+0} = \left( \frac{2 \, \Gamma(K^+_{e 3
(\gamma)}) \, M_{K^0}^5 \, I_{K^0}}{\Gamma(K^0_{e 3 (\gamma)}) \,
M_{K^+}^5 \, I_{K^+}} \right)^{1/2} & = & 
\displaystyle\frac{|f_{+}^{K^+ \pi^0} (0)|}{|f_{+}^{K^0 \pi^-} (0)|} ~.
\end{eqnarray}  
This ratio is independent of $f_{p^6}$ in Eq.~(\ref{eq:fplusexp}) and
it can therefore be predicted quite accurately \cite{Cirigliano:2004pv,
Descotes-Genon:2005pw}:
\begin{equation}
r_{+0}^{\rm th} = 1.023 \pm 0.003 ~,
\end{equation}
to be compared with the experimental value \cite{HN}:
\begin{equation}
r_{+0}^{\rm exp} =  1.036 \pm 0.008~. 
\end{equation}
What could be the origin of a possible discrepancy that is also
suggested by the compilation of recent data in Fig.~\ref{fig:Kl3data}?
\begin{itemize} 
\item[$\bullet$] In the past, radiative corrections have not always 
  been state of the art. Nowadays, also experimentalists should only 
  use the modern CHPT treatment \cite{Cirigliano:2004pv}.
\item[$\bullet$] Measurements of the $K^{+}$ and $K_{L}$ lifetimes should 
  still be improved. 
\item[$\bullet$] On the theory side, an unlikely but in principle
  still possible explanation could be that the error due to effects of 
  $O[(m_u - m_d) p^4,e^2\,p^4]$ is underestimated.
\end{itemize} 

The contribution $f_{p^6}$ is the sum of a loop and of a tree-level
part:
\begin{eqnarray}
f_{p^6}^{L=1,2} (M_\rho) &=&  0.0093 \pm 0.0005 \hspace*{3.7cm} 
{\rm Bijnens, Talavera} ~~[46]  \label{eq:fp6loop} \\[.2cm]
f_{p^6}^{\rm tree} (M_\rho) &=& 
8 \frac{\left( M_K^2 - M_\pi^2 \right)^2}{F_\pi^2}  
\, \left[\frac{\left(L_5^r (M_\rho) \right)^2}{F_\pi^2} - 
C_{12}^r (M_\rho) - C_{34}^r (M_\rho) \right] \no \\[.1cm]
&=& -  \frac{\left( M_K^2 - M_\pi^2 \right)^2}{2 \, M_S^4}  
\, \left( 1 - \frac{M_S^2}{M_P^2} \right)^2 ~.
\end{eqnarray}

\vspace*{-1.6cm}
\hspace*{10cm} large-$N_c$ matching \\[.1cm] 
\hspace*{11cm}  Cirigliano et al. \cite{Cirikl3} 

\begin{center}  
\begin{figure}[H]
\centering
\begin{picture}(200,125)  
\put(70,30){\makebox(50,50){\epsfig{figure=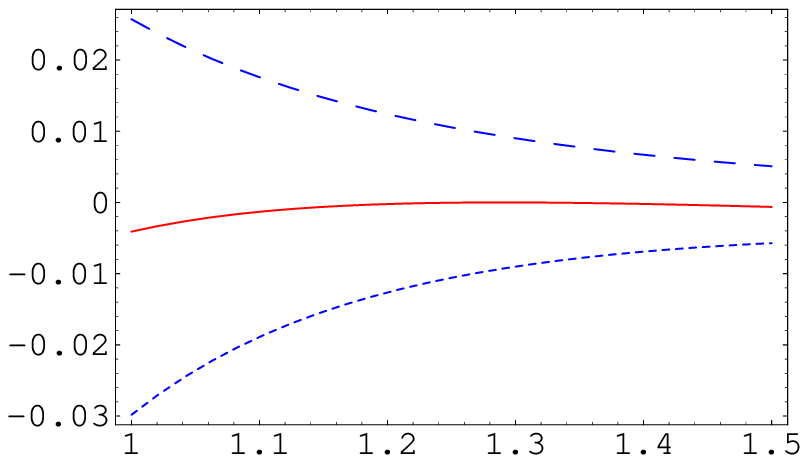,width=8cm}}}
\put(215,-10){$M_S \ ({\rm GeV})$ }
\put(-40,125){$ f_{p^6}^{\rm tree} (M_\rho)  $ }
\put(220,60){$M_P=$ 1.3 GeV }
\put(110,90){{\small
$L_5 \times L_5 / F_\pi^2 $
}}
\put(100,30){{\small
$- (C_{12} + C_{34})$
}}
\end{picture}

\vspace*{.2cm} 
\caption{Tree-level contribution of $O(p^6)$ to $f_+^{K^0 \pi^-} (0)$
and its two components \cite{Cirikl3}.}
\label{fig:fp6tree}
\end{figure}
\end{center}

\vspace*{.3cm} 
\noindent 
The last equation is based on a large-$N_c$ estimate of
Ref.~\cite{Cirikl3}. As can be seen in Fig.~\ref{fig:fp6tree}, two
terms interfere destructively weakening the overall dependence on the
scalar resonance mass $M_{S}$. The same
interference leads to a rather modest scale dependence of the result
for $M_{\eta} \le \mu \le 1$ GeV.

The final result, including the
uncertainty due to a possible second pseudoscalar multiplet
$P^{\prime}$, is \cite{Cirikl3}
\begin{eqnarray} 
f_{p^6}^{\rm tree} (M_\rho) &=& - 0.002  \pm 0.008_{\, 1/N_c} \pm 
0.002_{\, M_S} \,\mbox{}_{- 0.002}^{+0.000} \,\mbox{}_{\, P^\prime}
\nn[.1cm] 
f_{p^6} &=& 0.007 \pm 0.012 \no \\[.1cm] 
f_{+}^{K^0 \pi^-} (0) &=&  0.984 \pm 0.012~.
\end{eqnarray}


\noindent 
With our large-$N_c$ estimate of the tree-level contribution of
$O(p^{6})$, $f_{p^6}$ is dominated by the loop part
(\ref{eq:fp6loop}). It exhibits less $SU(3)$ breaking than the
well-known result of Leutwyler and Roos \cite{LeutRoos} and a recent
lattice estimate \cite{Becirevic}. Taking the most recent value of
$V_{ud}$ and assuming unitarity of the CKM matrix, the predictions can 
be compared directly with the
experimental results as shown in Fig.~\ref{fig:Kl3data}. A new
result for the neutron lifetime \cite{Serebrov} would imply a shift
of all theoretical values in Fig.~\ref{fig:Kl3data} to the left but 
the corresponding accuracy of $V_{ud}$ is not yet competitive with 
super-allowed nuclear Fermi transitions. Another way to read 
Fig.~\ref{fig:Kl3data} is that the estimate of Ref.~\cite{Cirikl3} 
leads to a value of $V_{us}$, 
\begin{equation}
|V_{us}| = 0.2208 \pm 0.0027_{f_+(0)} \pm 0.0008_{\rm exp}~,
\end{equation}
that is a bit smaller than the unitarity prediction.

Finally, the slope of the scalar form factor that also depends on the
LECs $C_{12}, C_{34}$ can also be
predicted \cite{Cirikl3} and it is in 
good agreement with the recent most precise experimental value
\cite{KTeV}: 
\begin{equation}  
\begin{tabular}{lcl}
$\lambda_0 = (13 \pm 3)\cdot 10^{-3}$ & \mbox{} \hspace*{2cm}  & 
Cirigliano et al. ~\cite{Cirikl3} \\[.2cm] 
$\lambda_0 = (13.72 \pm 1.31)\cdot 10^{-3}$ &  & KTeV ~\cite{KTeV}
\end{tabular}  
\end{equation}

\section{Summary and epilogue}
There is an amazing richness contained in the simple Lagrangian that
we ``derived'' from the existence of colour and from the gauge
principle: 
\begin{eqnarray*}  
\cL_{\rm QCD} = 
- \displaystyle\frac{1}{2} {\rm tr}
(G_{\mu\nu} G^{\mu\nu}) + \displaystyle\sum_{f=1}^{N_F}
\overline{q}_f \left(i \slashed{D} - m_f \mathbbm{1}_c \right) q_f ~.
\end{eqnarray*} 
There is in fact no better summary for these lectures.

It may seem a long way from the naive quark model to QCD but it
all happened in less than ten years. On the asymptotically free side,
perturbative QCD is a complete success and it will be especially 
needed for understanding the background for new physics at the LHC and
beyond. 

There is much more left to be understood at the other end of the
scale. If confinement is really forever, we would witness the first
case in the history of physics when new constituents of matter have
been identified beyond reasonable doubt and yet they can never be
isolated. The question sounds preposterous but it is well supported:
have we already arrived at the innermost structure of hadrons? Or put in
a different way, is there no further structure to be expected before
strings and quantum gravity eventually take over? In a 
short time, once the LHC will produce first results, we may learn how 
to rephrase the question.

\section*{Acknowledgements}
I want to thank both the organizers and the participants for creating
such a lively and inspiring atmosphere during the School. Special
thanks to Toni Pich for allowing me to use some of his figures in
Ref.~\cite{Pich:1999yz}.

\section*{Bibliography}

M.E. Peskin and D.V. Schroeder, \emph{An Introduction to Quantum Field
  Theory} (Addison-Wesley, 1995). \\[.1cm]
J.F. Donoghue, E. Golowich and B.R. Holstein, \emph{Dynamics of the
  Standard Model} (Cambridge University Press, Cambridge,
  1994). \\[.1cm]
M.H. Seymour, \emph{Quantum Chromodynamics}, Lectures given at the 2004
  European School of High-Energy Physics, arXiv:hep-ph/0505192. \\[.1cm]    
A. Khodjamirian, \emph{QCD and Hadrons: an Elementary Introduction}, 
  Lectures given at the 2003 European School of High-Energy Physics, 
  arXiv:hep-ph/0403145. \\[.1cm]
Yu. L. Dokshitzer, \emph{QCD Phenomenology}, Lectures given at the
  2002 European School of High-Energy Physics, 
  http://doc.cern.ch/yellowrep/2004/2004-001/p1.pdf.

\end{document}